\documentclass[letterpaper,12pt]{article}
\pdfoutput=1 
\usepackage[bottom]{footmisc} 
\usepackage{amssymb}
\usepackage{amsthm}
\usepackage{newtxmath}
\usepackage{microtype} 
\usepackage[utf8]{inputenc}
\usepackage{newtxtext} 
\usepackage{setspace}
\usepackage{lineno,xcolor}
\usepackage{graphicx}
\usepackage{float}
\usepackage{printlen}
\usepackage{relsize}
\usepackage[hidelinks]{hyperref}
\usepackage{pdfpages}
\usepackage{makecell} 
\usepackage{placeins} 
\usepackage{rotating} 
\usepackage{tablefootnote} 

\usepackage[top=1.5in, bottom=1.5in, left=1in, right=1in]{geometry}
\usepackage[square,numbers,sort&compress]{natbib}
\usepackage[font={small}]{caption}
\usepackage{multirow}
\usepackage{url}

\usepackage{abstract} 

\usepackage{fancyhdr}
\fancyheadoffset{0cm}
\usepackage{titlesec} 
\usepackage{authblk} 

\def\fwXmachina{\href{http://fwx.pitt.edu}{\texttt{\textsc{fwXmachina}}}}
\def\fwX{\fwXmachina}
\def\TMVA{\href{http://root.cern/manual/tmva/}{\texttt{\textit{T}\textsc{MVA}}}}

\begin{document}

\thispagestyle{empty}
\title{
    \vspace{-15mm}
    \begin{flushright}{\large PITT-PACC-2103-v3.1}\end{flushright}
    \vspace{10mm}
    \fontsize{20pt}{10pt}\selectfont\textbf{
    Nanosecond machine learning event classification with
    boosted decision trees in FPGA for high energy physics}
}

\author[]{\fontsize{16pt}{10pt}\selectfont T.M.\ Hong\thanks{Corresponding author, tmhong@pitt.edu}}
\author[]{\ B.T.\ Carlson}
\author[]{B.R.\ Eubanks}
\author[]{S.T.\ Racz}
\author[]{S.T.\ Roche}
\author[]{\\ \vspace{-14pt}J.\ Stelzer}
\author[]{D.C.\ Stumpp}
\affil[]{\large Department of Physics and Astronomy\\
University of Pittsburgh
}
\date{\today}

\maketitle
\vspace{-5mm}
\begin{abstract}
\noindent
    We present a novel implementation of classification using the machine learning
    / artificial intelligence method called boosted decision trees (BDT) on field programmable gate arrays (FPGA).
    The firmware implementation of binary classification requiring $100$ training trees with a maximum depth of $4$ using four input variables gives a latency value of about $10\,\textrm{ns}$,
    independent of the clock speed from $100$ to $320\,\textrm{MHz}$ in our setup.
    The low timing values are achieved by restructuring the BDT layout and reconfiguring its parameters.
    The FPGA resource utilization is also kept low at a range from $0.01\%$ to $0.2\%$ in our setup.
    A software package called \fwXmachina\ achieves this implementation.
    Our intended user is an expert in custom electronics-based trigger systems in high energy physics experiments or anyone that needs decisions at the lowest latency values for real-time event classification.
    Two problems from high energy physics are considered,
    in the separation of electrons vs.\ photons and
    in the selection of vector boson fusion-produced Higgs bosons vs.\ the rejection of the multijet processes.
\end{abstract}
\vspace{18pt}
    Trigger concepts and systems (hardware and software).
\vfill

\newpage
\tableofcontents

\setlength\linenumbersep{15pt}
\renewcommand\linenumberfont{\normalfont\footnotesize\sffamily\color{gray}}
\modulolinenumbers[1]


\section{Introduction}

Modern high energy physics experiments are saving more data at a faster rate than ever before.
The combination of the Large Hadron Collider (LHC) \cite{Evans:2008zzb} that collides proton bunches at $40\,\textrm{MHz}$ and an apparatus to record the energy deposits and their patterns coming from the collisions,
such as the ATLAS and CMS experiments 
\cite{Aad:2008zzm,
    Chatrchyan:2008aa},
leads to a large data volume at a high rate.
Without a reduction of incoming data the volume is prohibitively large.
Furthermore,
most data acquisition systems have limitations on the readout electronics that do not allow the data to be saved at the collision rate.
The combined issues of data storage and data acquisition readout limitations coupled with the fact that the collisions of interest occur at a small fraction of the total number of collisions,
typically one part in a few thousand,
necessitate the need of a multi-level trigger to help throttle the data 
\cite{Aad:2012xs,
    Aaboud:2016leb,
    Khachatryan:2016bia}.\footnote{
    A list of abbreviations and technical terms is given in appendix \ref{appendix:terminology}.
}

The first level (level-1) trigger has stringent demands that it produces a decision within a few microseconds depending on the system requirements,
or a fraction of the total latency budget for a given algorithm within the system.
The level-1 triggers typically consist of custom electronics boards with field programmable gate array (FPGA) and/or application specific integrated circuit (ASIC) chips that process a reduced amount of the total data generated by the collision 
\cite{Achenbach:2008zzb,
    CMS:2017gbu,
    Sirunyan:2020zal,
    Aad:2020hnu}.
The algorithms on the chips are often simplified versions of what is implemented in software.
For example,
in the ATLAS experiment,
the sliding window algorithm \cite{Achenbach:2008zzb} is implemented on the chip for hadronic jet reconstruction compared to the anti-k$_{t}$ algorithm \cite{Cacciari:2008gp} that is implemented in software.
In many of the cases,
algorithms on the chip involve cut-based requirements when multiple variables are computed for a given algorithm,
such as the isolation requirements for electrons \cite{Aad:2019wsl}.
There are also examples of increasingly sophisticated algorithms in the level-1 trigger.
For example,
the ATLAS level-1 topological trigger computes the invariant mass of the two-jet system \cite{Aad:2021cqq}.
More recently,
machine learning (ML) / artificial intelligence (AI) algorithms have started to make an appearance at level-1
\cite{Neuhaus:2014yma,
    Acosta:2018hjs}.

Among the popular ML algorithms in high energy physics are boosted decision trees (BDT) and neural networks.
For the past few decades,
such methods have been used in the analysis of the data recorded by experiments
\cite{Verkerk:1990bh,
    Verkerk:1992gx,
    Abazov:2006gd,
    Bevan:2014iga},
e.g.,
in the discovery of the Higgs boson 
\cite{Aad:2012tfa,
    Chatrchyan:2012ufa}.
Furthermore, ML algorithms have been utilized in the reconstruction of low-level detector information to produce physically meaningful variables \cite{Aad:2015ydr,
    Aaboud:2018xwy,
    ATLAS:2019fxb,
    ATLAS:2020efs,
    ATLAS:2019uhp}.
A variety of ML algorithms have been implemented in the level-1 and the subsequent software-based high level trigger of the ATLAS and CMS experiments
\cite{Aaboud:2016leb,
    Khachatryan:2016bia,
    Acosta:2018hjs}.
This demand for ML is driven by the need for higher signal sensitivity for event classification in an environment where the signal-to-background ratios are lower.
Furthermore, with the recent advancements in the FPGA size,
as well as progress in ML,
implementing more advanced algorithms on FPGA and ASIC chips has become an active area of research in high energy physics 
\cite{Neuhaus:2014yma,
    Duarte:2018ite,
    Acosta:2018hjs,
    Summers:2020xiy,
    DiGuglielmo:2020eqx,
    Iiyama:2020wap,
    Heintz:2020soy,
    Coelho:2020zfu,
    John:2020sak,
    Fahim:2021cic,
    Aarrestad:2021zos,
    Hawks:2021ruw}.
This area is also lively outside of high energy physics
\cite{NorthwesternBDT,
    DT-Sea,
    Decision-Char,
    DT-Elkanishy,
    jlpea1010045,
    DT-Tong,
    DT-Owaida,
    ML-Kara,
    DT-Lin}.
It is important to note that the implementations in trigger systems focus on the \emph{classification}
(also known as \emph{application}, \emph{evaluation}, or \emph{inference})
of ML to use in real-time systems,
rather than in the training of the ML.

We present a novel implementation of the evaluation of BDT on FPGA that focuses on speed and design simplicity.
The low timing values are achieved by restructuring the BDT layout and reconfiguring its parameters after the ML training step.
The software package called \fwXmachina\ produces the optimized BDT for High Level Synthesis (HLS),
which it subsequently converts to firmware in hardware description language.

Two binary classification problems are considered to give realistic use cases as well as to illustrate the performance of the design.
The first problem is in object identification in order to discriminate electrons vs.\ photons.
The second problem is in the event classification of the vector boson fusion-produced Higgs boson vs.\ the multijet process.

The paper is organized as follows.
The rest of this section gives the overview of \fwXmachina.
The next three sections
describe the stages of implementing a BDT in firmware and validating its results.
Sections \ref{sec:train},
\ref{sec:nano_opt},
and \ref{sec:firmware}
discuss the ML training,
Nanosecond Optimization,
and firmware design,
respectively.
Section \ref{sec:perf_cost} gives the physics performance and the FPGA cost of running the algorithm on the physical FPGA.
Finally, section \ref{sec:conclusion} concludes.
Within the paper,
the following parts may be of particular interest to some readers.
The set of benchmark parameters are defined in section \ref{sec:benchmark} and parameter scans in section \ref{sec:perf_scan}.
Comparisons with existing tools are made in section \ref{sec:compare} and appendix \ref{appendix:compare}.

\subsection{\fwXmachina, software package for ML/AI classification on FPGA}

\begin{figure}[htbp!]
\centering
\includegraphics[width=0.90\textwidth]{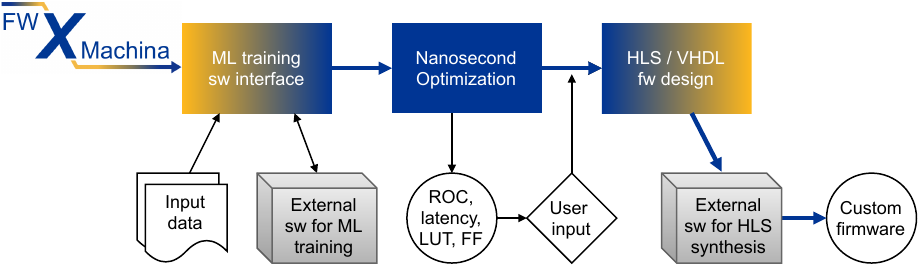}
\caption{
    \label{fig:fwx} 
    Workflow diagram for the \fwXmachina\ package.
    The flowchart reads from left to right following the thick arrows that connect the three main stages:
    ML training with the software (SW) interface, Nanosecond Optimization, and firmware (FW) design.
    The interactions with external inputs (shown in flat white figures)
    and external software packages (shown in gray cubes)
    are shown by thin vertical arrows.
    Nanosecond Optimization is shown in more detail later in figure \ref{fig:nano_opt}.
    The information flow for the user input is diagrammed in the appendix (figure \ref{fig:testbench}).
}
\end{figure}

\fwXmachina\ is a software package that consists of three sequential stages:
ML training,
Nanosecond Optimization,
and firmware design.\footnote{
    The source code and the technical documentation are available at \href{http://fwx.pitt.edu}{fwx.pitt.edu}.
}
The workflow showing the structure and interaction with external data and external input is shown in figure~\ref{fig:fwx}.

There are various approaches in the literature to optimize the result of the ML training for efficient firmware implementation.
One approach is to consider a penalty term in the feedback loop during the training process to balance the FPGA cost with physics performance \cite{Duarte:2018ite}.
Another approach is to optimize the result of the training process with firmware implementation in mind,
such as to choose the precision of the input variable values \cite{Summers:2020xiy}.
We expand on the latter approach in Nanosecond Optimization.

The first stage is ML training,
where the structure of the ML and its parameters are determined.
The training is done with available open source external software packages.
In this paper, 
we use BDT as the ML method and \emph{Toolkit for Multivariate Data Analysis} (\TMVA) for the external software package \cite{Hocker:2007ht}.\footnote{
    We plan to support {\footnotesize\texttt{scikit-learn}} \cite{Pedregosa:2012toh} and other such packages in the near future.
}
The current version of the code supports binary classification.
The user provides \fwX\ the data samples with input variables that characterize each event.
The output of \TMVA\ is handed off to the next stage.

The second stage is Nanosecond Optimization,
in which the structure of the BDT and its parameters are optimized for both physics performance and FPGA cost.
The physics performance is evaluated by considering the receiver operating characteristics (ROC) curves.
The ``FPGA cost'' is evaluated by the timing values and the resource utilization using Xilinx Vivado HLS \cite{xilinx-hls:2020}.
At this point,
the user chooses the working point that best suits the problem at hand.
The performance considerations (center circle in figure \ref{fig:fwx}) and user input (diamond) is part of Nanosecond Optimization.

The third stage is firmware design.
The inputs to Xilinx Vivado are created 
\cite{xilinx:2020,Ghanathe:2017dpm}.
The inputs are a combination of HLS and hardware description language (HDL).
We target VHDL, a type of HDL, for the output.
The output,
after synthesizing with Vivado,
is the firmware in bitstream format to be programmed on to the FPGA.
After the programming,
the FPGA is prepared to repeatedly execute the algorithm on incoming unclassified events that are fed to it.

\section{ML training}
\label{sec:train}

The ML training stage is executed by external packages as described in the previous section.

For the problem of binary classification of signal vs.\ background in a \emph{supervised learning} environment,
a given ML method needs to be trained using samples containing events labeled as ``signal'' or ``background.''
The training process starts with an initial set of parameters for the chosen ML architecture,
such as the decision tree structure for BDT and layer structure for neural network,
that is iteratively improved by a feedback loop consisting of a metric.

We emphasize that in the level-1 trigger for high energy physics,
the training step is typically done before the real-time evaluation.
The latency requirement of the level-1 trigger is not a constraint for the training step that uses training samples that are prepared beforehand.
In contrast,
for the operating conditions involving incoming data at high speeds,
e.g., 
$40\,\textrm{MHz}$ at the LHC,
the algorithms of the level-1 trigger must make decisions at time scales of microseconds,
or a fraction thereof,
depending on the algorithm's requirements within the level-1 trigger.

While the ML method of interest of this paper is BDT,
comparisons are made to the cut-based method.
Cut-based classification,
also referred to as ``rectangular cuts,''
is discussed for two reasons.
The first is that the BDT result is compared against cut-based results.
The second is that we have implemented it in \fwX.

Two problems are considered.
The ``object identification'' and the ``physics trigger'' are used to evaluate the firmware performance in different ways.
The former is used to define the benchmark configuration (section \ref{sec:benchmark}) and to scan one parameter at a time starting from the benchmark  (section \ref{sec:perf_scan}).
In contrast,
the latter is used to evaluate configurations with many parameters far from the benchmark  (section \ref{sec:far_benchmark}).

\subsection{Electron vs.\ photon}

The example considered for the ``object identification'' problem is to separate electrons vs.\ photons.
Studies are done with the \fwX\ software to make comparisons of approximations that are made for the FPGA implementation.
For example,
ROC curves for varying number of bits for the input variable values are compared.

The electron-photon problem is interesting in high energy physics,
especially for the level-1 trigger.
Due to latency constraints the level-1 trigger typically receives only a fraction of the calorimeter data containing localized energy deposits without much,
if any,
additional tracker data.
This is a challenge because whereas the electron leaves ionizing energy deposits in the tracker due to its electrical charge,
the photon leaves no such pattern as it passes through the tracker until it hits the calorimeter.
In the calorimeter,
both an electron and a photon leave similar patterns of energy deposits.
Some differences arise where electrons and photons deposit energy in the calorimeter as the electromagnetic shower develops laterally through the detector material.
An electron deposits energy primarily by bremsstrahlung radiation and a photon deposits energy mainly by pair production \cite{Fabjan:2020ixe}.
Therefore,
an electron tends to deposit its energy towards the beginning of its entry to the calorimeter, 
while a photon deposits its energy at least one radiation length later in its passage through the detector material.
These small differences are difficult to distinguish using the traditional cut-based method and may be better suited for ML \cite{Andrews:2018nwy,deOliveira:2018lqd}.

\begin{figure}[htbp!]
\centering
\includegraphics[width=0.50\textwidth]{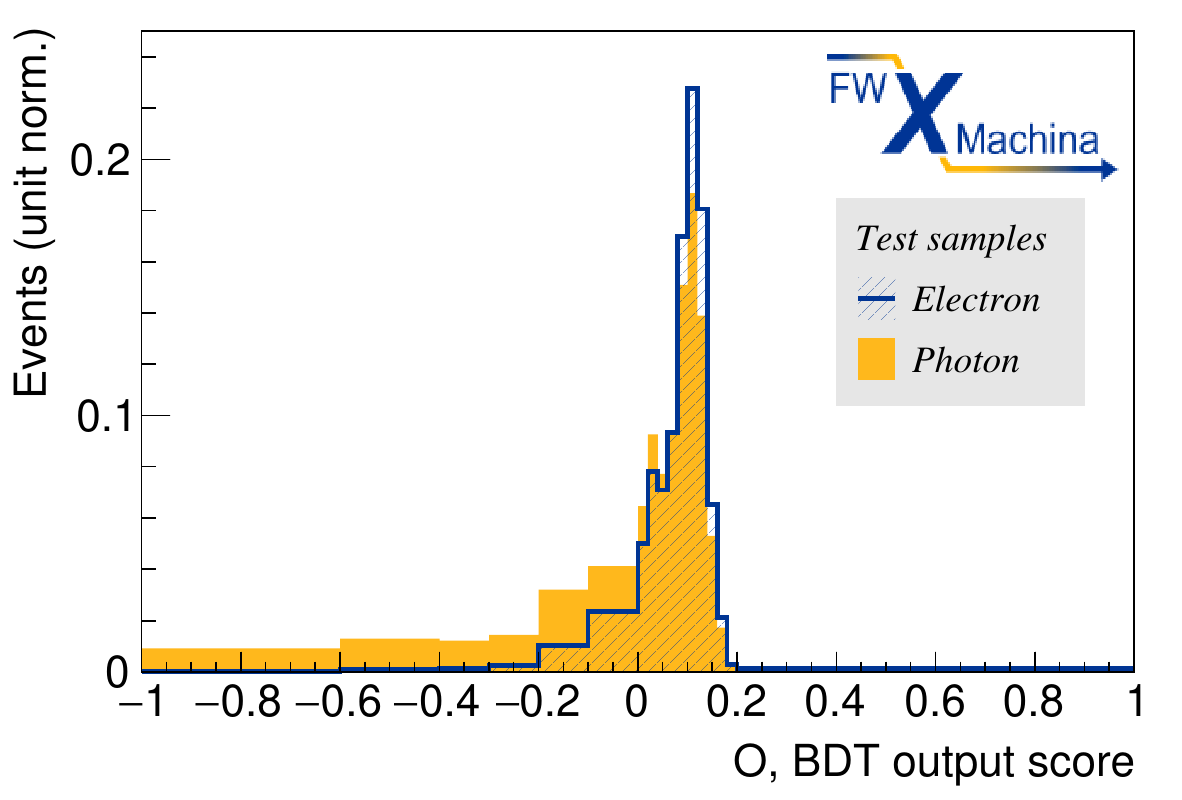}%
\includegraphics[width=0.50\textwidth]{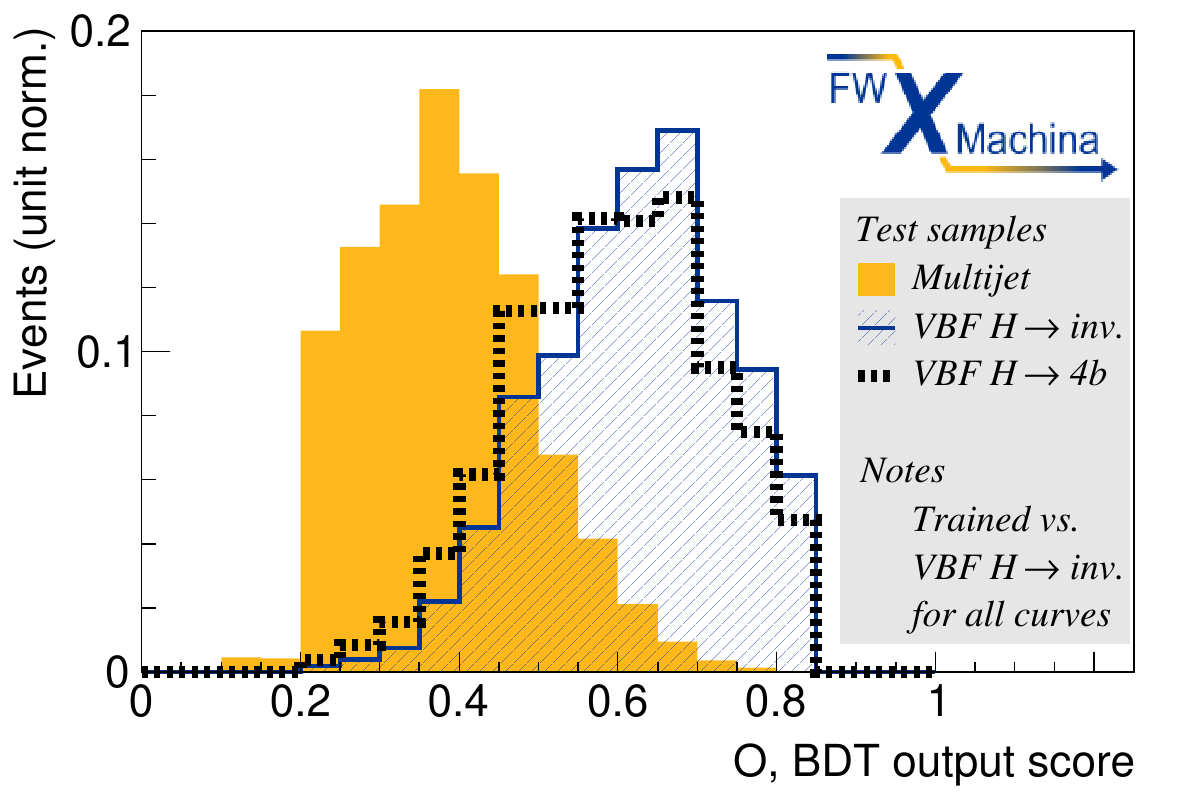}
\caption{
    \label{fig:train_bdt} 
    BDT output score distribution for electron vs.\ photon (left)
    and VBF Higgs vs.\ multijet process (right).
    The left figure uses the yes/no leaf metric that results in a range between $-1$ and $1$.
    The right figure uses the purity metric that results in a range between $0$ and $1$.
    For the right plot, the following details are notable.
    One BDT is trained for the binary classification problem of VBF $H\rightarrow\textit{invisible}$ vs.\ multijet.
    This BDT is then used to evaluate the output score of events in the three samples.
    Events with less than two jets are not plotted,
    but they are considered when making the ROC curves (figure \ref{fig:roc_ey}).
}
\end{figure}

The BDT architecture is trained using the adaptive boost (AdaBoost) metric \cite{Freund:1997xna}.
Since we compare our BDT results against the cut-based method,
the latter is trained using the genetic algorithm \cite{Shapiro:2001} .
The data samples of simulated events are obtained from ref.\ \cite{Mendeley:2017},
with one half being used for training and the remaining half is used for testing.
The BDT output score distributions for the test sample are shown in the left plot of figure \ref{fig:train_bdt} for the benchmark configuration described later (section \ref{sec:benchmark}).
The similarity in the trend for the two distributions in the plot demonstrates that this is a difficult problem.
However,
the shape differences in the tails of the distributions allow for a superior separation compared to the cut-based method.

The details of the training setup,
the data samples,
and the input variables are given in appendix \ref{appendix:train_ey}.

\subsection{VBF Higgs bosons vs.\ multijet process}

The example considered for the ``physics trigger'' problem is to discriminate between an event containing a vector boson fusion-produced Higgs boson (VBF Higgs) vs.\ multijet process.
The former is considered ``signal'' and the latter is considered ``background.''

The VBF Higgs vs.\ multijet problem is challenging at the LHC,
especially in the level-1 trigger.
We consider an algorithm to identify the signal by using the two ``VBF jets,''
typically with a large gap in pseudorapidity,
that emerge from the process.
The challenge for such an algorithm is that the two-jet sample is dominated by the multijet process at the LHC.
Such an algorithm is of interest to the level-1 trigger,
because it allows the selection of events containing Higgs bosons in a way that is agnostic to the decay pattern of the Higgs boson
\cite{Aaboud:2018sfi,
    Sirunyan:2018owy,
    ATLAS:2020cjb,
    CERN-LHCC-2020-004}.
In particular,
ref.\ \cite{CERN-LHCC-2020-004} considers this problem using a neural network.

The training setup is similar to the setup for electron-photon,
with $100$ training trees at a maximum depth of $4$.
We generated the data samples using publicly available tools (appendix \ref{appendix:train}).
Half of the data is used for training and the remaining half is used for testing.
Later in section \ref{sec:far_benchmark},
we compare the result of our BDT analysis against results using the cut thresholds for the ATLAS experiment.
The BDT is trained for VBF $H\rightarrow\textit{invisible}$ vs.\ multijet,
but it is used to categorize 
VBF $H\rightarrow\textit{invisible}$,
and multijet.
As the goal is to use only the VBF jets to identify the production of Higgs bosons,
we also test the trained BDT on the signal process VBF $H\rightarrow 4b$,
which contains jets in the decay process in addition to the one from the production process. 
The BDT output score distributions for the test sample is shown in the right plot of figure \ref{fig:train_bdt} for the optimized configuration described later (section \ref{sec:benchmark}).
The clear shape differences between the two signal distributions and the background distribution indicate that there are powerful input variables that separate them.
Moreover,
the distributions demonstrate that VBF Higgs events can be separated from multijet background in a way that retains sensitivity for two different Higgs boson decays.
There is a slight degradation in performance when the Higgs boson decays to a final state with jets, such as $4b$,
as these jets can be miscategorized as VBF jets.

\section{Nanosecond Optimization}
\label{sec:nano_opt}

Given the experimental constraints at the LHC and other real-time systems,
such as latency and resource limitations in FPGA-based triggers,
our question is how can we best implement ML in firmware to achieve the most optimal physics performance?
In this section we describe the restructuring of the BDT layout as well as the reconfiguration of its parameters.

\begin{figure}[htbp!]
\centering
\includegraphics[width=0.80\textwidth]{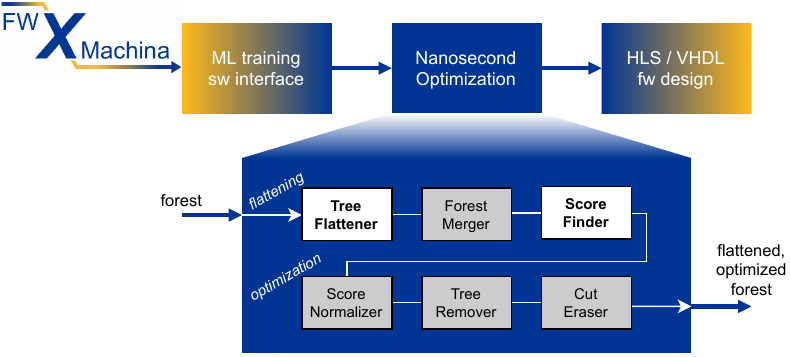}
\caption{
    \label{fig:nano_opt} 
    Workflow of Nanosecond Optimization.
    The four boxes shaded in grey denote steps that are not required to have functional firmware, but are often critical in improving the firmware's performance.
}
\end{figure}

Nanosecond Optimization is done in six sequential steps:
\textsc{Tree Flattener},
\textsc{Forest Merger},
\textsc{Score Finder},
\textsc{Score Normalizer},
\textsc{Tree Remover},
and \textsc{Cut Eraser}.
These steps are illustrated in figure \ref{fig:nano_opt}.
The first three steps can be grouped conceptually to achieve ``flattening'' and the second three to achieve ``optimization.''
The first two steps drive the firmware design whereas the remaining four steps are not as critical to the design.
Therefore, the two in the former group are explained in more detail in the rest of the section,
while those in the latter group are described in appendix \ref{appendix:nano_opt}.

\subsection{\textsc{Tree Flattener}}
\label{sec:tree_flatten}

The \textsc{Tree Flattener} converts a tree with maximum depth $D$ to a tree with depth $1$.

\begin{figure}[htbp!]
\centering
\includegraphics[width=0.95\textwidth]{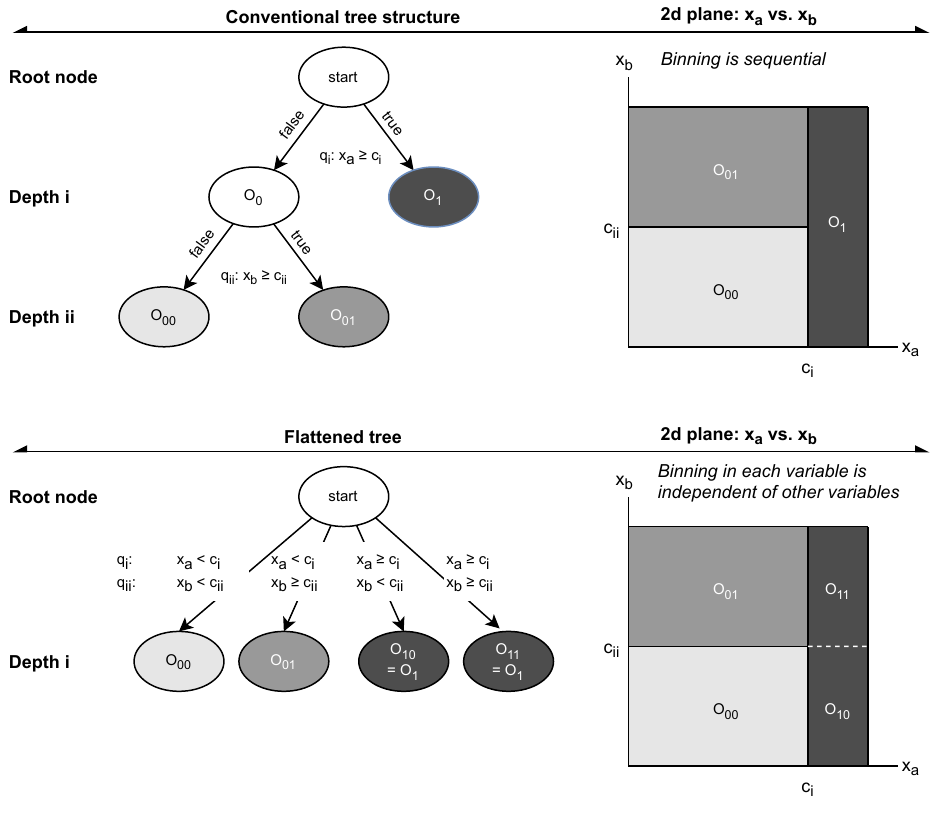}
\caption{
    \label{fig:tree_flatten}
    \textsc{Tree Flattener} example.
    A tree is flattened by compressing the vertical structure into a tree of depth of $1$.
    The conventional tree (top row) is flattened to the one depth structure (bottom row) by the insertion of ``ghost'' dotted lines that allow for the thresholds in each variable to be considered independently.
    The diagram considers an example of a one tree structure with two variables ($x_a, x_b$) and one cut threshold for each variable ($c_{i}, c_{ii}$).
    The node structure (left column) shows the decision making process, where the final terminating leaves contain the BDT output score $O$.
    The conventional node structure is an iterative procedure according to the number of depth of the tree.
    The graphical representation (right column) on the two-dimensional plane of ($x_a$ vs.\ $x_b$) shows the range defined by each threshold.
    For the graphical representation of the flattened tree (bottom right), a ``ghost'' dotted line is inserted---corresponding to the $O_{10}$ and $O_{11}$ terminating leaves in the node structure (bottom left).
}
\end{figure}

A decision tree is characterized by a binary structure that considers
$V$ variables, $\vec{x}=\{ x_0, \cdots, x_{V-1} \}$, per event.\footnote{
    More detail on the notation is given in table \ref{table:notation} in appendix \ref{appendix:terminology}.
}
An event recurses down the tree starting at the root node and ends in one of the terminating leaf nodes.
An example graphical representation of a tree with $D=2$ and $V=2$ is shown in the top-left diagram of figure \ref{fig:tree_flatten}.
Two comparisons $q_{i}$ and $q_{ii}$ place an event into one of three leaves with the corresponding BDT output scores $O$ in the following if-then-else structure:
\begin{itemize}
    \itemsep 0pt
    \item If $x_a \ge c_i$, then terminate at $O_{1}$
    \item Else
    \begin{itemize}
        \itemsep 0pt
        \item If $x_b \ge c_{ii}$, then terminate at $O_{01}$
        \item Else terminate at $O_{00}$.
    \end{itemize}
\end{itemize}
We note that the second if-then-else block must wait for the decision of the first if-then-else block in a sequential manner.
The distribution of the events for the two variables $x_a$ and $x_b$ can be visualized as dots on a two-dimensional plane in the top-right diagram in figure \ref{fig:tree_flatten},
where the cut thresholds $c_i$ and $c_{ii}$ are represented by vertical and horizontal boundaries,
respectively.
In this plane three rectangular regions define $O_{1}$, $O_{01}$, and $O_{11}$.

The decision tree is flattened to be optimized for firmware implementation.
The flattener extends every cut threshold \emph{through} the entire $V$-dimensional hyperspace to form $V$-rectangular bins.\footnote{
    \label{footnote:MLtheory}
    References \cite{Giabbanelli:2015,Strecht:2015} describe a similar algorithm in ML theory.
}
The example above with $D=2$ and $V=2$ leads to the following if-then-else structure:
\begin{itemize}
    \itemsep 0pt
    \item If      $x_a\ge c_i$ and $x_b\ge c_{ii}$, then terminate at $O_{11}$,
    \item Else if $x_a\ge c_i$ and $x_b <  c_{ii}$, then terminate at $O_{10}$,
    \item Else if $x_a <  c_i$ and $x_b\ge c_{ii}$, then terminate at $O_{01}$,
    \item Else if $x_a <  c_i$ and $x_b <  c_{ii}$, then terminate at $O_{00}$.
\end{itemize}
We note that,
in contrast to the above if-then-else,
the four comparisons are made simultaneously in parallel.
There are four ``bins,''
$N_\textrm{bin}=B=4$,
that correspond to each BDT output score.
The scores $O_{11}$ and $O_{10}$ correspond to the same value as $O_1$ in the above example.
The bottom two diagrams of figure \ref{fig:tree_flatten} illustrates the impact of flattening.

In summary,
the flattener transforms the operation from a recursion problem into a binning problem.
This is efficient on an FPGA because each variable can be binned in parallel.
Now that a tree is flattened into one with a single depth with $B$ bins,
two binning algorithms are available in \fwX.
The details of the binning algorithms are given in appendix \ref{appendix:binning}.

\subsection{\textsc{Forest Merger}}
\label{sec:forest_merge}

The \textsc{Forest Merger} combines multiple decision trees into one tree.{\footnotesize $^{\ref{footnote:MLtheory}}$}

\begin{figure}[htbp!]
\centering
\includegraphics[width=1\textwidth]{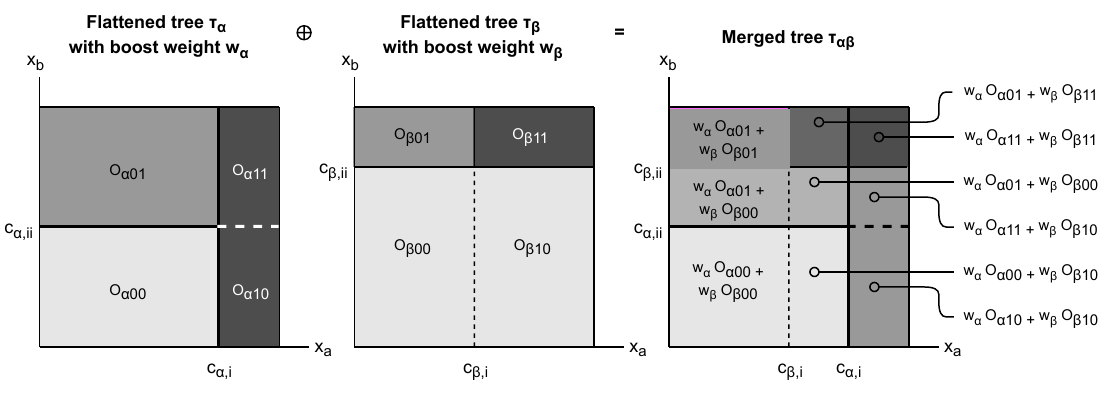}
\caption{
    \label{fig:forest_merge} 
    \textsc{Forest Merger} example.
    Visual binning example for two variables ($x_a$ vs.\ $x_b$) to demonstrate the merging of two flattened tree structures into one tree.
    The example given here continues the two-dimensional representation of figure~\ref{fig:tree_flatten}.
    Two trees are considered ($\tau_\alpha$, $\tau_\beta$),
    corresponding to the left and middle diagram, respectively.
    In each region defined by the thresholds $c$,
    the output scores are printed as,
    for instance,
    $O_{\alpha 00}$ that correspond to tree $\tau_\alpha$ and bin $00$.
    The vertical and horizontal lines of $\tau_\alpha$ and $\tau_\beta$ are superimposed on to one two-dimensional plane in the merged $\tau_{\alpha\beta}$ diagram on the right.
    The output score in each region of the merged tree is the weighted average in the rectangular region defined by the combined thresholds.
    The colors are shaded to suggest the output score value. 
}
\end{figure}

Consider a forest $\phi$ that contains $T$ decision trees,
$\vec{\tau} = \{\tau_0, \cdots, \tau_{T-1} \}$,
each with a set of corresponding boost weights,
$\{W_0, \cdots, W_{T-1}\}$.
It is easier to describe the procedure with an example.
We consider two trees, $T=2$, each considering two variables,
$V=2$.
Suppose the first tree $\tau_\alpha$ is the same one as in figure \ref{fig:tree_flatten} and the second one $\tau_\beta$ is similar with different cut threshold values.
The graphical representation of the two-dimensional distribution is given in the first two diagrams of figure \ref{fig:forest_merge}.

The merged tree $\tau_{\alpha\beta}$ considers the cut thresholds from both of the trees simultaneously.
The combined BDT output score in each bin is the weighted sum using the normalized boost weights $w_t = W_t / \sum_{t'}W_{t'}$.
The two-dimensional representation of $\tau_{\alpha\beta}$ is given on the right hand side of figure \ref{fig:forest_merge}.

The flattened structure of a merged tree is indistinguishable from a single tree.
They both consist only of the set of cut thresholds and a set of event-independent boost weights.
In practice,
we typically merge a forest down to a handful of trees instead of a single tree,
i.e.,
$\phi \rightarrow \{\tau_0, \ldots, \tau_{F-1}\}$ with $F\le T$.
For example,
a forest of $100$ trees is often divided into ten groups of $10$ trees;
with each group of tree merged into a single tree.
This smaller set of $10$ ``final trees'' is fed to the \textsc{Score Finder},
which associates an output score to each bin of the tree.

\section{Firmware design}
\label{sec:firmware}

\begin{figure}[htbp!]
\centering
\includegraphics[width=0.77\textwidth]{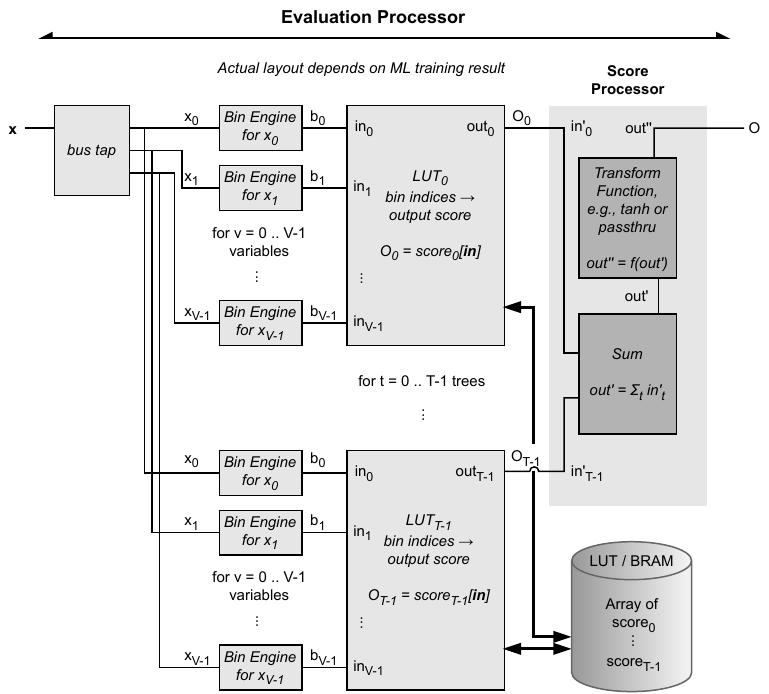}
\caption{
    \label{fig:eval_proc}
    Example layout of the \textsc{Evaluation Processor} that implements the BDT.
    The dataflow is left to right with a set of $N$-bit integers $\vec{x}$ as input.
    Look up table LUT$_t$ corresponds to decision tree $\tau_t$ ($t=0,\ldots,T-1$);
    a \textsc{Bin Engine} obtains the bin index $b_v$ for one input variable value $x_v$ for variables $v=0,\ldots,V-1$.
    The \textsc{Score Processor} combines the output scores of each decision tree and transforms the result,
    if necessary.
    The thick lines and arrows indicate the latency incurred by accessing external memory,
    either external LUT or BRAM.
}
\end{figure}

The \textsc{Evaluation Processor} encodes the entire forest of the BDT as illustrated by figure \ref{fig:eval_proc}.
The diagram flows from left to right with the vector of $V$ input variables $\vec{x}$ being the input to the \textsc{Processor}.
On the other side the BDT output score $O$ exits.

First,
the bus tap does the fanout of the vector $\vec{x}$ into $V$ individual values.
Each flattened tree $\tau_t$ is represented by a look up table $\textrm{LUT}_t$ that considers the result of $V$ instances of \textsc{Bin Engines}.
Each \textsc{Bin Engine} processes one of the input variables $x_v$ whose result $b_v$ is the bin index of the flattened tree $\tau_t$.
Each $\textrm{LUT}_t$ associates the list of bin indices $\{b_0, \ldots, b_{V-1}\}$ with the output score of the tree $\tau_t$,
$O_t$.
The set of output scores from all $T$ trees is combined and transformed,
if necessary,
in the \textsc{Score Processor},
which is the firmware implementation of \textsc{Score Normalizer}.
The transformation is a trivial one for AdaBoost whereas gradient boost (GradBoost) applies the $\tanh$ function to the sum of individual response scores \cite{Friedman2001GreedyFA}.
A description of the $\tanh$ implementation is given in appendix \ref{appendix:score_normalize} (figure \ref{fig:piecewise}).
Finally,
each set is duplicated $T$ times for $T$ trees in the forest.

The heart of the \textsc{Evaluation Processor} is the \textsc{Bin Engine},
whose design localizes the data so that the output score can be assigned.
Being the most important and computationally taxing component of the circuit,
two \textsc{Bin Engines} are employed and optimized for specific situations,
corresponding to the two bin algorithms mentioned in the rest of the section.

We note here that the cut-based method is implemented separately.
The details are given in appendix \ref{appendix:fw_cutbased}.

\subsection{\textsc{Bit Shift Bin Engine}}
\label{sec:fw_bsbe}

The \textsc{Bit Shift Bin Engine} (BSBE) relies on bit shifting to achieve fast binning.\footnote{
    A short refresher of the bit shifts introduces some of the nomenclature.
    The bit-shift operator is equivalent to an arithmetic shift right (ASR) in standard nomenclature,
    i.e.,
    the binary representation of a number is shifted to the right by a specified amount and discards the least significant bit (LSB).
    This operation divides the number by a factor $r^n$, where $r$ is the radix and $n$ is the number of shifts.
    This is already common intuition in radix-10.
    That is,
    when one wishes to divide a number by a power of $10$,
    one simply has to discard that many least significant digits.
    In CPU and FPGA architecture,
    integers are represented in radix-2,
    and all the same can be said about bit-shifting being used to divide by a power of $2$.
    }
The input is the value of one input variable $x_v$.
The output is the bin index $b_v$ for that variable.

The input value is localized by decomposing a bin into the set of ``grid spaces'' that cover the range of the final bin.
To do this the binning algorithm begins by shifting the input coordinate by various amounts to calculate its index in the many layers of grids.
The BSBE yields the coordinate of the input when evaluated on the grid corresponding to the amount shifted.

\begin{figure}[htbp!]
\centering
\includegraphics[width=0.77\textwidth]{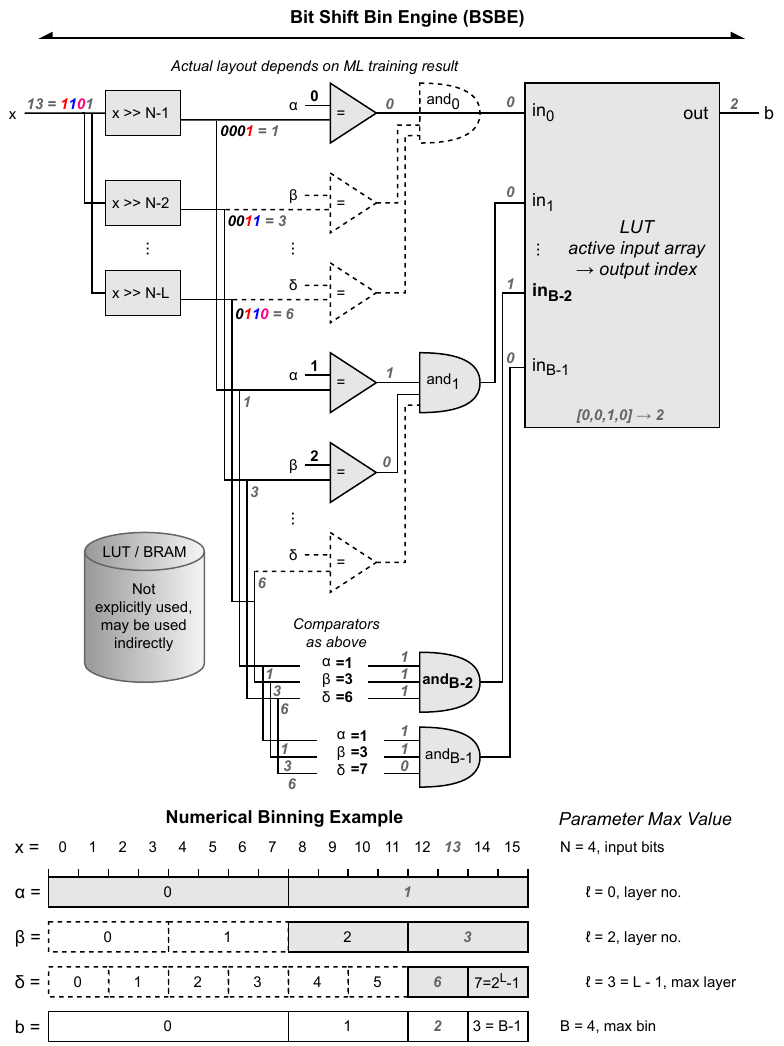}
\caption{
    \label{fig:bsbe} 
    Example gate-level diagram of the \textsc{Bit Shift Bin Engine}.
    The dataflow is left to right with an $N$-bit integer $x$ as input.
    The $x$ is binned in $L$ binary layers via bit-shift,
    comparator,
    and \texttt{AND} gates.
    The dotted elements are not present for the example considered,
    but are drawn for completeness.
    The comparator constants that correspond to each layer ($\ell=0, \ldots, L-1$) are denoted as $\alpha, \ldots, \delta$,
    respectively.
    There are $B$ copies of \texttt{AND} corresponding to the $B$ bins.
    Since only one \texttt{AND} gate (say, at position $b$) uniquely returns $\texttt{in}_b=1$ while all others return $0$,
    the list of $\texttt{in}$ is converted in a LUT via an active array to $\texttt{out}=b$.
    }
\end{figure}

The gate-level diagram at the top of figure \ref{fig:bsbe} shows an example of BSBE.
The numerical binning example that accompany it illustrates this process.
We follow the dataflow for an input value of $x=13$.
The Engine parameters are the number of bits $N=4$ and the number of layers $L=3$,
with three layers $\alpha, \beta,$ and $\delta$.
The \textsc{Cut Eraser},
in our example,
eliminates many of the possible bins to arrive at total number of bins $B=4$ for bins $b=0, 1, 2, 3$.
The erased bins are denoted by the dotted lines in the numerical example and correspond to the dotted lines in the diagram.
In the diagram, we boldface $\texttt{AND}_2$ and $\texttt{in}_2$ to highlight the activated path that leads to output $b=2$.

In general, the location of the input is then represented as an $L$-dimensional vector,
where $\ell<L$ is the grid index in which the event input is evaluated.
Every bin can thus be represented with a unique set of vector components.
The components are compared in one of the $L\cdot B$ comparators,
where $B$ represents the total number of bins for a given decision tree.
The outputs of each $L$ group of comparators are fed in to an AND operation that indicate the correct bin index in a ``one-hot'' style,
where only one AND gate returns a $1$ result and the rest return a $0$ result.
Finally,
the vector of $B$ inputs to the LUT converts the one-hot results into an output index.
In the example in the figure,
the array of $[0,0,1,0]$ is converted to index $2$ since it represents the number $0010$ in binary format.

The major benefit of this approach is the extremely low latency.
The parallel structure of the bit-shift divisions and of the vector comparisons trades a more quickly growing area usage for the decreased latency that it enjoys.
Another reason as to why the output is ready in such a short time is the fact that no memory is directly accessed.
All of the constant values that are compared to the vector components are distributed throughout the FPGA fabric and accessible without incurring access latency. 

A potential limitation,
depending on the use case,
is that the bin boundaries,
i.e.,
the cut threshold values,
are restricted to binary representation.
For users who desire more flexibility in the bin boundary values,
the threshold approach using the look up approach is presented next.


\subsection{\textsc{Look Up Bin Engine}}
\label{sec:fw_lube}

\begin{figure}[htbp!]
\centering
\includegraphics[width=0.78\textwidth]{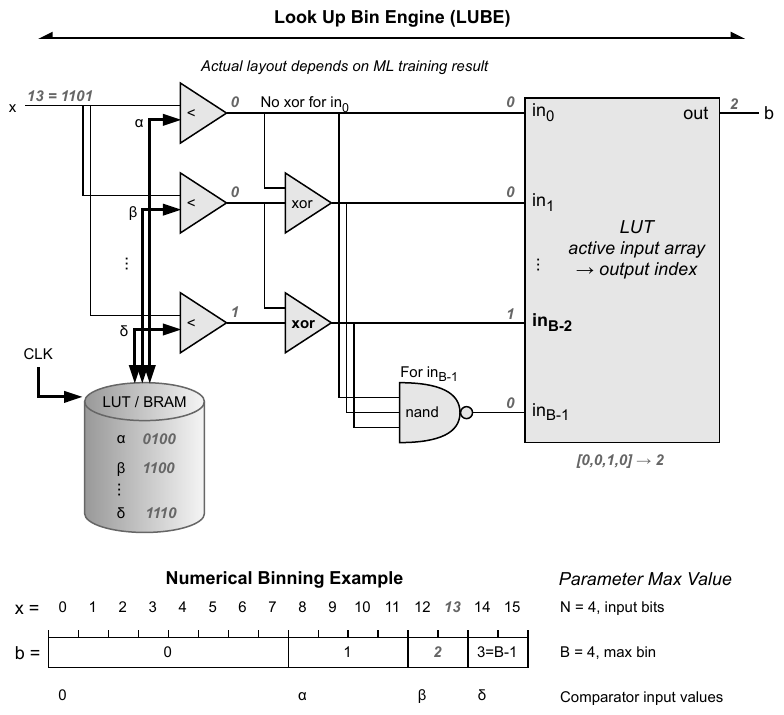}
\caption{
    \label{fig:lube} 
    Example diagram of the \textsc{Look Up Bin Engine}.
    The dataflow is left to right with an $N$-bit integer $x$ as input.
    The $x$ is binned by comparison with the threshold values $\alpha, \ldots \gamma$ in the comparators gates.
    The \texttt{NAND} gate of all of the \texttt{XOR} results gives the result of the last $B-1$ bin.
    Thick arrows indicate the latency incurred by accessing memory,
    either LUT or BRAM.
}
\end{figure}

The \textsc{Look Up Bin Engine} (LUBE) utilizes a modified linear search algorithm optimized for parallel implementation on an FPGA.
A commonly used conceptual approach to the bin problem is to loop over each bin,
comparing the event value to the bin edges,
and breaking out of the loop when the matching bin is found.
This implementation is not well suited for hardware acceleration because the number of loops performed varies based on the event and it requires a large number of comparisons.
Instead we propose a solution that uses two problem-specific optimizations to achieve better performance.
The gate-level diagram at the top of figure \ref{fig:lube} shows an example of the implementation.

The first optimization for binning recognizes that LUBE always receives the bin edges in sorted order.
Therefore,
the problem can be reduced from a generic binning problem to a more simple search for a bin edge that is larger than the event value.
When that edge is found,
it is then known that the event falls in the smaller bin adjacent to that edge.
This reduces the total number of comparisons required by half. 

The second optimization addresses the issue of an indeterminate loop count.
Instead of returning after the correct bin is found,
the algorithm should always loop over the full number of bin edges.
This allows more of the loop to be implemented in parallel,
leading to lower latency and higher throughput.
To achieve this we simply loop through the entire length of the bin edge vector and set a flag when the first valid bin is found.
The rest of the bins,
while valid because their upper edge is larger than the event value,
will not be indicated as such due to the flag being set. 

In comparison to BSBE,
LUBE typically has a higher latency value due to the fact that the LUBE method is designed to accept arbitrary bin edges,
whereas the BSBE method requires bin edges on certain set intervals.
The LUBE approach also affords more flexibility in which resources are used to implement the \textsc{Bin Engine}.

Another advantage of LUBE is its ability to produce more easily synthesizable source code.
The BSBE method is known to,
at times,
produce code that HLS struggles to synthesize.
This issue arises when HLS is asked to perform excessive loop unrolling or is provided with unmanageably large source code. These cases can result from large values of $N_\textrm{bin}$ or a high dimensionality problem,
e.g.,
a large number of input variables.   
Due to its structure,
source code generated using the LUBE method can often be synthesized in these cases where BSBE struggles.
This is a key advantage and it provides firmware designers with the added flexibility when implementing their design.

\subsection{Firmware verification and validation}
\label{sec:valid}

The firmware is checked using C-synthesis and RTL-level HLS co-simulation of the C code from \fwX.
Two checks are done.
First,
the firmware co-simulation results are verified against the results using a physical FPGA.
We consider different FPGA choices,
clock speeds,
and versions of Vivado HLS.
Second,
the firmware co-simulation results are validated against the results using software simulation of the algorithm.
All tests use the same benchmark configuration described in the next section (table \ref{table:benchmark}).

The summary of the tests are given below.
More details are given in appendix \ref{appendix:valid},
in which the test benches are also illustrated (figure \ref{fig:testbench}).

\subsubsection*{Verification against physical FPGA}

For verification,
one input test vector is considered by the synthesized bitstream file,
which is programmed on the physical FPGA.
Two different FPGA setups are considered.
\begin{itemize}
    \itemsep 0pt
    \item Virtex UltraScale+ FPGA VCU118 Evaluation Kit (EK-U1-VCU118-G), our benchmark,
    \item Artix-7 FPGA on Zynq-7020 System on Chip (SoC), a smaller FPGA.
\end{itemize}
The Ultrascale+ is run on three clock speeds---$320\,\textrm{MHz}$,
$200\,\textrm{MHz}$, and
$100\,\textrm{MHz}$---while the Artix-7 is run on $100\,\textrm{MHz}$.
In all scenarios,
the actual BDT output score from the physical FPGA matches the expected result from co-simulation.
The actual latency matches the estimated result.
Moreover,
the latency is constant for Ultrascale+ at around $10\,\textrm{ns}$ and does not depend on the clock choice.
The plot of the clock independence is given in appendix \ref{appendix:valid} (figure \ref{fig:latency_clock}).

One notable observation regarding latency is made regarding Vivado HLS versions.
For the $320\,\textrm{MHz}$ clock speed,
the latency using version 2019.2 confirmed the $3$ clock ticks while version 2018.2 resulted in a slightly higher value at $4$ clock ticks.
As the BDT output score is verified for the two versions,
the difference in latency is likely due to improvements in the more recent version.
The actual-to-estimated ratios matched for the two tests.

The resource utilization on the actual results is generally equal to or smaller than the C-synthesis estimate.
The actual-to-estimate ratio for the look up table usage range from $0.1$ to $0.4$.
The ratios for flip flop usage range from $0.7$ to $1.1$ with the exception of one case for Ultrascale+ at $100\,\textrm{MHz}$,
where the actual number of flip flops matches that of the other clock speeds,
but is estimated to use a negligible amount.
The ratios for block RAM range from $0.4$ to $0.7$.
No ultra RAM is used and digital signal processor usage remains negligible.

In summary,
the latency results match exactly while the actual resource utilization is generally smaller than what is reported by C-synthesis. 
This suggests that our synthesis results are conservative estimates and that the actual usage is likely to be smaller.

As for the BDT output score for a few test vectors,
no difference is seen in all tests.

\subsubsection*{Validation against software simulation}

For validation,
$10^5$ input test vectors are considered for $200$ different BDT configurations.

\fwX\ provides a software simulation that includes the conversion of floating point values to bit integers.
A detailed diagram is given in appendix \ref{appendix:valid} (figure \ref{fig:testbench}).

The co-simulation results are compared to the software simulation.
No difference is seen for the BDT output score.

\section{Physics performance and FPGA cost}
\label{sec:perf_cost}

The physics performance and FPGA cost are described for a given BDT configuration.

Physics performance is measured by considering the ROC curve with two metrics,
depending on the problem.
The ROC curve is a representation of operating points in a two-dimensional space to display the acceptances of signal and background events.
The \emph{efficiency} (also called \emph{acceptance}) of signal events ($\varepsilon_S$)
is defined as the fraction of the number of signal events that pass the algorithm criteria divided by the total number of considered signal events.
The background acceptance ($\varepsilon_B$) is similarly defined for background events.
The axes of the ROC curves are chosen to suit the problem at hand.
The $x$-axis shows $\varepsilon_S$ in linear scale,
but the $y$-axis can be either the background acceptance $\varepsilon_B$,
the background \emph{veto} ($1-\varepsilon_B$),
or the background \emph{rejection factor} ($1/\varepsilon_B$).\footnote{
    In statistical terminology,
    the four quantities correspond to the true positive rate (TPR),
    false positive rate (FPR),
    true negative rate (TNR),
    and the inverse false positive rate (1/FPR),
    respectively.
    The FPR and FNR are also called type I error and type II error,
    respectively.
    The correspondence of the terms are listed in appendix \ref{appendix:terminology} (table \ref{table:abbrev}).
}

The first metric is the area under the ROC curve (AUC) when plotting in the $(x,y)$ plane with $(\varepsilon_S, 1-\varepsilon_B)$.
The second metric is to measure the relative change in $\varepsilon_S$ for a fixed value of $\varepsilon_B$,
i.e.,
a ``fixed background rate'' study.
The physics performance results are obtained using the \fwX\ simulation software to mimic the treatment of integers in firmware,
such as the $N_\textrm{bit}$ parameter.

``FPGA cost'' is the collective term we use to refer to the resource utilization estimates and the timing results as estimated by C-synthesis and the co-simulation of the firmware, respectively.
The timing result refers to the latency and interval measurements.
The resource utilization refers to the look up tables (LUT),
flip flops (FF),
block RAM (BRAM),
ultra RAM (URAM),
and digital signal processor (DSP) usage.
The FPGA cost results are obtained from C-synthesis and the co-simulation.
HLS programs are,
in our experience,
designed to be conservative with their timing and resource estimations.
Timing constraints are upheld by the synthesizer by ensuring the predicted maximum possible clock speed for a given design is higher than the target clock speed by a certain uncertainty factor.
This uncertainty factor is relatively large and maximum clock estimates are made conservatively,
which ensures that the target clock speed is realistic for the design.
In cases where the estimated clock period exceeds the target period,
\fwX\ reports an error to the user.
No such errors are reported in any studies in this paper.

The final considerations of the choice of the configuration parameters depend on the the physics goals as well as the experimental constraints of the user.
In general,
the better the physics performance the higher the FPGA cost.
We explain the conceptual trends by giving examples of how the optimization can be done,
with an emphasis on the trade-offs between physics performance and FPGA cost.

This section is organized into four subsections.
First,
the benchmark configuration is defined using the electron-photon problem.
Second,
using this problem we scan one parameter at a time starting from the benchmark.
Third,
we consider configurations far from the benchmark using the VBF Higgs problem.
Finally,
we end the section with a comparison to previous results.

In particular,
the technical terminology used for the FPGA cost are defined in appendix \ref{appendix:terminology} (table \ref{table:cost_defn});
the performance metric and the BDT settings that are varied are listed in appendix \ref{appendix:terminology} (table \ref{table:fpga_param}).
The values reported in this section are obtained using a setup that is illustrated in more detail in appendix \ref{appendix:valid} (figure \ref{fig:testbench}).

\subsection{Benchmark parameters}
\label{sec:benchmark}

A benchmark configuration is defined in order to compare  variations with respect to the benchmark.
We choose the electron-photon problem for the benchmark.
Our choice for the configuration,
such as a BDT with $100$ training trees and $4$ input variables,
is intended to reflect a realistic implementation in a high energy physics environment.

Table \ref{table:benchmark} lists the configuration parameters in three groups followed by the final group that gives the FPGA cost.
The first group is for the FPGA setup.
It details the hardware and development environment.
It also mentions the interface protocol,
a changeable option.
This choice in interface can impact the reported latency,
so a ``none'' interface is chosen to most accurately estimate the latency of the algorithm itself.
The second group is for the ML training setup,
which configures \TMVA.
The third group is the firmware-related parameters for the Nanosecond Optimization.
For example,
the user may choose different numbers of bits for different quantities.
This variability is useful when some variables need more precision to perform well.
For the benchmark,
$8$ bits are used for all quantities.
Although the number of bins is not a configurable parameter,
but rather a result of the configuration,
we note the value in the table as it will be seen to scale with FPGA cost.

\begin{table}[p!]
\caption{\label{table:benchmark}
    Benchmark configuration and the FPGA cost.
    Four groups of information are given.
    The top-most group defines the FPGA setup.
    The second group defines the ML training used for the electron-photon problem.
    The third group defines the Nanosecond Optimization,
    The final group gives the results.
}
\centering
{\small
\begin{tabular}{
    p{0.32\textwidth}
    p{0.20\textwidth}
    p{0.40\textwidth}
    }
\hline
Parameter                                   & Value & Comments \\
\hline
FPGA setup \\
    \quad Chip family                       & \multicolumn{2}{l}{Xilinx Virtex Ultrascale+} \\
    \quad Chip model                        & \multicolumn{2}{l}{xcvu9p-flga2104-2L-e} \\
    \quad Vivado version                    & \multicolumn{2}{l}{2019.2} \\
    \quad Synthesis type                    & \multicolumn{2}{l}{C-Synthesis} \\
    \quad HLS or RTL                        & \multicolumn{2}{l}{HLS} \\
    \quad HLS interface pragma              & \multicolumn{2}{l}{None} \\
    \quad Clock speed                       & $320\,\textrm{MHz}$     & Clock period is $3.125\,\textrm{ns}$ \\
\hline
ML training configuration \\
    \quad ML training method                & Boosted decision tree   & Binary classification \\
    \quad Boost method                      & Adaptive                & AdaBoost with yes/no leaf\\
    \quad No.\ of trees used for training   & \multicolumn{1}{l}{100} & Maximum depth of $4$ \\
    \quad No.\ of input variables           & \multicolumn{1}{l}{4} \\
\hline
\multicolumn{2}{l}{Nanosecond Optimization configuration} \\
    \quad \textsc{Bin Engine} type          & \multicolumn{2}{l}{\textsc{Bit Shift Bin Engine} (BSBE)} \\
    \quad No.\ of bits for input variables  & 8 bits for each   & Same for cut thresholds \\
    \quad No.\ of bits for BDT output score & 8 bits            & User configurable \\
    \quad No.\ of trees after merging       & 10                & \textsc{Tree Merger} via ordered list \\
    \quad No.\ of final trees       & 10, none removed          & \textsc{Tree Remover} by truncation \\
    \quad No.\ of bins              & $26\,132$                 & \textsc{Cut Eraser} not used \\
\hline
FPGA cost \\
    \multicolumn{3}{l}{\quad Estimated timing values by HLS co-simulation and resource usage by HLS C synthesis} \\
    \qquad Latency                   & $3$ clock ticks           & $9.375\,\textrm{ns}$,
    see, also, appendix \ref{appendix:valid} (figure \ref{fig:latency_clock}) \\
    \qquad Interval                  & $1$ clock tick            & $3.125\,\textrm{ns}$ \\
    \qquad Look up tables            & $1903$ out of $1\,182\,240$& $< 0.2\%$ of available \\
    \qquad Flip flops                & $138$ out of $2\,364\,480$& $< 0.01\%$ of available \\
    \qquad Block RAM                 & $8$ out of $4320$         & $< 0.2\%$ of available \\
    \qquad Ultra RAM                 & $0$ out of $960$          & - \\
    \qquad Digital signal processors & $0$ out of $6840$         & - \\
    \multicolumn{3}{l}{\quad Actual timing values and resource usage by RTL synthesis and implementation} \\
    \qquad Latency                   & $3$ clock ticks           & $9.375\,\textrm{ns}$ \\
    \qquad Interval                  & $1$ clock tick            & $3.125\,\textrm{ns}$ \\
    \qquad Look up tables            & $717$ out of $1\,182\,240$& $0.06\%$  of available \\
    \qquad Flip flops                & $147$ out of $2\,364\,480$& $< 0.01\%$  of available \\
    \qquad Block RAM                 & $5.5$ out of $4320$      & $0.1\%$ of available \\
    \qquad Ultra RAM                 & $0$  out of $960$         & - \\
    \qquad Digital signal processors & $2$  out of $6840$        & $0.03\%$ of avaailable \\
\hline
\end{tabular}
}
\end{table}

The final group in the table reports the FPGA cost.
Algorithm latency is $3$ clock ticks and the interval is $1$ clock tick,
which corresponds to about $10\,\textrm{ns}$ and $3\,\textrm{ns}$, respectively.
The other five parameters are the amount of logical units used on the FPGA.
A minimal amount of LUT and BRAM is used at less than $0.2\%$ of the available resources.
A negligible amount of FF is used at $0.01\%$ of the available resources.
No URAM or DSP is used.

\subsection{Results of scanning from the benchmark}
\label{sec:perf_scan}

Starting from the benchmark point,
we scan one parameter to observe its effect on the physics performance and the FPGA cost.
For the physics performance study we vary $N=N_\textrm{bit}$,
the number of bits used for the cut thresholds,
input variable values, and
the BDT output score.
For the other parameters used to scan FPGA cost,
given below,
the plots of physics performance considering the AUC are given in appendix \ref{appendix:perf_auc} (figure \ref{fig:auc_ey}).

For the FPGA cost study we vary four parameters:
$N_\textrm{final tree}$, the number of final trees after the \textsc{Tree Merger} and \textsc{Tree Remover} steps;
$D$, the maximum depth of a tree;
$N_\textrm{bit}$, the number of bits;
and $V=N_\textrm{var}$, number of input variables.
It is important to note that while $N_\textrm{final tree}$ can vary,
the number of trees used for the ML training step is kept constant at $T=100$.
Throughout our discussion we compare the results using the two \textsc{Bin Engines}.
Additionally,
we report dependence on two derived quantities,
the number of bins $N_\textrm{bin}$ and the BRAM usage.

\subsubsection*{Physics performance}

\begin{figure}[htbp!]
\centering
\includegraphics[width=0.65\textwidth]{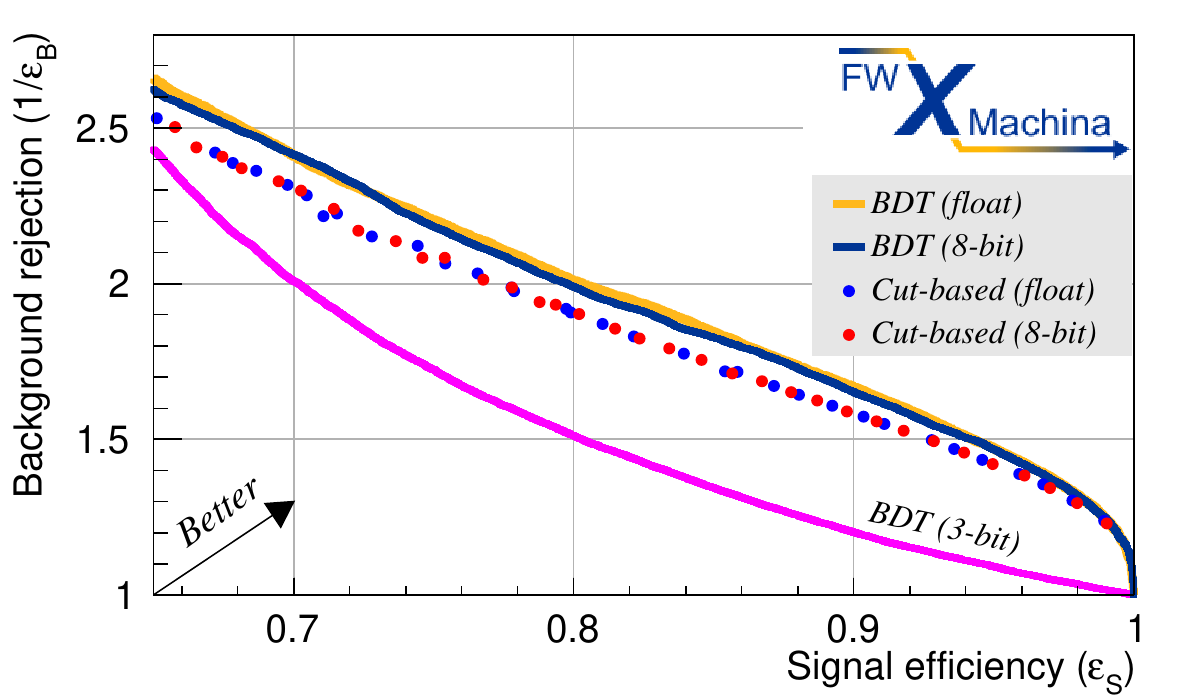}
\caption{
    \label{fig:roc_ey}
    ROC curves for electron (signal) vs.\ photon (background) classifiers.
    The pair of BDT curves (curves at the top) outperform the pair of cut-based curves (dots in the middle).
    The $x$-axis shows the signal efficiency or TPR.
    The $y$-axis shows the background rejection factor,
    defined as the inverse of background acceptance or 1/FPR,
    in linear scale,
    so higher values are better as indicated by the arrow.
    For both BDT and cut-based,
    no noticeable difference is seen between floating point precision for input variables compared to $8$-bit precision.
    Both sets outperform the BDT using $3$-bit precision (curve at the bottom).
}
\end{figure}

ROC curves are considered for $N_\textrm{bit}=3$ and $N_\textrm{bit}=8$,
and are compared to the results using floating point precision.
Both BDT and cut-based results are shown in figure \ref{fig:roc_ey}.
The highest performing result uses the BDT using floating point values,
as expected.
This is followed closely by the BDT using $8$-bit values.
It is notable to see that the $8$-bit BDT outperform the cut-based result using floating point values.
Lastly,
the $3$-bit BDT is the lowest performing result.

We note that this plot shows $(x,y)$ values of $(\varepsilon_S,1/\varepsilon_B)$.
The latter shows the inverse of background acceptance because,
for this physics scenario,
small changes in background rejection are important.

Whereas the physics performance degradation in considering $N_\textrm{bit}=8$ instead of floating point values is negligible,
the FPGA cost is greatly reduced.
This is discussed next.

The studies on the effect on AUC are given in appendix \ref{appendix:perf_auc}.

\subsubsection*{FPGA cost}

\begin{figure}[htbp!]
\centering
\includegraphics[width=0.5\textwidth]{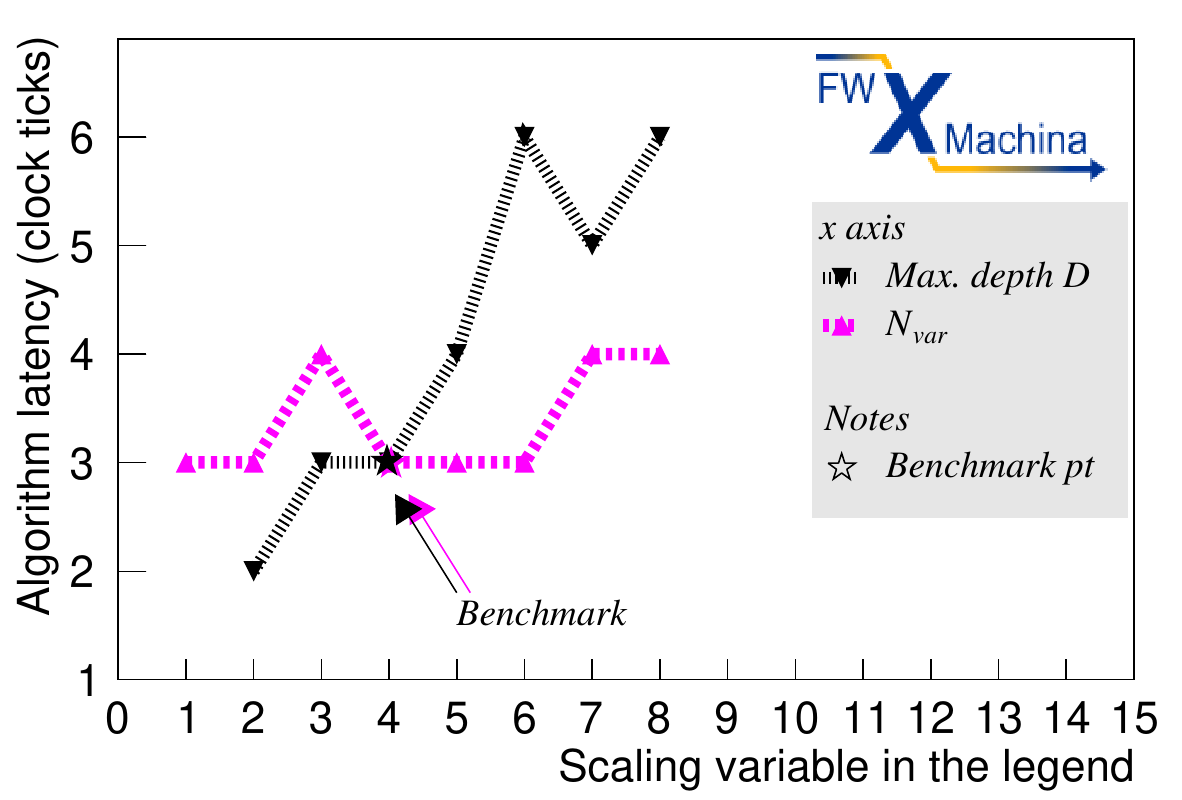}%
\includegraphics[width=0.5\textwidth]{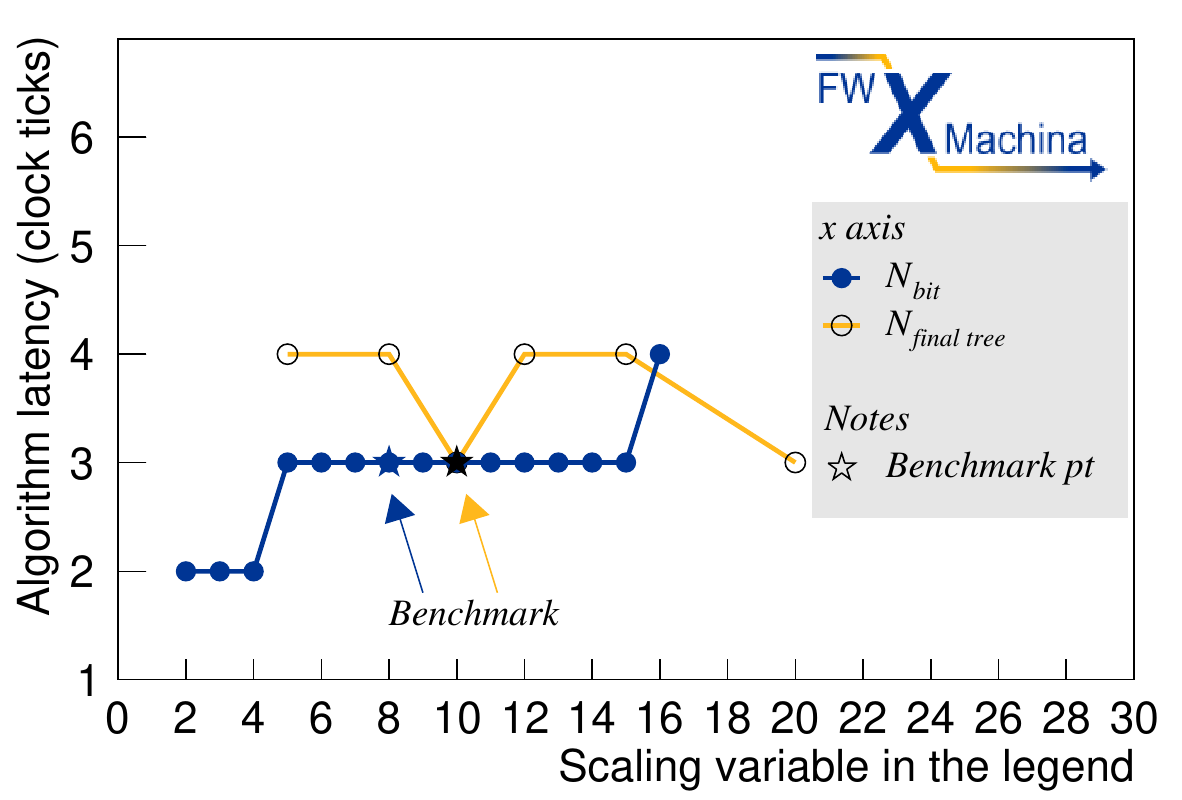}%
\caption{
    \label{fig:timing_nvar}
    Latency scaling.
    The left figure shows the dependence on the maximum depth $D$
    (upside down triangle with dotted line) and
    $N_\textrm{var}$ (upright triangle with dashed line).
    The right figure shows the dependence on $N_\textrm{bit}$ (solid circle) and $N_\textrm{final tree}$ (open circle).
    The lines connecting the symbols serve as a visual guide and do not represent interpolations.
    The latter two curves are the same ones in the left plot of figure \ref{fig:timing_be} for BSBE.
}
\end{figure}

\begin{figure}[htbp!]
\centering
\includegraphics[width=0.5\textwidth]{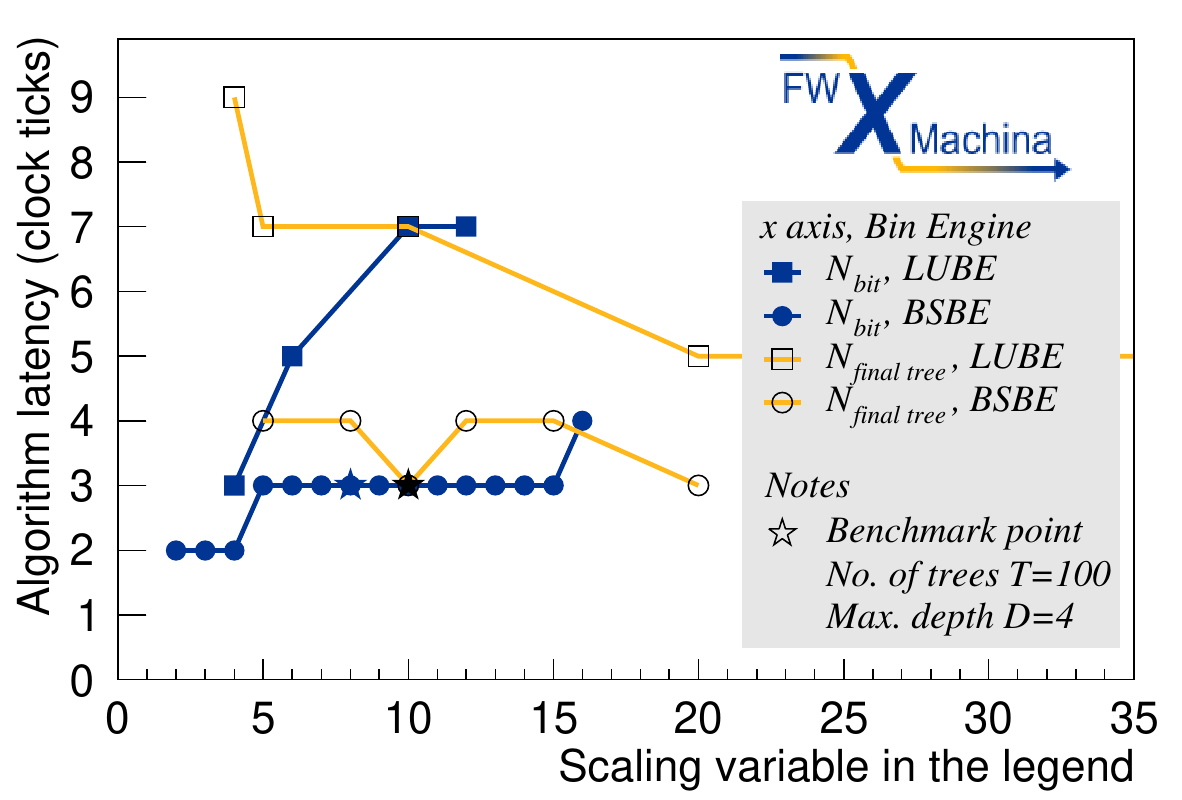}%
\includegraphics[width=0.5\textwidth]{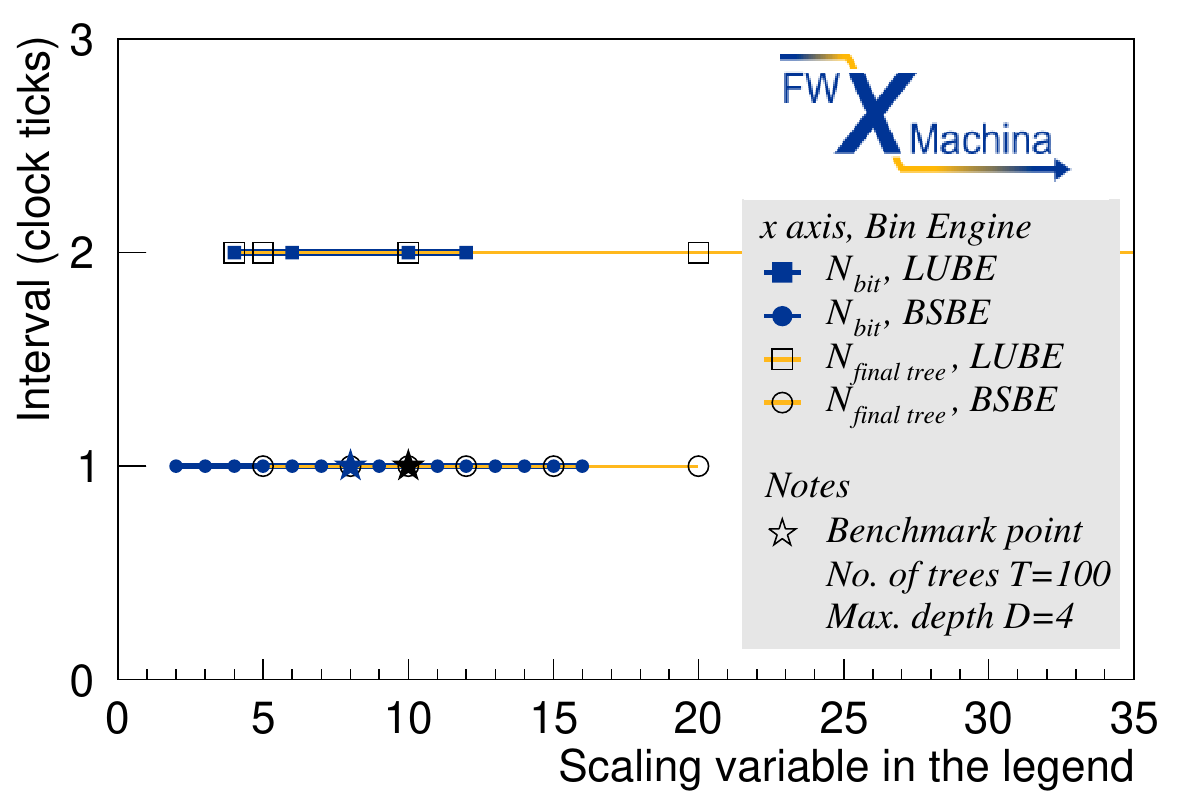}
\caption{
    \label{fig:timing_be}
    Latency (left) and interval (right) comparison for the two \textsc{Bin Engines}.
    For each plot,
    the dependence is shown for four combinations of 
    $N_\textrm{bit}$ (solid symbols) and 
    $N_\textrm{final tree}$ (open symbols)
    with
    LUBE (square) and
    BSBE (circle).
    The lines connecting the symbols serve as a visual guide and do not represent interpolations.
    The benchmark configuration is marked by a star.
    Three data points of $N_\textrm{final tree}$ with LUBE
    (at $x=33, 50, 74$)
    are omitted in the plot to zoom-in on the lower range;
    the latency for those points remain constant at $5$,
    as suggested by the flat line going to the right.
    The interval for these points remain at $2$.
}
\end{figure}

The latency is shown as a function of the four parameters in figure \ref{fig:timing_nvar}.
There is an approximate linear dependence on the maximum depth $D$
while a flat distribution with fluctuations is observed for the other three parameters.
The interval value for all sets of parameters is $1$ clock tick as seen in the right plot of figure \ref{fig:timing_be}.

The latency results for the \textsc{Bin Engines} are compared for $N_\textrm{bit}$ and $N_\textrm{final tree}$ in figure \ref{fig:timing_be}.
The left plot compares the latency.
The two latency curves that correspond to BSBE were shown previously in figure \ref{fig:timing_nvar}.
Over the $x$-axis range,
the latency values from LUBE is a factor of $1.5$ to $2$ higher than those from BSBE.
The right plot compares the interval results.
The interval curve is constant at $1$ for BSBE,
as stated previously,
and is increased by one clock tick at $2$ for LUBE.

\begin{figure}[htbp!]
\centering
\includegraphics[width=0.5\textwidth]{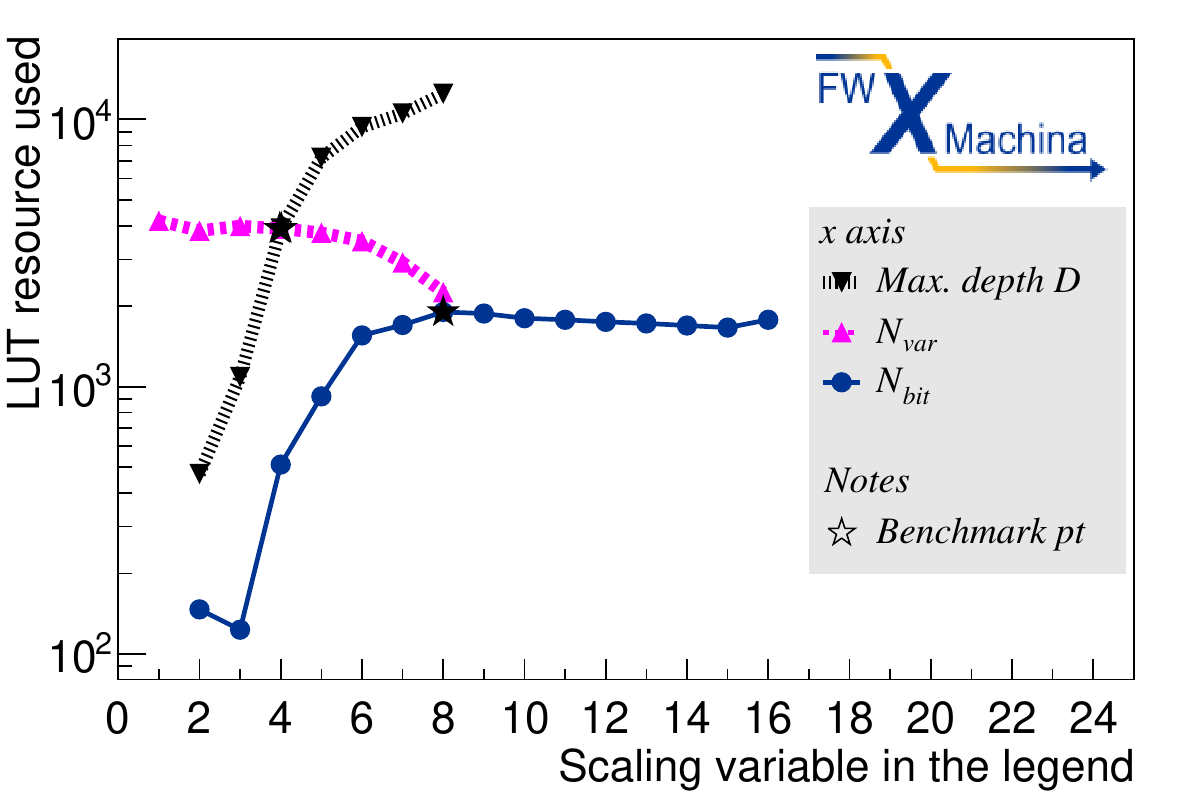}%
\includegraphics[width=0.5\textwidth]{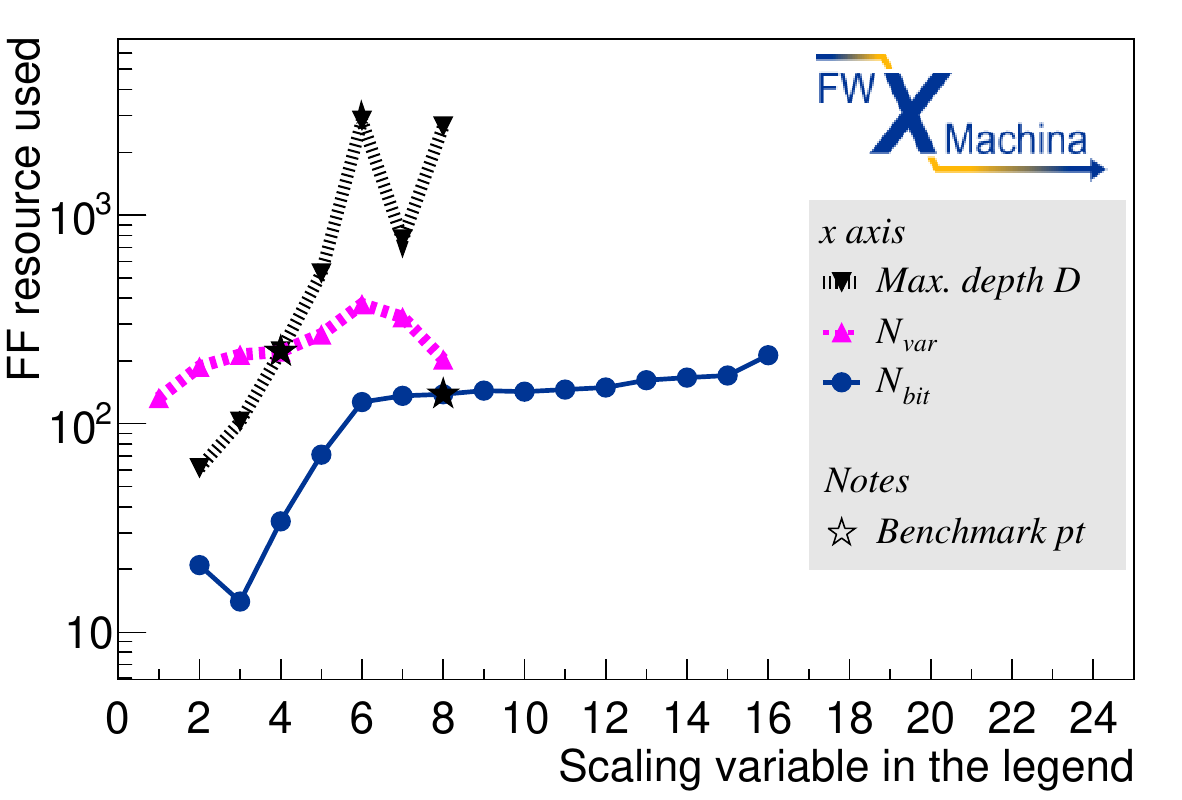}
\caption{
    \label{fig:resource_nvar}
    Resource scaling for LUT (blue lines with solid symbols) and FF (gold lines with open symbols) for three parameters:
    maximum depth (triangle), number of input variables (circle), and number of bits (square).
    The lines connecting the symbols serve as a visual guide and do not represent interpolations.
    The benchmark point (table \ref{table:benchmark}) is marked by a star.
}
\end{figure}

The LUT and FF dependence on three parameters---maximum depth $D$, $N_\textrm{var}$, and $N_\textrm{bit}$---is shown in figure \ref{fig:resource_nvar};
the dependence on $N_\textrm{final tree}$ is discussed separately below.
Since both the LUT and FF show similar behavior for a given parameter,
it helps to discuss them in pairs.
The dependence on $D$ is strong because $N_\textrm{bin}$ grows as $2^D$.
For this operating point, the dependence on $N_\textrm{var}$ is weak with minor variations
because it does not directly impact $N_\textrm{bin}$.
The dependence on $N_\textrm{bit}$ increases until around $6$,
then saturates.

\begin{figure}[htbp!]
\centering
\includegraphics[width=0.5\textwidth]{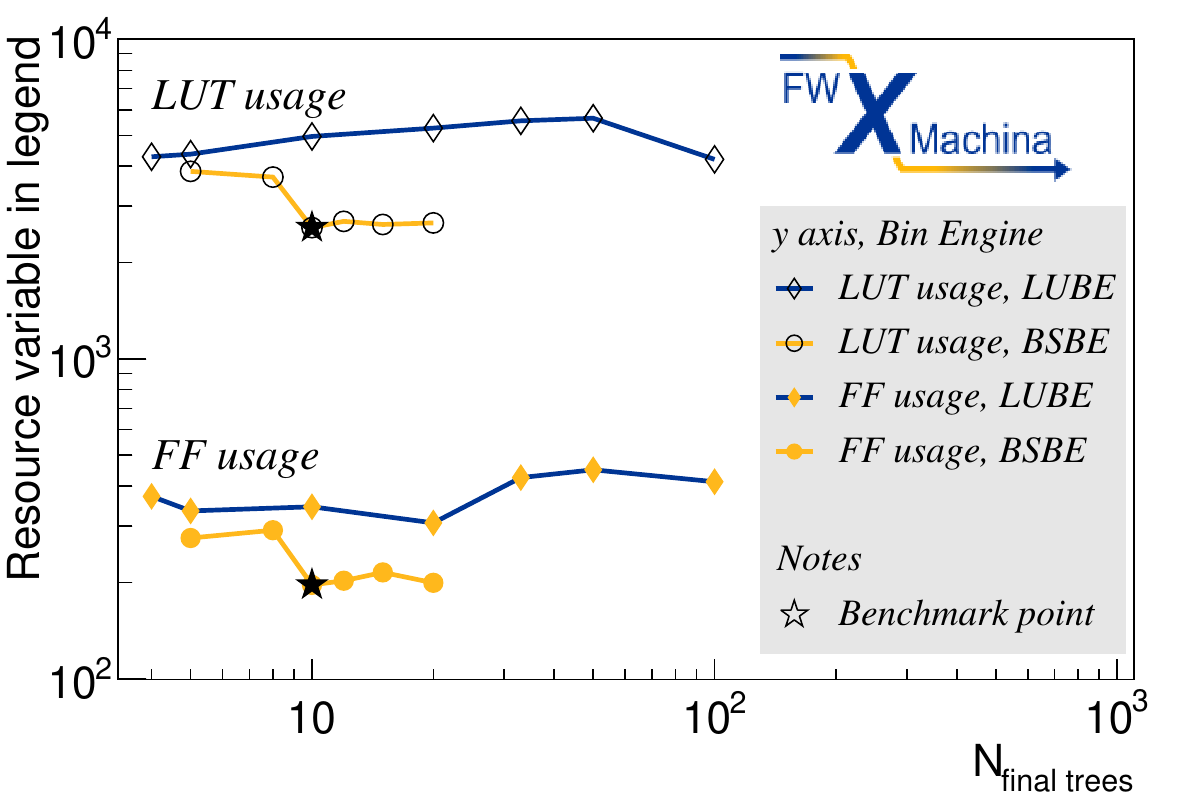}%
\includegraphics[width=0.5\textwidth]{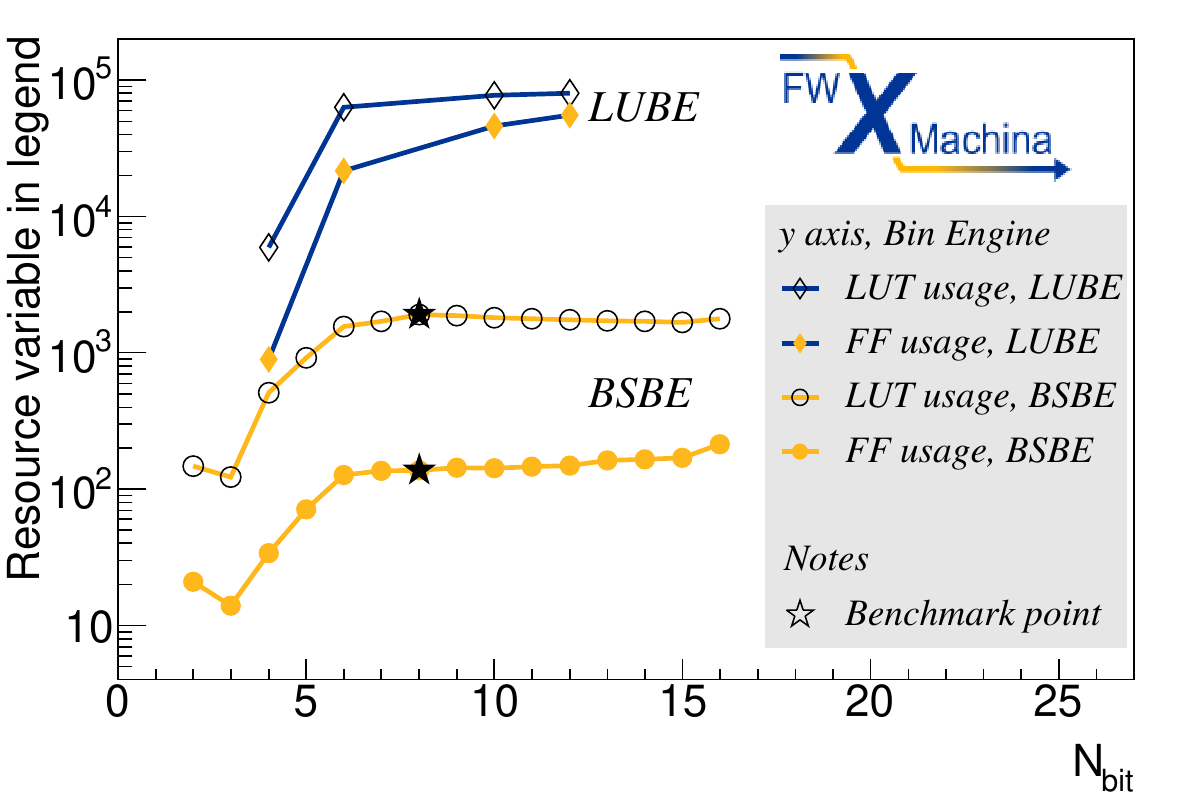}
\caption{
    \label{fig:cost_nvar}
    Resource usage comparison for the two \textsc{Bin Engines}.
    LUT and FF are shown vs.\ number of final trees (left plot) and number of bits (right plot).
    The lines connecting the symbols serve as a visual guide and do not represent interpolations.
    The benchmark point (table \ref{table:benchmark}) is marked by a star.
}
\end{figure}

The resource utilization for the \textsc{Bin Engines} are compared for $N_\textrm{final tree}$,
which was missing above,
and on $N_\textrm{bit}$ in figure \ref{fig:cost_nvar}.
The plots show that BSBE uses less LUT and FF by a factor of a few compared to LUBE.
Both BSBE and LUBE show no dependence on $N_\textrm{final tree}$ within a factor of two.
This last feature is because $N_\textrm{final tree}$ does not directly impact $N_\textrm{bin}$.
The dependence on $N_\textrm{bin}$ is discussed next.

We also varied the FPGA chip choice,
the clock speed,
and Vivado version,
all of which was discussed with the firmware verification in section \ref{sec:valid}.
We found the latency to be constant around $10\,\textrm{ns}$ after changing the clock speed.

\subsubsection*{Dependence on number of bins and BRAM}

\begin{figure}[htbp!]
\centering
\includegraphics[width=0.5\textwidth]{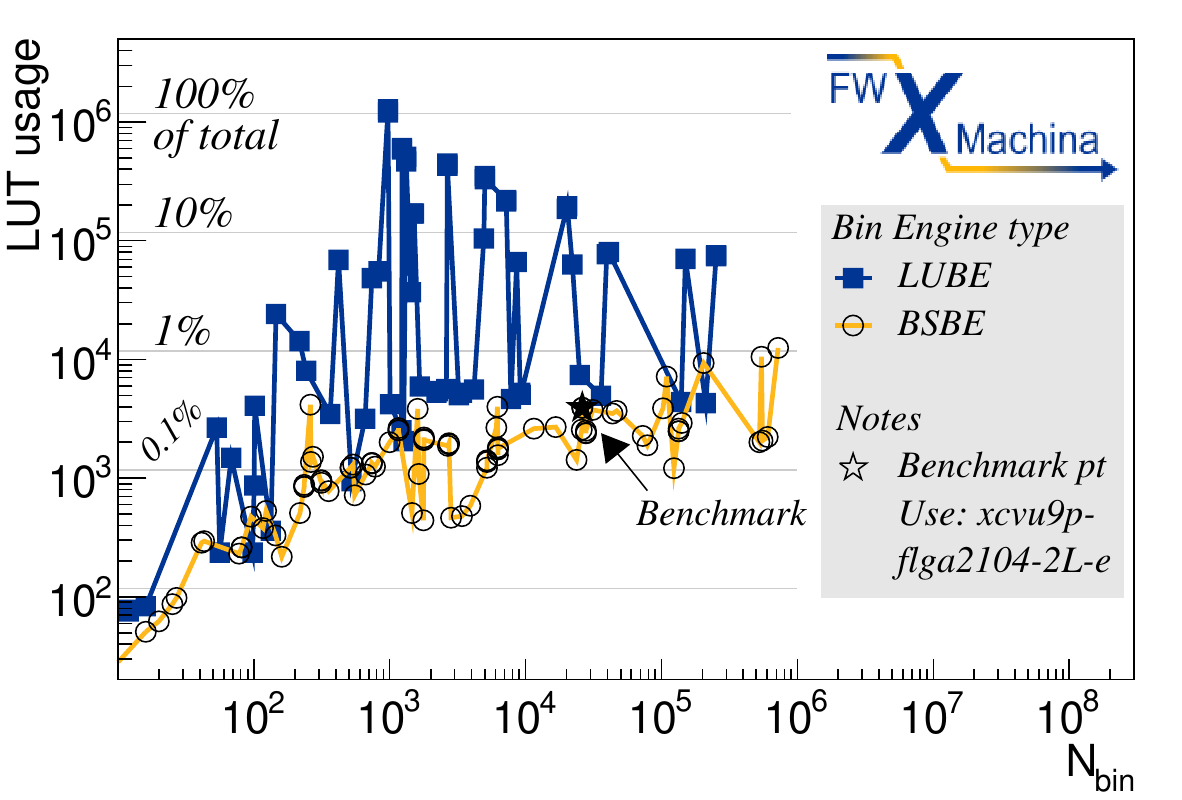}%
\includegraphics[width=0.5\textwidth]{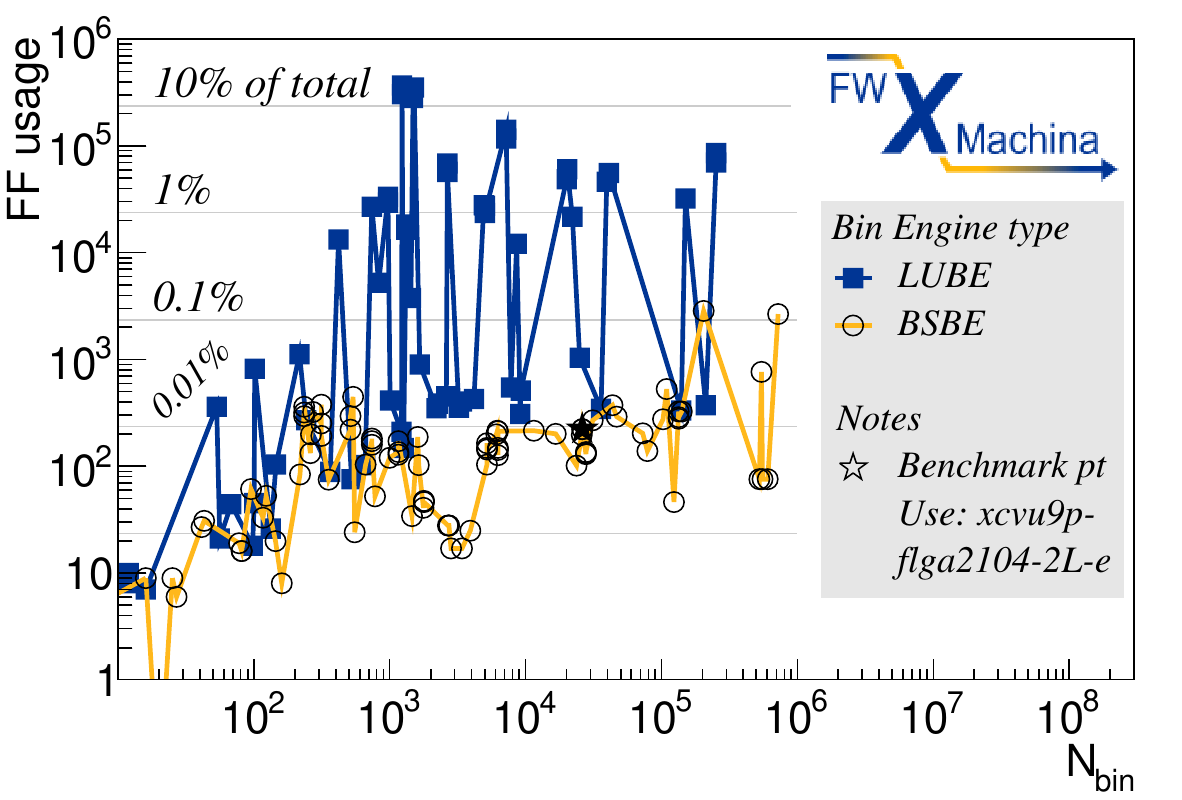}\\
\includegraphics[width=0.5\textwidth]{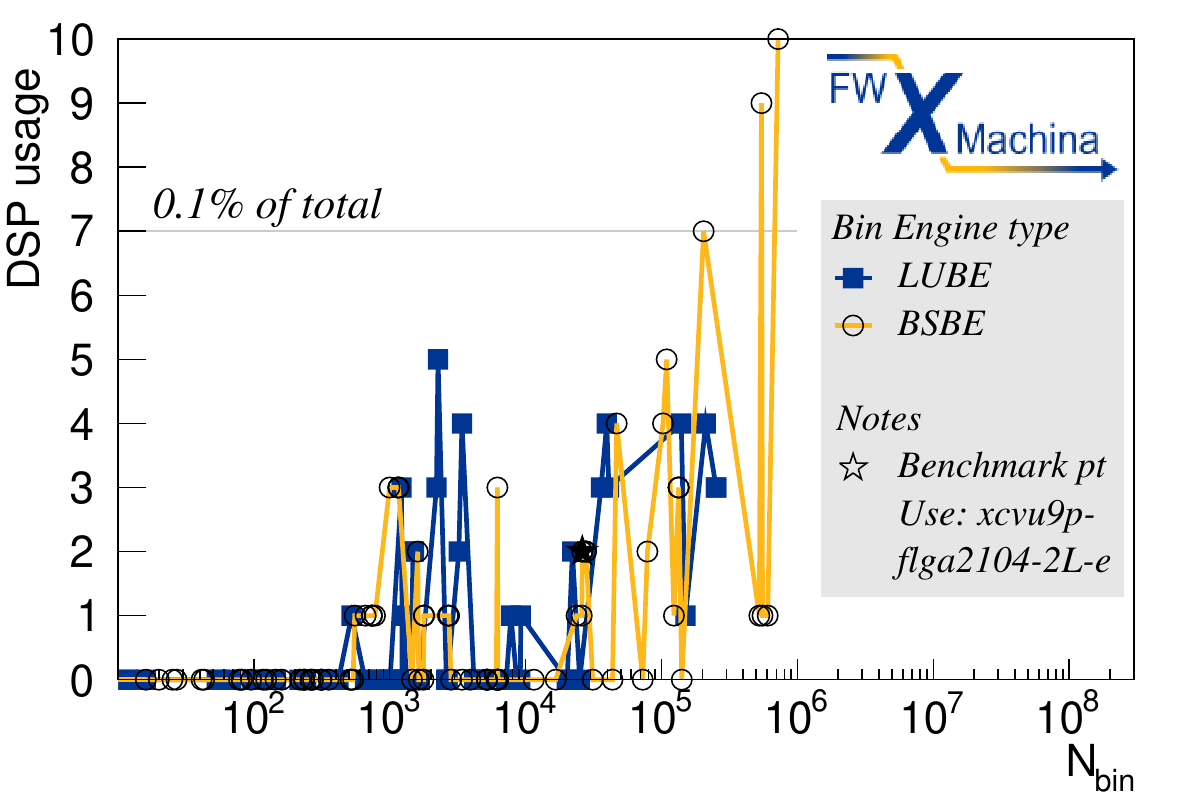}%
\includegraphics[width=0.5\textwidth]{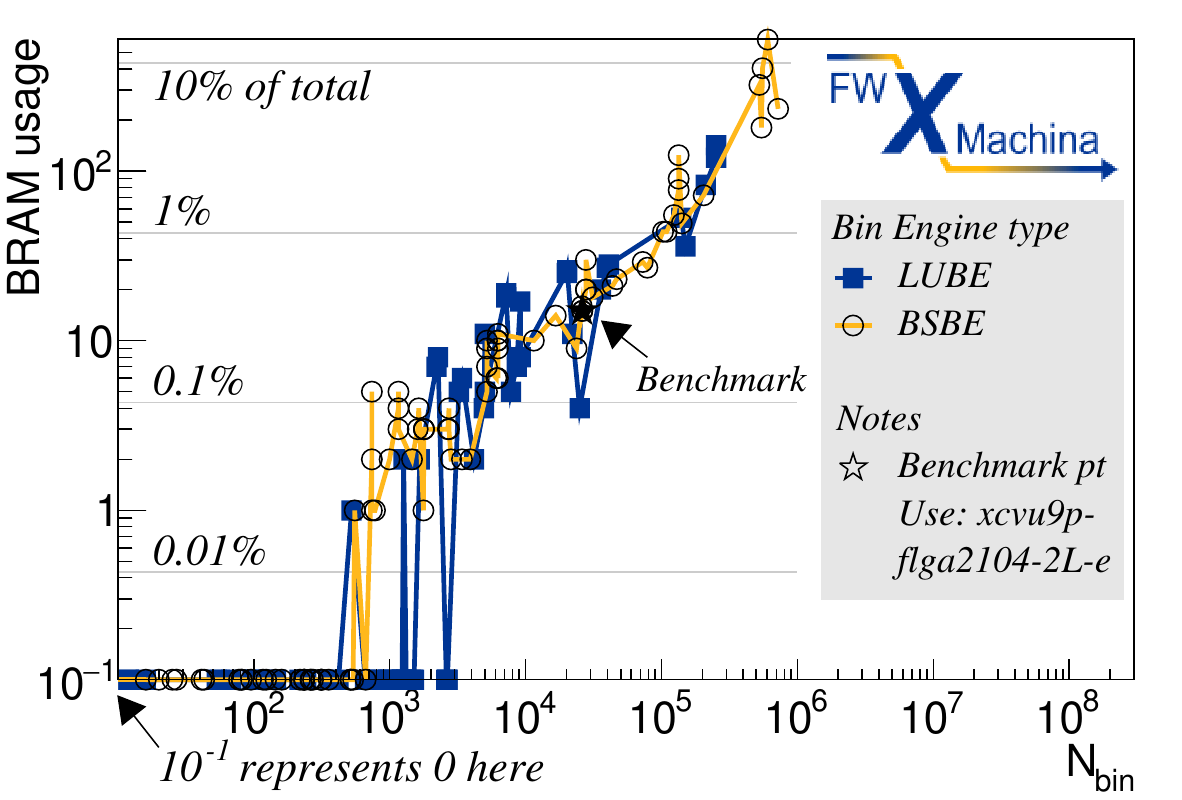}
\caption{
    \label{fig:resource_nbin}
    Resource usage vs.\ number of bins for the two \textsc{Bin Engines}.
    Plots are shown for LUT (top left),
    FF (top right),
    DSP (bottom left), 
    and BRAM (bottom right).
    URAM usage is not plotted because it is $0$ for all configurations.
    One set of data from LUBE (square) is compared to
    the other set from BSBE (circle).
    The percentages are noted in plot given with respect to the total available resource for the benchmark FPGA,
    whose model number is also stated in the legends.
    The lines connecting the symbols serve as a visual guide and do not represent interpolations.
    The benchmark configuration is marked by a star.
}
\end{figure}

The resource utilization (LUT, FF, DSP, and BRAM usage) vs.\ $N_\textrm{bin}$ is given in figure \ref{fig:resource_nbin} for the two \textsc{Bin Engines}.
First,
the top-left plot shows that LUT usage has a slow dependence on $N_\textrm{bin}$.
The LUBE result fluctuates by two orders of magnitude because the results of various configurations are shown together,
but the usage typically is less than $10\%$.
The fluctuation in LUT usage between the points with similar bin counts is largely due to the differences in the integer bit precision.
All points using over $15\%$ of the available LUT used configurations with $10$ bits or higher.
The BSBE result fluctuates less and grows to about $1\%$ at a million bins.
Second,
the top-right plot shows that FF usage also shows the same behavior as the LUT usage,
although the percentages are an order of magnitude lower.
The BSBE result grows to about $0.1\%$ at a million bins.
Third,
the bottom-left plot shows that DSP usage is zero until around a thousand bins then it starts to grow after around $10^5$.
However,
even at a million bins the usage is very low at the $0.1\%$ level.
Lastly,
the bottom-right plot shows that BRAM usage shows a stronger dependence on $N_\textrm{bin}$ and,
in contrast to the other three,
the same result is seen for LUBE and BSBE.
The same dependence and scale on $N_\textrm{bin}$ for BRAM usage for the two \textsc{Bin Engines} make us wonder whether latency is affected in a similar manner.

\begin{figure}[htbp!]
\centering
\includegraphics[width=0.5\textwidth]{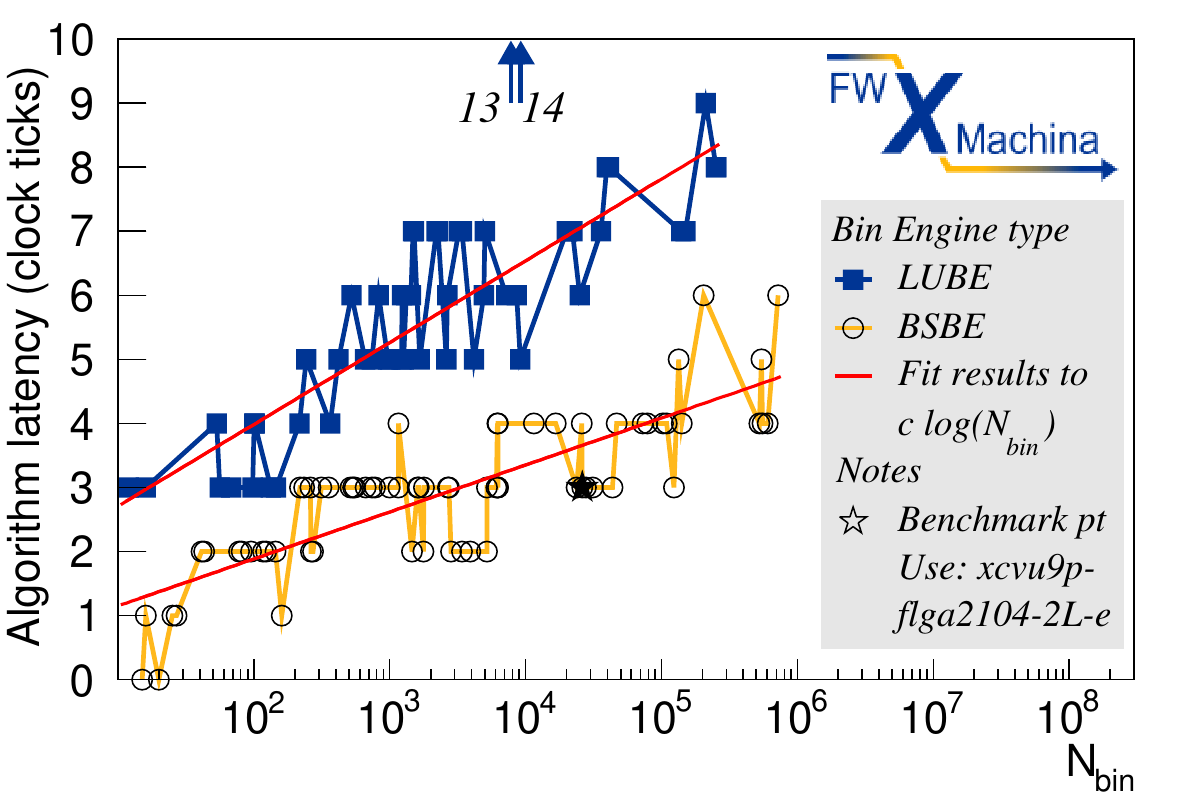}%
\includegraphics[width=0.5\textwidth]{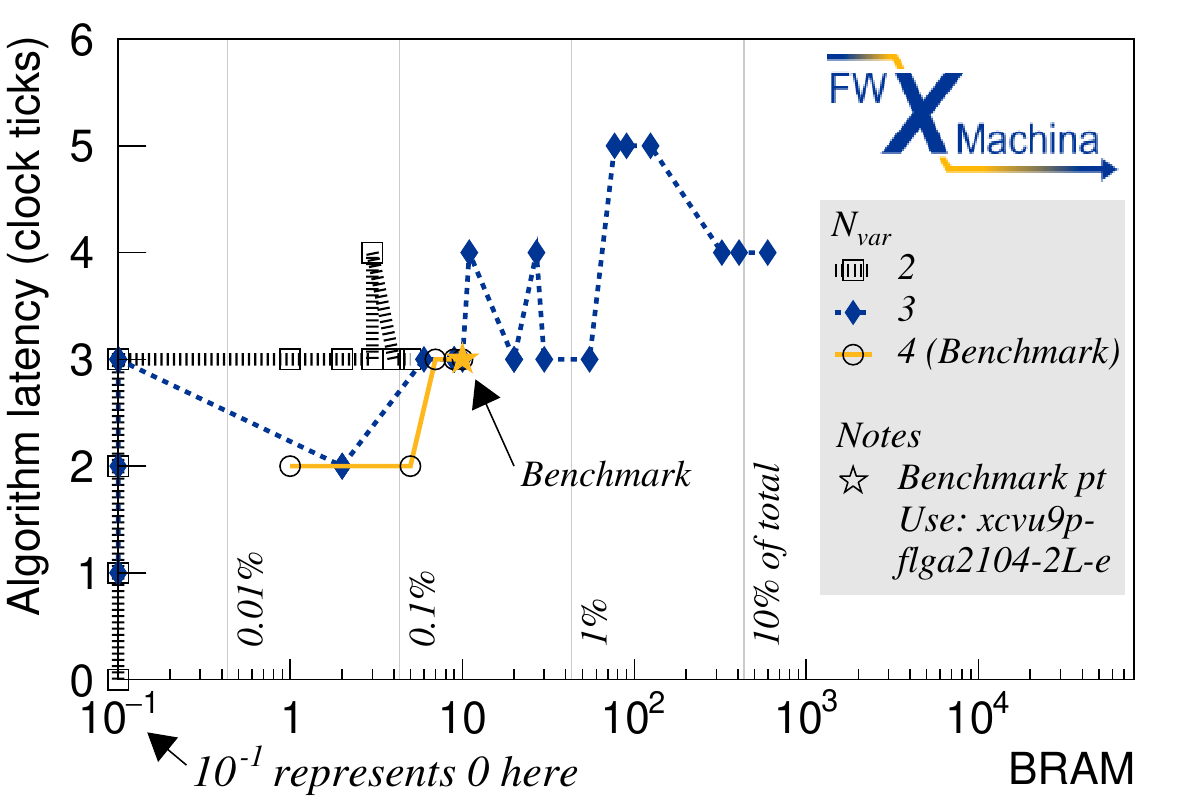}
\caption{
    \label{fig:perf_timing_bins}
    Latency dependence on $N_\textrm{bin}$ (left) and BRAM usage (right).
    On the left,
    both \textrm{Bin Engines} show a logarithmic dependence.
    The BSBE shows a slightly lower dependence than LUBE,
    but there is a gap of about $2$ to $4$ clock ticks between the methods.
    On the right,
    no BRAM usage is marked by the points at $x=10^{-1}$;
    data is separated for the number of input variables.
    The lines connecting the symbols serve as a visual guide and do not represent interpolations.
    Two outliers are indicated by the arrows corresponding to the LUBE data.
    The benchmark configuration is marked by a star.
}
\end{figure}

The latency dependence on $N_\textrm{bin}$ is given in figure \ref{fig:perf_timing_bins}.
The plot on the left shows the logarithmic dependence on $N_\textrm{bin}$ with LUBE having an offset of about $2$ to $4$ clock ticks higher than BSBE.
The fit model is
\begin{equation}
    \textrm{latency in clock ticks} = c\log(N_\textrm{bin})
\end{equation}
with $c$ around two thirds for LUBE and a third for BSBE.

We check the latency dependence on BRAM with the plot on the right of figure \ref{fig:perf_timing_bins}.
The latency ranges from $0$ to $3$ clock ticks for no BRAM usage,
while it fluctuates between $2$ to $5$ for $N_\textrm{bin}$ up to a million.
This alleviates our potential worry that large BRAM usage blows up the latency value;
no such behavior is seen.
We also checked the dependence on LUT,
FF,
and DSP,
which show no such pattern.

\subsection{Results far from the benchmark}
\label{sec:far_benchmark}

Whereas the previous subsection varied one parameter at a time with respect to the benchmark point,
we now consider configurations with many parameters that are far from the benchmark.
For this we consider the VBF Higgs problem.

Two BDT configurations are considered.
One is an optimized example and the other is a non-optimal example.
The latter is given to show the impact on FPGA cost for an extreme case of a non-optimal choice made by the user.
The configurations are listed in table \ref{table:far_benchmark} and described below.

The optimal configuration uses LUBE with $5$ input variables,
$8$-bits for the cut thresholds and input variable values,
and $16$-bits for the BDT output score.
\textsc{Cut Eraser} is used with a threshold of $5\%$,
which resulted in about $4\cdot 10^4$ bins.
The non-optimal configuration uses BSBE with $7$ input variables,
$12$-bits for the cut thresholds and input variable values,
and $16$-bits for the BDT output score.
\textsc{Cut Eraser} is not used,
which resulted in about $10^6$ bins.

\begin{table}[H]
\caption{
\label{table:far_benchmark}
    FPGA cost examples for the VBF Higgs problem.
    Two operating points are chosen far from the benchmark configuration.
    An example of the optimized configuration (left column) is compared to an example of the non-optimized configuration (right column).
    The parameters for the configuration at given at the top group of rows;
    the FPGA cost is given in the middle group of rows;
    and the user experience is given at the bottom.
}
\centering
{\small
\begin{tabular}{
    p{0.35\textwidth}
    p{0.20\textwidth}
    p{0.20\textwidth}
    p{0.15\textwidth}
}
\hline
Quantity
    & Optimized example
    & Non-opt.\ example
    & Ratio
    \\
\hline
Configuration
    \\
    \quad \textsc{Bin Engine}
    & \multicolumn{1}{l}{LUBE}
    & \multicolumn{1}{l}{BSBE}
    & -
    \\
    \quad No.\ of input variables                   & $5$   & $7$   & $1.4$ \\
    \quad No.\ of bits for input variable values    & $8$   & $12$  & $1.5$ \\
    \quad No.\ of bits for cut threshold values     & $8$   & $12$  & $1.5$ \\
    \quad No.\ of bits for BDT output score         & $16$  & $16$  & $1$ \\
    \quad Max.\ depth           & $4$                   & $4$       & $1$\\
    \quad No.\ of training trees& $100$                 & $100$     & $1$\\
    \quad No.\ of final trees   & $100$                 & $50$      & $0.5$\\
    \quad \textsc{Cut Eraser},
            threshold $\Delta$  & Yes, $5\%$            & -         & -\\
    \quad No.\ of bins          & $39\,308\approx 40$k  & $996\,710\approx 1$M & $25$\\
    \quad \textsc{Tree Remover} & -                     & -         & - \\
\hline
FPGA cost \\
    \quad Latency (clock ticks) & $5$       & $6$       & $1.2$\\
    \quad Interval (clock ticks)& $1$       & $1$       & $1$\\
    \quad LUT                   & $1.0\%$   & $1.5\%$   & $1.5$ \\
    \quad FF                    & $<0.1\%$  & $<0.2\%$  & - \\
    \quad BRAM 18k              & $2.3\%$   & $32\%$    & $14$\\
    \quad URAM                  & $0$       & $0$       & - \\
    \quad DSP                   & $0$       & $<0.3\%$  & - \\ 
\hline
User experience \\
    \quad Time to synthesize\tablefootnote{
    The setup is a typical commercial PC with the following specifications.
    Intel Core i9 $10^\textrm{th}$ generation processor with
    10 core,
    20M cache, and
    $3.7\,\textrm{GHz}$ to $5.3\,\textrm{GHz}$.
    RAM with specification of $64\,\textrm{GB}$,
    $2$x$32\,\textrm{GB}$,
    DDR4,
    $2933\,\textrm{MHz}$.
    }
    & $<15\,$minutes & $<30\,$minutes & $\approx 2$\\
\hline
\end{tabular}
}
\end{table}

\subsubsection*{Physics performance}

The VBF Higgs problem contains $J$ jet pairs per event,
for which a corresponding BDT output score is produced for each pair $\{O_0,\ldots,O_{J-1}\}$.
This poses a challenge because the $J$ copies of the BDT would have to be implemented in firmware for the maximum \emph{expected} number of jet pairs per event,
in order for the processing to occur in parallel.
We consider the ideal scenario followed by a more realistic scenario.

\begin{figure}[htbp!]
\centering
\includegraphics[width=0.65\textwidth]{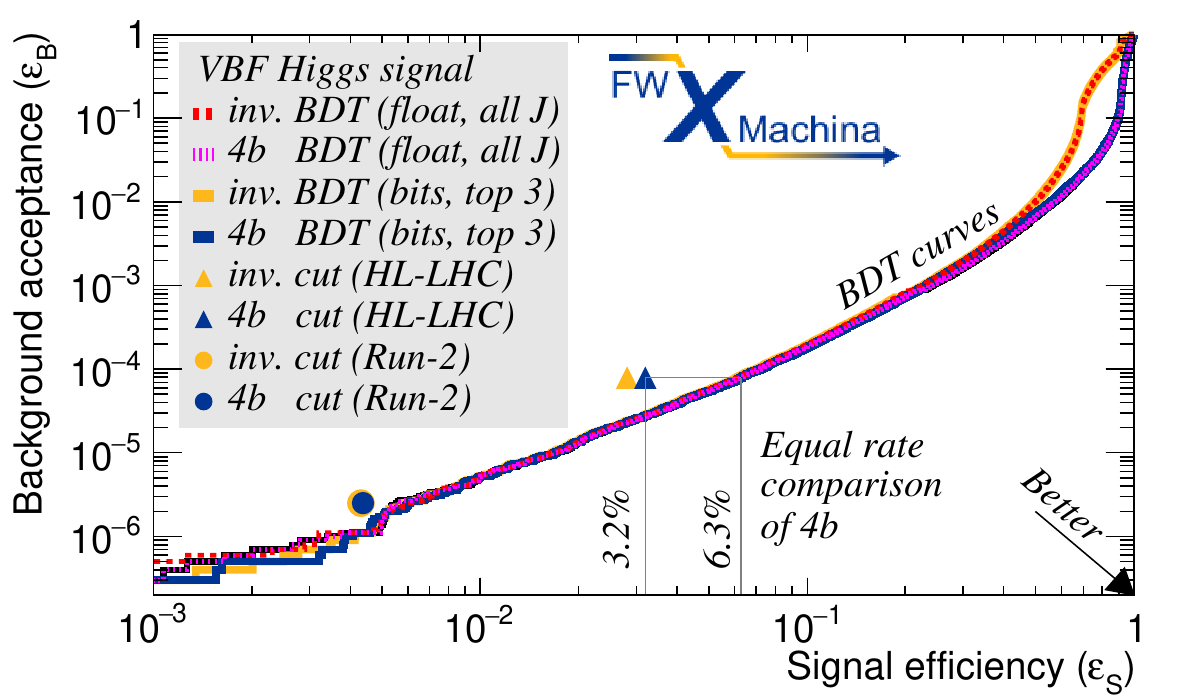}%
\caption{
    \label{fig:roc_vbf}
    ROC curves for the VBF Higgs (signal) vs.\ multijet (background) classifier.
    The BDT is trained for \emph{invisible} vs.\ multijet for both curves.
    The $x$-axis shows the signal efficiency or TPR;
    the $y$-axis shows the background acceptance or FPR.
    The BDT results (four curves) are compared to the result of the cut-based results (four symbols);
    see table \ref{table:cuts_vbf} for the cut values.
    For the BDT,
    see the text for the details on the consideration of jet pairs (all $J$ and ``top $3$'')
    and the precision of input variable values
    (floating points and bit integers).
    For cut-based,
    the ATLAS-inspired Run-2 operating points \cite{ATLAS-DAQ-PUB-2019-001} are given on the left bottom corner (circles)
    and the ATLAS-inspired HL-LHC operating points \cite{CERN-LHCC-2017-020} are given on the middle top area (triangles).
    Better operating direction is indicated by the arrow.
    The BDT curves outperform all of the cut-based operating points.
    Comparison is made for the ``equal rate'' value,
    i.e., the same value of $\varepsilon_B$ at around $10^{-4}$,
    corresponding to the HL-LHC $4b$ cut-based with the BDT result.
    For this equal rate value,
    the $\varepsilon_S$ result for the cut-based point is $3.2\%$ while for the BDT it is two-fold larger at $6.3\%$.
    A more complete table of value are given in table \ref{tab:roc_vbf}.
}
\end{figure}

First,
we consider the ideal scenario in considering all jet pairs in the event using floating point precision.
From the list of BDT output scores,
we choose the highest value in the list to represent the event's BDT output score,
i.e.,
$O_\textrm{event}=\max\{O_0,\ldots,O_{J-1}\}$.
This ideal scenario with floating point values is used to produce the plot in figure~\ref{fig:roc_vbf}.
The plot shows that the realistic scenario of limiting the number of $J$ to three and using bit-integer approximations do not degrade the result.
This is discussed below after a discussion of the cut-based results.

Our BDT curves are compared to two cut-based results.
The latter uses the ATLAS-inspired cut thresholds from ATLAS documents,
the Run-2 thresholds in \cite{ATLAS-DAQ-PUB-2019-001} and the proposed HL-LHC thresholds in \cite{CERN-LHCC-2017-020}.
The list of thresholds are given in the appendix (table \ref{table:cuts_vbf}).
We implement the cut-based approach using the same samples as described for the BDT training.
For the cut-based classifiers,
an event passes if any of the jet pairs pass the cut requirements.

The $y$-axis shows $\varepsilon_B$ in a logarithmic scale because,
for this physics scenario,
one is interested in a very large rejection of the multijet process.
A small acceptance of multijet background is needed due to the relatively large inelastic $pp$ cross section of approximately $80\,\textrm{mb}$ \cite{Aaboud:2016mmw},
combined with relatively high instantaneous luminosity values reaching $2\cdot 10^{34}\,\textrm{cm}^{-2}\textrm{s}^{-1}$ \cite{ATLAS-CONF-2019-021}.
In comparison,
the expected inclusive signal cross sections of VBF Higgs is ten orders of magnitude smaller than the background at approximately $3.8\,\textrm{pb}$ \cite{deFlorian:2016spz}.
For typical use cases the most optimal combination is a point with the high value of signal acceptance and the low value of background acceptance,
$\varepsilon_B$ typically less than $10^{-4}$.
In this representation of $\varepsilon_S$ vs.\ $\log(\varepsilon_B)$,
an operating point towards the right-bottom corner of the plot offers the optimal combination.

The BDT results (curves) are compared to the results from the cut-based approach (symbols).
Following the above color scheme,
the right symbol in the pair in red represents the VBF $H\rightarrow$\emph{invisible} training for the cut-based approach while the left symbol in blue represents VBF $H\rightarrow 4b$.
As opposed to the cut-based result with a particular operating point due to the set of fixed cut values,
the BDT result offers a continuous curve because of the ability to scan the BDT output score.

It is notable that the BDT curves outperform the results from the ATLAS-inspired cut-based selections.
For instance,
considering the background acceptance level for VBF $H\rightarrow 4b$ for HL-LHC cut-based (corresponding to signal efficiency of $3.2\%$),
the corresponding BDT result yields a signal efficiency of $6.6\%$,
a two-fold increase.
This example shows the performance enhancement potential that can be realized in the level-1 trigger.

\begin{table}[htbp!]
\centering
\caption{
    \label{tab:roc_vbf}
    Physics performance for VBF Higgs vs.\ multijet background for various configurations shown in figure \ref{fig:roc_vbf}.
    Two groups of values are given corresponding to a fixed value of background rate,
    i.e.,
    a fixed value of $\varepsilon_B$,
    for the ATLAS-inspired Run-2 and HL-LHC comparisons.
    Within each group,
    the cut-based values are obtained by our ATLAS-inspired implementation of the study done by the cited references.
    Two BDT results are given for varying precision of the input variables and the output score,
    as well as the number jet pairs considered.
    For all BDT results shown in this table,
    the training is done for binary classification of VBF $H_\textit{inv}$ vs.\ multijet.
    In the case of VBF $H_\textit{4b}$ the above-mentioned BDT is applied on the signal sample;
    no BDT is trained for VBF $H_\textit{4b}$ vs.\ multijet nor is any multiclassification done.
    We note that the statistical fluctuations caused a minor dip so that the bits result is slightly higher than the float result for the Run-2 VBF $H_{4b}$ efficiency.
   }
{\small
\begin{tabular}{
    p{0.12\textwidth}
    p{0.08\textwidth}
    p{0.10\textwidth}
    p{0.10\textwidth}
    p{0.45\textwidth}
    }
    \hline
    Configuration
    & Precision
    & \multicolumn{2}{l}{Signal efficiency $\varepsilon_S$}
    & Comments
    \\
    &
    & VBF $H_\textit{inv}$
    & VBF $H_{4b}$ \\
    \hline
    \multicolumn{4}{l}{Run-2 equal background rate\dotfill}
    & Operating point at FPR value of $\varepsilon_B=2.5\cdot 10^{-6}$ \\
    \quad Cut-based & float & $0.43\%$ & $0.44\%$ & ATLAS-inspired from ref.\ \cite{ATLAS-DAQ-PUB-2019-001}, see text\\
    \quad BDT       & float & $0.66\%$ & $0.57\%$ & Trained VBF $H_\textit{inv}$ vs.\ multijet; evaluate all $J$ pairs\\
    \quad BDT       & bits  & $0.64\%$ & $0.65\%$ & Trained VBF $H_\textit{inv}$ vs.\ multijet; eval.\ top $3$ pairs\\
    \multicolumn{4}{l}{HL-LHC equal background rate\dotfill}
    & Operating point at FPR value of $\varepsilon_B=7.9\cdot 10^{-5}$ \\
    \quad Cut-based & float & $2.81\%$ & $3.20\%$ & ATLAS-inspired from ref.\ \cite{CERN-LHCC-2017-020}, see text\\
    \quad BDT       & float & $6.16\%$ & $6.28\%$ & Trained VBF $H_\textit{inv}$ vs.\ multijet; evaluate all $J$ pairs\\
    \quad BDT       & bits  & $6.16\%$ & $6.29\%$ & Trained VBF $H_\textit{inv}$ vs.\ multijet; eval.\ top $3$ pairs\\
    \hline
\end{tabular}
}
\end{table}

Second,
a more realistic firmware implementation would not consider all $J$ jet pairs in the event
Evaluating the BDT for each jet pair has the potential to be very resource-intensive on the FPGA, particularly in events with relatively high numbers of jets,
as the number of jet pairs is $J = N_\textrm{jet}(N_\textrm{jet}-1)/2$.
In order to reduce the value of $J$ considered on firmware,
we considered examining only the three jet pairs with the highest $m_{jj}$ values.
We found that,
for the same $\varepsilon_S$ value,
the $\varepsilon_B$ varied by less than $2\%$ (figure \ref{fig:roc_vbf}),
i.e.,
$\big|1-\mbox{\small$\varepsilon_B^{\textrm{all-}J}/\varepsilon_B^{\textrm{top-3}}$}\big|< 2\%$.
We also considered using top six jet pairs,
but only find a negligible improvement.
The details of this study are given in appendix \ref{appendix:njet}.
Finally,
the performance results for the floating point study is compared to the results from the optimized firmware configuration (table \ref{table:far_benchmark});
we find negligible differences between the two setup (figure \ref{fig:roc_vbf}).

Therefore,
we conclude that a threefold increase in the BDT implementation is sufficient to achieve the ideal scenario.
Furthermore,
we assume that the implementation of the $\max$ function does not add much to the relative FPGA cost and that the $m_{jj}$ values are provided in sorted order.

\subsubsection*{FPGA cost}

The latency is $5$ clock ticks,
corresponding to about $19\,\textrm{ns}$ for our setup,
which is the optimal configuration.
The non-optimal configuration added one clock tick to bring the latency to $6$ clock ticks.

The resource utilization is low for LUT, FF, and DSP at a range of $0.2\%$ to $1.5\%$.
No URAM is used.
However,
a large difference in BRAM usage is seen.
The optimized configuration uses $2.3\%$ whereas the non-optimized uses $32\%$.

Lastly,
we note aside the user experience for the time to synthesize.
The optimized configuration takes less than $15$ minutes and the non-optimized less than $30$ minutes for our setup.

The bottom of table \ref{table:far_benchmark} summarizes the above findings.

\subsection{Comparisons with previous results}
\label{sec:compare}

The hls4ml group implemented a method for evaluating the BDT on FPGA \cite{Summers:2020xiy},
with the BDT project now called Conifer.
This method retains the tree structure and recurses down the binary splits.
An advantage to this method is that it does not face the issue of maintaining a reasonable number of bins,
thus it can handle arbitrarily high precision and numbers of variables.
Another advantage is the symmetric tree configuration that allows the user to reuse the firmware structure.

A comparison of our result with that of hls4ml/Conifer is difficult to achieve due to structural differences in the architecture of the implementation.
Moreover,
hls4ml/Conifer considers a five-class problem whereas \fwX\ is designed for binary classification problems.
Therefore,
we ran the out-of-the-box version of the hls4ml/Conifer \cite{conifer-github} on our benchmark configuration.
The results of this study are given in appendix \ref{appendix:compare}.
    
\section{Conclusions}
\label{sec:conclusion}

We presented a novel implementation of machine learning / artificial intelligence method of boosted decision trees in FPGA.
The firmware-oriented Nanosecond Optimization stage includes restructuring the tree layout and reconfiguring the BDT parameters.
Our design philosophy is to remove clocked operations in favor of combinatoric logic.
The resulting design is realized by using a software package called \fwXmachina,
available at \href{http://fwx.pitt.edu}{fwx.pitt.edu}.

We used \fwX\ to investigate two physics problems for use in the real-time triggers in high energy physics experiments.
The first problem is for electron vs.\ photon separation.
The second problem is the selection of the vector boson fusion-produced Higgs boson selection and the rejection of the multijet process.
It is a binary classification problem that serves as a challenging case of where ML/AI can make improvements beyond the cut-based approach.
These two problems are considered as potential use cases at the LHC.

For our benchmark point of $100$ training trees with a maximum depth of $4$ using four input variables,
we report a latency value of around $10\,\textrm{ns}$,
or $3$ clock ticks at $320\,\textrm{MHz}$.
The resource utilization is minimal at less than $0.2\%$ of look up tables and block RAM usage,
less than $0.01\%$ of flip flop usage,
and no ultra RAM or digital signal processor usage.
We studied the dependence of performance on configurable parameters.
Far from the benchmark using more variables and more bit precision,
we report a latency value of $16\,\textrm{ns}$ with $1$ to $2\%$ resource utilization.
We find that latency results in nanoseconds do not depend on the clock speed,
as it is compensated by clock ticks.
By considering the physics performance and FPGA cost at the same time,
the user of \fwX\ has the ability to consider the trade-offs between the two for their own use case.

\fwX\ gives users the versatility to work within various experimental constraints.
In particular,
this flexibility may be beneficial for trigger systems for the HL-LHC operation of the ATLAS and CMS experiments,
or any such systems,
that are under strict timing requirements.

\section*{Acknowledgments}
We thank Lin Yao for early engineering discussions.
We thank Gracie Jane Gollinger for computing infrastructure support.
We thank Chris R.\ Hayes for the discussions on VBF simulations.
We thank the University of Pittsburgh for the support of this project,
especially for both STRs and DCS.
TMH and BTC were supported by the US Department of Energy [award no.\ DE-SC0007914].
BTC was supported by the \textsc{PITTsburgh Particle physics Astrophysics and Cosmology Center} (PITT PACC).
JS was supported by the US Department of Energy [award no.\ DE-SC0012704].
BRE was supported by the US National Science Foundation [award nos.\ PHY-1948993 and PHY-1624739].
DCS was supported by the NASA Pennsylvania Space Grant Consortium.
Patent pending.

\appendix

\section{List of terminology used in the paper}
\label{appendix:terminology}

This paper uses abbreviations and technical terminology from many different fields of study.
They are listed in table \ref{table:abbrev}.
The notation used for Nanosecond Optimization in section \ref{sec:nano_opt} is given in table \ref{table:notation}.
The definitions of the technical terms for firmware design in section \ref{sec:firmware} are given in table \ref{table:cost_defn}.
The parameters used in the performance and FPGA cost evaluation in section \ref{sec:perf_cost} are described in table \ref{table:fpga_param}.

\begin{table}[htbp]
\caption{
    \label{table:abbrev}
    List of terms used in this paper in five categories.
    BRAM, DSP, FF, LUT, URAM,
    latency,
    and interval are described in table \ref{table:cost_defn}.
    The latter two are illustrated in figure \ref{fig:interval_schematic}.
}
\centering
{\small
\begin{tabular}{
    p{0.15\textwidth}
    p{0.75\textwidth}
}
\hline
Terms & Description \\
\hline
\multicolumn{2}{l}{Physics}\\
\quad 2HDM & Two-Higgs doublet model, a BSM theory that extends the Higgs boson sector \\
\quad BSM & Beyond the Standard Model, i.e., hypotheses that extend the Standard Model \\
\quad $H$ & Higgs boson \\
\quad HEP & High energy physics \\
\quad HL-LHC & High Luminosity LHC, the upgrade of the LHC to start around 2026 \\
\quad LHC & Large Hadron Collider \\
\quad Multijet & Physics process wherein multiple ``jets'' are produced from the collision \\
\quad VBF & Vector boson fusion interaction process from the collision \\
\hline
\multicolumn{2}{l}{Machine learning} \\
\quad AdaBoost & Adaptive boost method for BDT \\
\quad BDT & Boosted decision tree, the type of ML used in this paper \\
\quad GradBoost & Gradient boost method for BDT \\
\quad ROC & Receiver operating characteristics \\
\hline
\multicolumn{2}{l}{Statistics} \\
\quad TPR & True positive rate, i.e., signal efficiency ($\varepsilon_S$), corresponds to correct signal inference \\
\quad FNR & False negative rate, i.e., signal veto ($1 - \varepsilon_S$); also called type II error ($\beta$) \\
\quad TNR & True negative rate, i.e., background veto ($1-\varepsilon_B$) \\
\quad FPR & False positive rate, i.e., background acceptance ($\varepsilon_B$); also called type I error ($\alpha$) \\
\quad 1/FPR & Inverse false positive rate, i.e., background rejection factor ($1/\varepsilon_B$) \\
\hline
\multicolumn{2}{l}{Engineering} \\
\quad ASIC & Application specific integrated circuit \\
\quad ASR & Arithmetic shift right \\
\quad Bitstream & A file that contains the programming information for an FPGA \\
\quad Bus tap & Taps a subset of values stored in the input bus and fans them out \\
\quad FPGA & Field programmable gate array \\
\quad HDL & Hardware description language \\
\quad HLS & High Level Synthesis \\
\quad ILA & Internal Logic Analyzer \\
\quad IP & Intellectual Property core \\
\quad LSB & Least significant bit \\
\quad RTL & Register transfer level \\
\quad VHDL & Very High Speed Integrated Circuit (VHSIC) Hardware Description Language  \\
\hline
\multicolumn{2}{l}{This paper} \\
\quad BSBE & \textsc{Bit Shift Bin Engine} \\
\quad FPGA cost & Timing (latency, interval) and resource usage (LUT, FF, BRAM, URAM, DSP) \\
\quad LUBE & \textsc{Look Up Bin Engine} \\
\hline
\end{tabular}
}
\end{table}

\begin{table}[htbp!]
\caption{
    \label{table:notation}
    Notation used for Nanosecond Optimization.
    When appropriate the lowercase variables represent the index of the corresponding uppercase versions in this table,
    e.g.,
    $t$ as a tree index for a value from $0$ to $T-1$.
}
\centering
{\small
\begin{tabular}{
    p{0.20\textwidth}
    p{0.70\textwidth}
}
\hline
    Symbol                    & Description \\
\hline
    $\tau$, $\vec{\tau}=\phi$ & Single tree, Ordered list of trees = Forest \\
    $x$, $\vec{x}$            & Single input variable, Ordered list of variables \\
    $D$                       & Depth of the tree \\
    $V=N_\textrm{var}$        & Number of input variables \\
    $T=N_\textrm{tree}$       & Number of trees \\
    $B=N_\textrm{bin}$        & Number of bins \\
    $N=N_\textrm{bit}$        & Number of bits for input variables \\
    $L$                       & Number of binary layers \\
    $O$                       & BDT output score \\
    $W_t$                     & Boost weight for tree $t$ \\
    $w_t$                     & Normalized boost weight, i.e., $w_t=W_t/\sum_{t'}{W_{t'}}$ \\
\hline
\end{tabular}
}
\end{table}

\begin{table}[htbp!]
\caption{
    \label{table:cost_defn}
    List of technical terms for FPGA cost and firmware test bench.
    Figure \ref{fig:interval_schematic} illustrates the timing.
}
\centering
{\small
\begin{tabular}{p{0.155\textwidth}p{0.775\textwidth}}
\hline
Term
    &
    Description
    \\
\hline
\makecell[tl]{Chip\\ information}
    &
    We use Xilinx Virtex Ultrascale+ VU9P.
    The Ultrascale+ series is among Xilinx's families released starting in 2016 and manufactured with a $16\,\textrm{nm}$ process.
    \\
\cline{2-2}
\makecell[tl]{High Level\\ Synthesis (HLS)}
    &
    Software tool provided by Xilinx for converting C (or C++) code into a register transfer level (RTL) implementation that can be synthesized to run on an FPGA. 
    The BDT implementation is largely written in HLS.
    \\
\cline{2-2}
Pragma
    &
    Special commands that can be placed in your C code to give the HLS compiler directives regarding the desired hardware implementation.
    For example,
    pragma can ``unroll'' a loop so that its results can be computed in parallel,
    as opposed to in series. 
    \\
\cline{2-2}
Clock speed
    &
    Target clock frequency.
    The maximum clock speed that an FPGA design can run at depends on the design itself.
    For example,
    every logic gate has a small delay and 
    gates strung together begin to stack up.
    This becomes a constraint on the maximum clock speed that the design can handle as the results of the logic need to be stored or passed to another part of the design.
    The synthesizer attempts to optimize the maximum clock to be higher than the target clock frequency by a certain margin.
    \\
\cline{2-2}
Clock ticks
    &
    For synchronous circuits,
    a ``clock'' is required to synchronize portions of the circuit.
    The clock is an input that periodically goes from logical high to logical low.
    A ``clock tick'' occurs when the clock moves from low to high.
    \\
\cline{2-2}
Latency
    &
    Time required for an algorithm to take an input and produce a valid output.
    \\
\cline{2-2}
Interval
    &
    Time required between two successive inputs to the algorithm.
    \\
\cline{2-2}
\makecell[tl]{Resource\\ utilization}
    &
    FPGA chips have various different circuit elements that can be used in a design, 
    e.g.,
    BRAM, LUT, and FF.
    Resource utilization is an inclusive term that refers to these various circuit elements.
    It can be reported in percentage of the available element or as the total number of each element.
    \\
\cline{2-2}
\makecell[tl]{Look up table\\ (LUT)}
    &
    Type of asynchronous memory.
    LUT has the benefit of quick memory accesses that do not depend on a clock,
    but they come at the expense of consuming more area on the FPGA die.
    FPGA chips have many LUT built-in for utilization in a design.
    \\
\cline{2-2}
Flip flop (FF)
    &
    One of the basic elements of data storage in electrical circuits.
    A flip flop stores one bit of information. 
    Compared to other memory technologies,
    flip flops have a higher access speed,
    but the circuit is very large leading to poor memory density.
    \\
\cline{2-2}
\makecell[tl]{Block RAM\\ (BRAM)}
    &
    Block of random access memory (RAM),
    which is a synchronous memory element.
    The RAM requires a clock to put data in or to get it out.
    Because memory can only be accessed on a clock edge,
    the data is not available immediately.
    RAM trades access speed for increased memory density.
    \\
\cline{2-2}
\makecell[tl]{Ultra RAM\\ (URAM)}
    &
    Ultrascale+ has two types of RAM: URAM, and BRAM.
    URAM is a block of memory similar to BRAM,
    but are much larger than BRAM.
    The larger size making them inefficient for smaller amounts of data.
    \\
\cline{2-2}
\makecell[tl]{Digital Signal\\ Processor (DSP)}
    &
    Many applications in the field of DSP require the use of a multiply and accumulate operation.
    As a result,
    multiply and accumulate circuits are often referred to as DSP.
    \\
\hline
\end{tabular}
}
\end{table}

\begin{figure}[htbp!]
\centering
\includegraphics[width=0.78\textwidth]{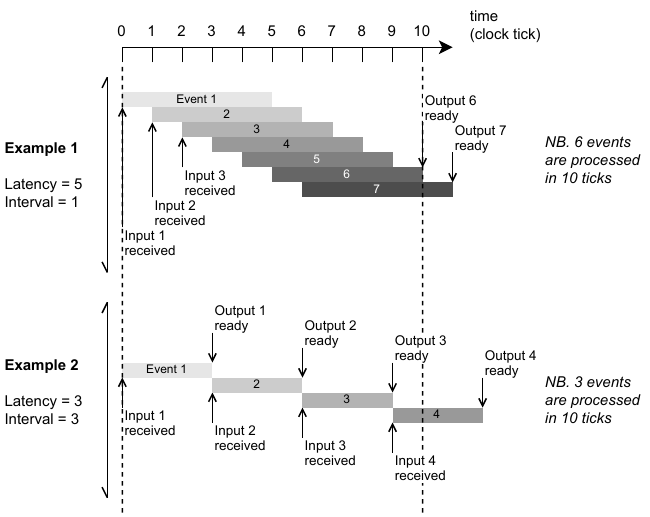}
\caption{
    Illustration of latency and interval.
    The top example shows that successive inputs can be sent after each clock tick with each event taking five ticks,
    so six events are processed in ten ticks.
    The bottom example shows that successive inputs can be sent every third  tick with each event taking three ticks,
    so three events are processed in the same period.
    The two examples illustrate the importance of interval.
}
\label{fig:interval_schematic}
\end{figure}

\begin{table}[htbp!]
\caption{
\label{table:fpga_param}
    Parameters considered when studying the FPGA cost.
    Some of the terms in this list correspond to terms defined in table \ref{table:notation}.
}
\centering
{\small
\begin{tabular}{
    p{0.150\textwidth}
    p{0.120\textwidth}
    p{0.62\textwidth}
}
\hline
Quantity
    & Type 
    & Description
    \\
\hline
AUC
    &
    \makecell[tl]{Performance\\ metric}
    &
    Area under the ROC curve of background veto vs.\ signal acceptance,
    i.e.,
    $1-\varepsilon_{B}$ vs. $\varepsilon_{S}$.
    This is similar to the curves in figure \ref{fig:roc_vbf},
    but with the $y$-axis being the complement.
    The higher AUC corresponds to a better performing BDT.
    \\
\cline{3-3}
$N_\textrm{var}=V$
    &
    Parameter
    &
    Number of input variables for the ML method.
    \\
\cline{3-3}
$N_\textrm{bit}=N$
    &
    Parameter
    &
    Number of bits used for cut threshold values $N_\textrm{bit,cut}$,
    input variable values $N_\textrm{bit,var}$,
    and BDT output score $N_\textrm{bit,score}$.
    Although the number of bits can set separately for the three types of values,
    the studies presented in this paper set them all to the same value,
    i.e.,
    $N_\textrm{bit}=N_\textrm{bit,cut}=N_\textrm{bit,var}=N_\textrm{bit,score}$.
    \\
\cline{3-3}
Max.\ depth $=D$
    &
    Parameter
    &
    When training the BDT,
    this is the maximum depth for each tree.
    For a maximum depth $D$,
    the most cuts a tree can have is $2^D -1$.
    Training will stop when either the maximum depth is reached,
    or when a small enough fraction of events are in a given node.
    That fraction can generally be set by the user,
    e.g., in \TMVA\ it is called \texttt{MinNodeSize} and has a default value of $5\%$.
    \\
\cline{3-3}
$N_\textrm{tree}=T$
    &
    Parameter
    &
    Number of trees used for the BDT training stage.
    Some of these may be removed by \textsc{Tree Remover}.
    \\
\cline{3-3}
$N_\textrm{final tree}$
    &
    Derived
    &
    Number of final trees after \textsc{Tree Merger} and \textsc{Tree Remover}.
    \\
\cline{3-3}
$N_\textrm{bin}=B$
    &
    Derived
    &
    Number of bins evaluated in the \fwX\ firmware implementation of the BDT.
    Tends to scale with the product of the $N_\textrm{var}$ and $N_\textrm{bit}$.
    \\
\hline
\end{tabular}
}
\end{table}

\section{Details of the ML training}
\label{appendix:train}

The details of the samples and the ML training are given for the two physics problems that are considered in this paper.

\subsection{Electron vs.\ photon}
\label{appendix:train_ey}

Electron and photon samples were obtained from ref.~\cite{Mendeley:2017} described elsewhere \cite{Paganini:2017dwg,Paganini:2017hrr}.

A sample of 400k positrons ($e^{+}$) is considered signal and 400k photons ($\gamma$) is considered as background.
The detector simulation is done with GEANT4~\cite{Agostinelli:2002hh},
a first principles simulation that describes the interaction of each particle with the detector,
to implement a $480\,\textrm{mm}^{3}$ section of a calorimeter \cite{deOliveira:2018lqd} that is inspired by the ATLAS liquid argon calorimeter \cite{Aad:2010ai}. 
The calorimeter has three alternating layers of lead absorber material and active liquid argon. 
The variables derived from measurements in the three layers are denoted with subscripts $0$, $1$, and $2$ in increasing distance from the collision axis.

\begin{figure}[htbp!]
\centering
\includegraphics[width=0.90\textwidth]{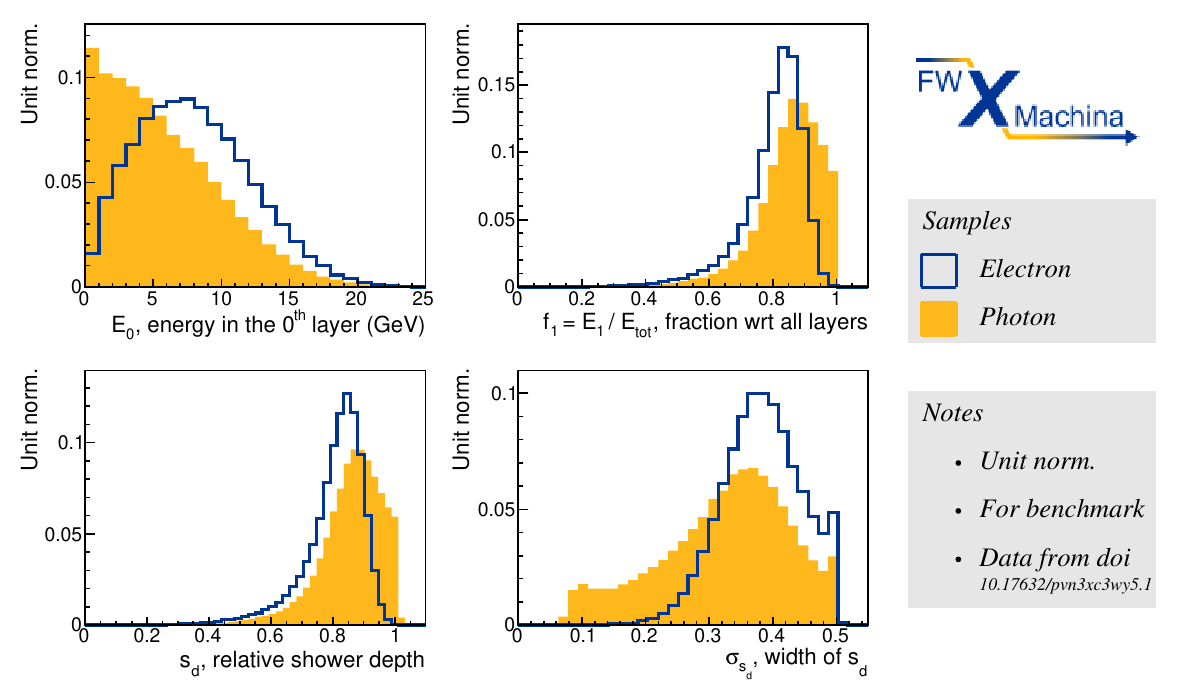}
\caption{
    \label{fig:vars_ey} 
    Input variable distributions for the benchmark BDT to classify electron vs.\ photon.
}
\end{figure}

We consider a total of eight input variables that capture the pattern of energy deposits in the calorimeter:
$E_0$, $E_1$, $f_0$, $f_1$, $s_d$, $\sigma_{s_d}$, $l_d$, and $l_{d^2}$.
These variables are listed in table \ref{table:vars_ey} \cite{deOliveira:2018lqd,lateralDepth2}.
A subset of four of the following variables are used for the benchmark configuration:
$E_0$, $f_1$, $s_d$, and $\sigma_{s_d}$.
The distributions are shown in figure \ref{fig:vars_ey}.

\begin{table}[htbp!]
\caption{
    \label{table:vars_ey}
    List of input variables for the classification of electron vs.\ photon.
    Four input variables are used for the benchmark BDT.
    Up to $8$ input variables are used to study the scaling of firmware performance later in section \ref{sec:firmware}.
    The variables are defined in ref.~\cite{deOliveira:2018lqd,lateralDepth2}.
}
\centering
{\small
\begin{tabular}{p{0.080\textwidth}p{0.40\textwidth}p{0.20\textwidth}p{0.210\textwidth}}
\hline
Variable & Description
    & Used as input variable for benchmark BDT
    & Used to study firmware cost scaling
    \\
\hline
    $E_{0}$          & Energy deposited in the $0^\mathrm{th}$ layer              & yes & BDT with $N_\textrm{var}\ge 1$ \\
    $E_{1}$          & Energy deposited in the $1^\mathrm{st}$ layer              & -   & BDT with $N_\textrm{var}\ge 7$ \\
    $E_{2}$          & Energy deposited in the $2^\mathrm{nd}$ layer              & -   & - \\
    $E_\mathrm{tot}$ & Energy deposited in all three layers                       & -   & - \\
    $l_{d}$          & Depth-weighted energy, $\sum_{i=0}^2 i\cdot E_i$           & -   & BDT with $N_\textrm{var}\ge 8$ \\
    $l_{d^2}$        & Squared-depth-weighted energy, $\sum_{i=0}^2 i^2\cdot E_i$ & -   & BDT with $N_\textrm{var}\ge 6$ \\
    $f_{0}$          & Ratio of $E_{0}$ and $E_\mathrm{tot}$                      & -   & BDT with $N_\textrm{var}\ge 5$ \\
    $f_{1}$          & Ratio of $E_{1}$ and $E_\mathrm{tot}$                      & yes & BDT with $N_\textrm{var}\ge 3$ \\
    $s_d$            & Shower depth, $(E_1+2\,E_2)/E_\mathrm{tot}$                & yes & BDT with $N_\textrm{var}\ge 4$ \\
    $\sigma_{s_d}$   & Shower depth width in standard deviations                  & yes & BDT with $N_\textrm{var}\ge 2$ \\
\hline
\end{tabular}
}
\end{table}

\subsection{VBF Higgs bosons vs.\ multijet process}
\label{appendix:train_vbf}

We generated $10$ million events for each of the samples below \cite{Mendeley:2021}
with MadGraph5$\_$aMC version 2.7.3 \cite{Alwall:2014hca}.
The computation is done at leading order in the strong coupling constant $\alpha_{s}$ with a minimum jet $p_\mathrm{T}$ threshold of $15\,\mbox{GeV}$ and minimum $\Delta R$ between the two jets of 0.4.
The samples are created by event generation that captures the theoretical process followed by smearing that accounts for the experimental effects.

For the signal sample, two subsamples for VBF-produced Higgs are generated.
Each subsample considers a different decay mode of the Higgs boson.
Both subsamples were generated with MadGraph5$\_$aMC version 2.7.3 and showered using Pythia8 \cite{Sjostrand:2014zea}.
The VBF $H\rightarrow\textit{invisible}$ subsample simulates the Higgs boson decays to the neutrino final state, which ends up with two visible ``VBF jets'' and are typically not accompanied by additional jets.
The VBF$\rightarrow 4b$ subsample simulates the Higgs boson decays involving a beyond-the-Standard-Model (BSM) two-Higgs doublet model (2HDM) with an additional spinless particle $a$
\cite{Ellwanger:2005uu,
    Curtin:2013fra,
    Aaboud:2018iil,
    Aad:2020rtv},
which also produces two VBF jets from the production and, in this case, four $b$-quark jets from the decay.
The decay with six total jets is an interesting test case to ensure that the non-VBF jets are misidentified as VBF jets whose quantities are inputs to the BDT.
After event generation,
smearing is done using Delphes version 3.4.2~\cite{Ovyn:2009tx,deFavereau:2013fsa}.
The Delphes model accounts for detector resolution and other experimental effects,
such as pileup.
In summary,
the two VBF signal subsamples are
\begin{itemize}
    \itemsep 0pt
    \item ``VBF $H\rightarrow\textit{invisible}$'' for the Higgs boson decay $H \rightarrow ZZ^{(\ast)} \rightarrow \nu\overline{\nu}\nu\overline{\nu}$ and
    \item ``VBF $H\rightarrow 4b$'' for the Higgs boson decay $H\rightarrow aa \rightarrow b\bar{b}b\bar{b}$,
    where $a$ is a hypothetical spinless particle with mass $m_{a}=50\,\mbox{GeV}$ from a 2HDM model.
\end{itemize}

For the signal sample,
we validate our setup by reproducing the VBF $H\rightarrow\textit{invisible}$ analysis published by the ATLAS Collaboration \cite{Aaboud:2018sfi} using our samples.
We find the same signal efficiency value $0.7\%$ as found by ATLAS for this validation.

For the background sample,
minimum bias pileup events are generated using the minimum bias tune ATLAS AZ corresponding to Pythia pp tune 17 \cite{Aad:2014xaa}.
Showering uses the Delphes CMS with pileup card described in the Delphes documentation.

For both the signal and the background samples,
a minimum bias pileup with mean pileup $\langle\mu\rangle=50$ is specified.
Jets were reconstructed after particle level smearing using the anti-k$_{t}$ algorithm with a radius parameter of $R=0.4$ and a minimum jet threshold of $p_\mathrm{T}=20\,\mbox{GeV}$ \cite{Cacciari:2008gp}.
Pileup jets are identified and removed using the CMS pileup jet removal scheme described elsewhere~\cite{CMS:2013wea}.

The classifier is trained on variables associated with dijet pairs,
with the goal to discriminate VBF jet pairs from background jet pairs.
For the signal,
the training sample was composed of the highest $m_{jj}$ pair from each VBF $H\rightarrow\textit{invisible}$ event;
this is assumed to be the correctly identified VBF jet pair in those events.
For the background,
every possible jet pairing is trained on,
as none of these are ``VBF jets.''
For example,
if a background event has three jets ($j_1$, $j_2$, and $j_3$),
then the three combinations $j_1j_2$, $j_1j_3$, and $j_2j_3$ 
are all considered as background pairs in the training.

\begin{figure}[htbp!]
\centering
\includegraphics[width=0.90\textwidth]{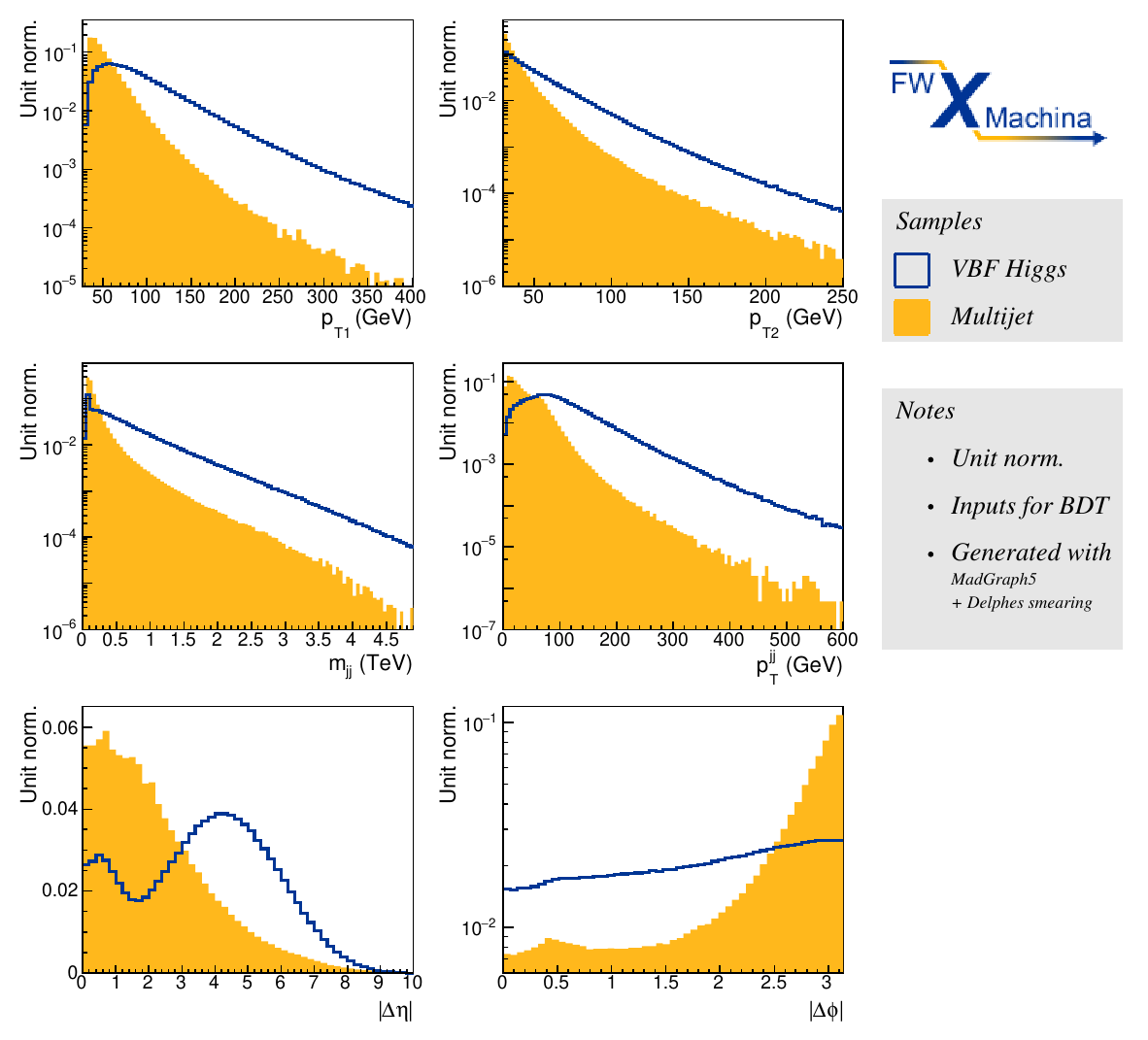}
\caption{
    \label{fig:vars_vbf} 
    Distributions of a few characteristic variables for VBF Higgs and multijet processes.
    The variables are defined in table \ref{table:cuts_vbf}.
}
\end{figure}

For each dijet pair,
$j_1$ is the higher $p_\mathrm{T}$ jet and $j_2$ is the other jet. 
Cylindrical $\eta$-$\phi$ coordinates are used with pseudorapidity $\eta$ and azimuthal angle $\phi$.
The ranges of the angles are $-4.9<\eta<4.9$ and $-\pi < \phi < \pi$, respectively.
These define the input variables listed in table \ref{table:cuts_vbf}.
The distributions are shown in figure \ref{fig:vars_vbf}.

\begin{table}[htbp!]
\caption{
    \label{table:cuts_vbf}
    List of input variables for the classification of the VBF Higgs boson vs.\ multijet process.
    Also given are the ATLAS-inspired cut-based offline thresholds for Run 2 \cite{ATLAS-DAQ-PUB-2019-001} and HL-LHC \cite{CERN-LHCC-2017-020}.
    For Run-2, differences arise with respect to the document when the $m_{jj}$ threshold is quoted as $1100\,\mbox{GeV}$ for  {\footnotesize\texttt{L1 MJJ-500-NFF}};
    we use the $>99\%$ offline efficiency point,
    which is achieved around $m_{jj}>1300\,\mbox{GeV}$.
    for others the offline thresholds are used.
    For HL-LHC, the single-level scheme values are quoted.
    The performance of the cut-based approach using these values is compared the performance to the BDT result in figure~\ref{fig:roc_vbf}.
    The non-optimized (non-opt) configuration includes the five variables from the optimized configuration.
}
\centering
{\small
\begin{tabular}{
    p{0.080\textwidth}
    p{0.210\textwidth}
    p{0.210\textwidth}
    p{0.240\textwidth}
    p{0.120\textwidth}
}
\hline
\makecell[tl]{Input\\ variable}
& Description
& ATLAS Run-2 offline cut \cite{ATLAS-DAQ-PUB-2019-001}, see caption
& ATLAS HL-LHC offline cut \cite{CERN-LHCC-2017-020}, see caption
& Used in BDT \\
\hline
$p_\mathrm{T1}$    & Leading jet $p_\mathrm{T}$        & $>90\,\mbox{GeV}$  & $>75\,\mbox{GeV}$ & - \\
$p_\mathrm{T2}$    & Subleading jet $p_\mathrm{T}$     & $>80\,\mbox{GeV}$  & $>75\,\mbox{GeV}$ & Optimized \\
$p_\mathrm{T12}$   & Sum $p_\mathrm{T1}+p_\mathrm{T2}$ & -                  & -                 & Optimized \\
$|\eta_1|$         & Leading jet $\eta$                & $<3.2$             & -                 & - \\
$|\eta_2|$         & Subleading jet $\eta$             & $<4.9$             & -                 & - \\
$\prod_\eta$       & Product $\eta_1\cdot\eta_2$       & -                  & -                 & Optimized \\
$|\Delta \eta|$    & Separation in $|\eta_2-\eta_1|$   & $>4.0$             & $>2.5$            & - \\
$|\Delta \phi|$    & Separation in $|\phi_2-\phi_1|$   & $<2.0$             & $<2.5$            & non-opt \\
$|\Delta R|$       & $\sqrt{(\Delta\eta)^2+(\Delta\phi)^2}$
                                                       & -                  & -                 & non-opt \\
$m_{jj}$           & Dijet invariant mass              & $>1300\,\mbox{GeV}$& -                 & Optimized \\
$p_T^{jj}$         & Dijet $p_\mathrm{T}$              & -                  & -                 & Optimized \\
\hline
\end{tabular}
}
\end{table}

The BDT was trained with $100$ trees each with a maximum depth of $4$.
Given the target operating point at very low background acceptance,
the background training tree was weighted by a factor of $10^5$ to strongly encourage the classifier to minimize erroneous background acceptance.
The signal and background events were evenly split between training and testing sets.
The BDT setup uses the AdaBoost metric in \TMVA\ with node purity as the output score.

\section{Details of the Nanosecond Optimization}
\label{appendix:nano_opt}

Details of the binning algorithms as well as for the four latter steps of the Nanosecond Optimization are described.

\subsection{Binning algorithms}
\label{appendix:binning}

Two algorithms are described.
The first is binning by bit shift.
The second is binning by look up.
These algorithms are implemented in the BSBE and LUBE \textsc{Bin Engine} firmware,
respectively.

\begin{figure}[htbp!]
\centering
\includegraphics[width=0.95\textwidth]{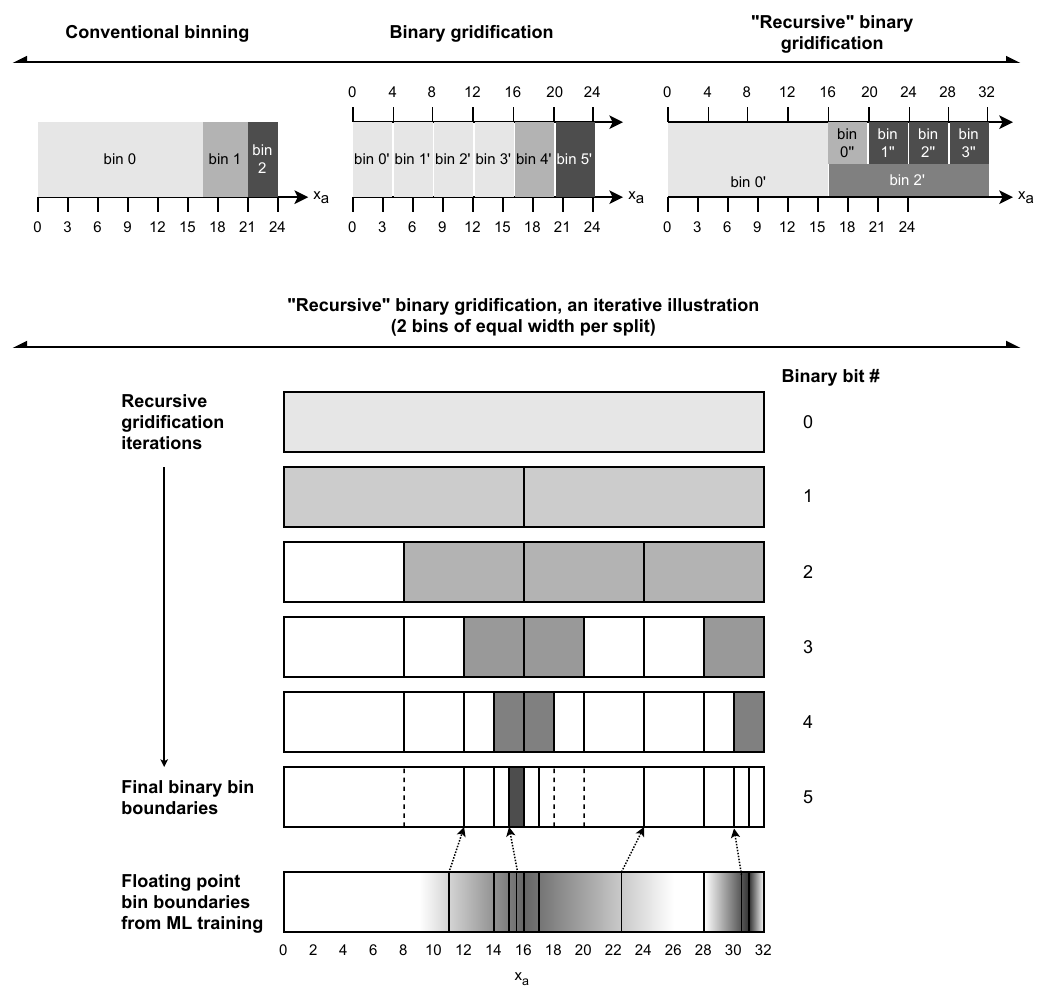}
\caption{
    \label{fig:bsbe_algo} 
    Example of the binning by bit shift algorithm for one variable $x_a$.
    At the top row of figures,
    the result of the training step are three bins in the conventional bin boundaries (left-most).
    If the bin boundaries occur in multiples of a power of two, then bit shifting can be employed to reduce the latency to find the bin index of an event.
    In our example,
    we choose $2^2=4$ as the fixed bin width and arrive at six primed bins (middle).
    However,
    the binary gridification approach potentially incurs a large resource usage in cases where the a large amount variation occurs in a small area.
    In that case, we can employ a ``recursive'' approach (right-most).
    We put quotation marks because the implementation is not recursive since it can be done by combinatoric logic as shown in figure~\ref{fig:bsbe}.
    At the bottom group of figures,
    the ``recursive'' process is expanded upon in $5$ layers.
    The $5^\textrm{th}$ layer is compared to the floating point boundaries from training.
}
\end{figure}

Binning by bit shift achieves fast binning by taking advantage of the binary representation of numbers in hardware.
For example,
figure \ref{fig:bsbe_algo} demonstrates this strategy for one variable $x$ in bit-shift layers.
The goal of this process is to best approximate the floating point bin boundaries from training shown at the bottom of figure \ref{fig:bsbe_algo}.
In the bit shift approach,
cuts are added by splitting the bins from the previous layer in half wherever there is a high density of floating point cuts. 
When evaluated in each layer,
the input value can be said to reside within a bin of a certain index.
Evaluating the bin index for an event in a given layer can be accomplished by integer-dividing by the number of bins in that layer.
An integer division is a standard division operation where the answer is rounded down to the nearest integer.
Since each grid layer is generated by a binary split of the previous,
the number of bins in every grid layer is equal to a power of two.
An integer division by a power of two is equivalent to a bit shift operation by that power.
Now that the bin index in each grid layer has been determined, a deterministic combination of these bin indices is used to determine the final bin index.

To begin setting up the grid layers,
the first layer divides $x_a$ into two ranges $0$--$15$ and $16$--$31$,
as seen in figure \ref{fig:bsbe_algo}.
Likewise,
the second layer subdivides the two ranges into two further bins.
The divisions start to get interesting with the third layer,
where the bins are subdivided \emph{except} for the first bin.
That is because the bin thresholds from training do not contain a division in the first bin of layer $\ell=3$.
This process continues until $\ell=L-1$,
depending on the user's specifications.
The arrows are meant to indicate that the binary boundaries will not be as exact as the boundaries from floating point precision from the training.
We call the process of splitting each layer as ``gridification.''
Since we repeat the process for each layer,
we call it recursive gridification,
but we note that it is recursive only on the software side of \fwX,
not the firmware side.
In the firmware representation the method employs combinatoric logic without recursion.
The gridification process creates a binary tree for each variable with cuts defined by powers of two.

Binning by look up uses a set of thresholds that were introduced and discussed in the text.
The benefit of this approach,
compared to by bit shift discussed above,
is that it is less focused on the numbers of bits used for the input variables and the cut thresholds.
This gives more flexibility for the user at the cost of of latency.

\subsection{\textsc{Score Finder}}

Once the bin boundaries are determined \textsc{Score Finder} associates an output score to it.
The output score for an event is the combined score using a set of weights $\{W_t\}$ for tree $t$,
which,
e.g.,
for the AdaBoost boost method,
is
\begin{equation}
    \label{eq:weighted_sum}
    O 
    = \frac{O_0\,W_0 + O_1\,W_1 + \cdots + O_{T-1}\,W_{T-1}}{W_0 + W_1 + \cdots + W_{T-1}} 
    = \sum_{t=0}^{T-1} O_t\,w_t,
\end{equation}
where $w_t = W_t\big/\sum_{t'=0}^{T-1} W_{t'}$ is the normalized boost weight for tree $t$.\footnote{
    The combined score formula for some of the other boost methods,
    such as GradBoost,
    uses different metrics.
    The treatment of the combined score for GradBoost is discussed in appendix \ref{appendix:score_normalize} (table \ref{table:metric} and figure \ref{fig:piecewise}).
}

For each bin in tree $t$,
the score $O_t$ and weight $W_t$ is associated to that bin.
The next step is to normalize these scores.

\subsection{\textsc{Score Normalizer}}
\label{appendix:score_normalize}

The \textsc{Score Finder} defined the boost-weight normalization in eq.\ (\ref{eq:weighted_sum}).
Now we discuss three aspects of normalizing the output scores by \textsc{Score Normalizer}.
First, we revisit the aspect of using less precision than floating point.
Second, we discuss choice of a score range because it plays an important role in the tree remover step discussed in the next subsection.
Third, we describe the piece-wise approximation of a continuous function.

\subsubsection*{Conversion of floating point to bit integer}

The initial values of the cut thresholds and the output scores from the ML training are floating point precision.
But in the FPGA it is beneficial to represent data as $N$-bit integers,
so convert the data to $N$-bit integers according to the user's specification.
The values are between $0$ and $2^N-1$ for $N$-bit integers,
e.g.,
$0$ to $1023$ for $10$-bit integers.

The conversion of cut threshold values from floating point $c_\textrm{float}$ to bit integers $c_\textrm{int}$ is straight forward using a linear function $f$:
\begin{equation}
    \label{eq:cuts}
    c_\textrm{int} = f(c_\textrm{float}) = \left\lfloor\frac{c_\textrm{float}-c_\textrm{min}}{c_\textrm{max}-c_\textrm{min}}
    \cdot\left( 2^N - 1 \right)\right\rfloor,
\end{equation}
where the ``floor'' operator is used in the final step.
The difference between the maximum and minimum values in the denominator represents the range of possible values.
By default for cut thresholds,
this is the range of the training data,
though this range can be set by the user.
The flexibility allows the user to ensure that the incoming data matches the convention used in the classifier.

The conversion of output scores from floating point $v_\textrm{float}$ to bit integers $v_\textrm{int}$ needs to be treated with more care so that the operation respects addition,
i.e., whether $f(v_1+v_2)=f(v_1)+f(v_2)$ holds.
It does not in the presence of a normalized boost weight $w$,
so the conversion is modified as follows with transformation $g$.
We take $v_\textrm{min}=0$ and $v_\textrm{max}=1$.
\begin{equation}
    \label{eq:boost}
    v_\textrm{int} = g(v_\textrm{float}) = \left\lfloor w\cdot v_\textrm{float}
    \cdot\left( 2^N - 1 \right)\right\rfloor,
\end{equation}
For output scores ranging from $-1$ to $1$,
this function provides an integer value ranging from $-(2^N +1)$ to $2^N -1$.
For output scores ranging from $0$ to $1$, this provides an integer value ranging from $0$ to $2^N-1$.

Let us check the addition for the purity metric with $O=S/(S+B)$ ranging from $0$ to $1$.
For example,
two sets of purity values $O_{\textrm{float},1}=\{0, 0.3\}$ and $O_{\textrm{float},2}=\{0.1, 0.7\}$
are converted into 10-bits 
with normalized boost weights $w_1=0.2$ and $w_2=0.8$.
Then,
using eq.\ (\ref{eq:boost}),
$O_{\textrm{int},1}=\{\lfloor 0.2\cdot 0\cdot 1023\rfloor, \lfloor 0.2\cdot 0.3\cdot 1023\rfloor\}=\{0, 61\}$
and
$O_{\textrm{int},2}=\{\lfloor 0.8\cdot 0.1\cdot 1023\rfloor, \lfloor 0.8\cdot 0.7\cdot 1023\rfloor\}=\{81, 572\}$.
The sum of the bit integers are $O_{\textrm{int},1}+O_{\textrm{int},2}=\{81,633\}$.
To compare,
the weighted sum of the floating point values are
$O_{\textrm{float},12}=\{0.2\cdot0+0.8\cdot0.1, 0.2\cdot0.3+0.8\cdot0.7\}=\{0.08, 0.62\}$.
The conversion of the sum,
with $w=1$ since there is no weight,
is
$g(O_{\textrm{float},12})
=\{\lfloor 0.16\cdot 1023\rfloor, \lfloor 0.62\cdot1023\rfloor\}
=\{81,634\}$,
which is almost,
but not exactly,
what we obtained before.

This example illustrates an important aspect of the function $g$ under addition.
Because of the floor function,
the difference between the sum of the conversion $g(w_1\cdot v_1)+g(w_2\cdot v_2)$ and the conversion of the sum $g(w_1\cdot v_1+w_2\cdot v_2)$ can give rise to a \emph{difference of up to one bit} due to rounding.
However,
the ability to maintain addition is an important aspect of speeding up the firmware.
This is because \fwX\ does all of the conversions to provide the set of $\{g(w_i\cdot v_i)\}$ values to the FPGA.
The FPGA simply has to add the scores without needing to apply any transformations.

\subsubsection*{Output score range}
ML classifiers often report an output score,
generally ranging from either $0$ to $1$ or from $-1$ to $1$,
with the lower value indicating a background-like event and the higher indicating a signal-like event.
Here we introduce three possible metrics for AdaBoost and one for GradBoost that may be chosen to optimize classifier and firmware performance.

For AdaBoost,
we first consider the purity output score.
Each tree provides a response value that ranges from $0$ to $1$ based on the ratio $S/(S+B)$ in the terminal node of a tree,
where $S$ and $B$ represent the number of signal and background events,
respectively.
The response is converted to $N$-bit integers by eq.~(\ref{eq:boost}) to yield
a value in the range from $0$ to $2^N - 1$.
Second,
we consider the ``adjusted purity.''
We define it as $(S-B)/(S+B)$,
which is just a twice the purity shifted by unity,
i.e.,
$2\cdot S/(S+B) - 1$.
The response ranges from $-1$ to $1$,
which is converted to $-2^N+1$ to $2^N-1$ after the bit integer transformation as was done for the purity response.
This adjusted value can be advantageous as it assigns indeterminate events a score near zero, which is beneficial for \textsc{Tree Remover} described in appendix~\ref{appendix:tree_remove}.
Lastly,
we consider the ``yes/no leaf.''
The response is determined for a terminal node under the following conditions.
If $S/(S+B)\geq\gamma$,
then the terminal node is assigned a score of $1$.
Otherwise,
the terminal node is assigned a score of $-1$.
The value of $\gamma$ is a user-configurable parameter;
in this paper,
it is set to $0.5$ when yes/no leaf is used.
For all three boosting algorithms mentioned above,
the combined output score is the weighted sum of the response values from each of the trees.
The weights are the normalized boost weights,
i.e.,
$O_\textrm{combined} = \sum_t O_t\cdot w_t$,
where $O_t$ is the response value for tree $t$.
Because the combination process is a simple sum,
each component of the sum is pre-computed to be hard-coded into firmware.
The firmware then executes the sum without having to multiply or convert.

For GradBoost,
an unbounded response value is provided for each terminal node.
Rather than taking the weighted average of the scores as is done for AdaBoost,
the combined output score is defined by $\tanh{(\sum_t O_t)}$.
Since $\tanh{(x_1 + x_2)} \neq \tanh{(x_1)} + \tanh{(x_2)}$,
this operation cannot be pre-computed as is done for AdaBoost.
Therefore,
we optimize it with a piece-wise approximation in firmware,
which is discussed next.

A summary of the above metrics is given in table~\ref{table:metric}.

\begin{table}[htbp!]
\caption{
    \label{table:metric}
    BDT score definitions for various output boost metrics.
    The quantity $S$ ($B$) represents the number of signal (background) events in the terminal node.
    Adjusted purity is defined in this paper.
    One version of yes/no is given in this table;
    in principle, the threshold for the comparison can be changed.
    The range for yes/no is continuous because of weighted score sums in eq.\ (\ref{eq:boost}).
    The scores from the trees in the forest are combined by using the normalized boost weight $w_t$.
}
\centering
{\small
\begin{tabular}{
    p{0.15\textwidth}
    p{0.25\textwidth}
    p{0.25\textwidth}
    p{0.22\textwidth}
}
\hline
Algorithm 
    & Response value per tree
    & Combined score
    & Combined score range\\
\hline
Purity
    & $O_t=S/(S+B)$
    & $O_\textrm{purity}=\sum_t O_t\cdot w_t$
    & $\phantom{-}0 \le O_\textrm{purity} \le 1$
\\
Adjusted purity
    & $O_t=2\cdot S/(S+B) - 1$
    & $O_\textrm{purity'}=\sum_t O_t\cdot w_t$
    & $-1 \le O_\textrm{purity'} \le 1$
\\
Yes/no leaf
    & $O_t=1$ if $S{>}B$, else $-1$
    & $O_\textrm{yes/no}=\sum_t O_t\cdot w_t$
    & $-1 \le O_\textrm{yes/no} \le 1$
\\
Gradient
    & $O_t$ provided by the algo.
    & $O_\textrm{grad}=\tanh\left(\sum_t O_t\right)$
    & $-1 \le O_\textrm{grad} \le 1$
\\
\hline
\end{tabular}
}
\end{table}

\subsubsection*{\boldmath Piece-wise approximation of $\tanh$}

\begin{figure}[htbp!]
\centering
\includegraphics[width=0.45\textwidth]{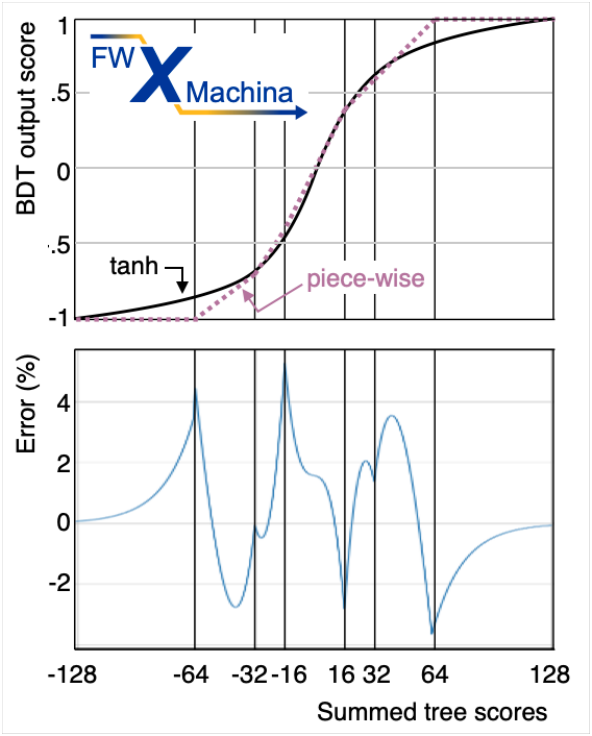}
\caption{
    \label{fig:piecewise} 
    Piece-wise approximation.
    The $\tanh$ function is described by a set of linear functions for the score conversion in the gradient boost method.
    The top plot shows the approximation in dotted red line superimposed on the solid black line.
    The bottom plot shows the percent error,
    which is accurate to $5\%$ over the range.
    The vertical lines denote the bin boundaries,
    which are in powers of two so that binning can be achieved by bit shifting.
}
\end{figure}

The $\tanh$ function can be approximated in a piece-wise manner \cite{piecewise} as shown in figure \ref{fig:piecewise}.
Our piece-wise implementation is to divide the $x$-axis range in powers of two,
so that binning can be done by bit shifting.
The plots show that this seven-piece piece-wise function approximates the exact value to $5\%$ over the range.
The piece-wise function becomes a look up table in the FPGA.

\subsection{\textsc{Tree Remover}}
\label{appendix:tree_remove}

Thus far in Nanosecond Optimization,
we have merged trees and converted the floating point values to bit integers.
Now we take a closer look at each merged tree in the forest to remove the ones that have no impact on physics performance.

At this stage
the tree remover receives the normalized score in each bin of a flattened tree $t$,
i.e.,
$O_t\cdot w_t$.
Since the sum of the normalized scores represents the final output score---%
with the exception of gradient boost that needs the $\tanh$ applied at the end---%
it is more convenient to think of the product
\begin{equation}
    \alpha_t = O_t\cdot w_t
    \label{eq:alpha}
\end{equation}
so that $O_\textrm{combined}=\sum_t \alpha_t$.

The key point is that $\alpha_t$ is small whenever $w_t$ is small independent of the value for $O_t$.
This is because $O_t$ is bounded to be between $0$ and $1$ or $-1$ to $1$.
So for trees with low relative boost weights $w_t$,
any floating point value $\alpha_t$ is also very small.
After the bit integer conversion step,
such an $\alpha_t$ will be rounded down to a bit integer output of $0$.
Since it can be known in advance that some trees will have only $\alpha_t$ values of $0$,
the tree will have no impact on the final output score so we remove the tree from the forest.

The choice of BDT boost metric is important because $0$ has a special meaning in the purity scheme,
and tree remover does not work there.
The $0$ values represent background-like events,
so removing them would influence the event distributions.
The adjusted purity is developed to counter this effect so that background-like events are near $-1$ and signal-like events are near $1$.
This $-1$ to $1$ range also holds for yes/no as well as gradient boosting.

In the latter three schemes the $0$ score represents the perfect ambiguity of signal or background.
The practical way of thinking about this is to see that the normalized weights $w_t$ range from $0$ to $1$.
The weights near $0$ are the least important and those near $1$ is the most important.

Removing entire trees saves the FPGA clock ticks associated with binning for these trees and pointlessly adding $0$ in the final summation.
We have developed three methods for tree remover that are listed in table \ref{table:tree_remove}.

\begin{table}[!hbtp]
\caption{
    \label{table:tree_remove}
    \textsc{Tree Remover} (TR) methods for Nanosecond Optimization.
    The default choice used in this paper is the first option.
}
\centering
{\small
\begin{tabular}{
    p{0.3\textwidth}
    p{0.6\textwidth}
}
\hline
\textsc{Tree Remover} Method
    & Description
    \\
\hline
\textsc{TR by Zero Suppression} 
    &
    All possible output scores associated with each tree are considered.
    If every score is $0$ in a tree,
    then that tree is removed.
    \\
\cline{2-2}
\textsc{TR by Bin Threshold} 
    &
    All possible scores associated with each tree are considered.
    If a user-defined fraction $f_\textrm{score}$ of them have a less than a user-defined impact $f_\textrm{impact}$,
    then the tree is removed.
    For example,
    if $f_\textrm{score}=90\%$ of the bins contributes to $f_\textrm{impact}<3\%$ of the output score,
    then the tree is removed.
    \\
\cline{2-2}
\textsc{TR by Boost Threshold} 
    & 
    If the boost-weight falls below a user-defined threshold fraction $f_\textrm{thr}$ of the average boost weight $f_\textrm{avg}$,
    the tree is removed.
    For example,
    if the average of 100 trees' boost weights is $f_\textrm{avg}=0.1$,
    the user-defined threshold fraction is set to $f_\textrm{thr}=0.02$,
    then the tree with a boost weight of $0.0001$ is removed since it is less than $f_\textrm{avg}\cdot f_\textrm{thr}$.
    \\
\hline
\end{tabular}
}
\end{table}

\subsection{\textsc{Cut Eraser}}
\label{appendix:cut_erase}

The \textsc{Cut Eraser} removes bin boundaries if they do not affect physics performance under the user specified threshold.

When trees are merged,
we may be left with what we can call ``useless'' cuts: 
cuts which have an impact in each tree independently,
but,
upon merging,
are rendered redundant.
The simplest case of this would be a cut location shared by two or more trees.
Moreover, there are cases in which cuts are not redundant, but may still be irrelevant.

For a cut to be a candidate for removal,
we must first show that removing it has a negligible impact on the output of the classifier.
For a flattened tree,
this means that there is no,
or minimal,
change in outcome between an event falling in the bin on the ``left'' or the ``right'' of that cut.
Therefore,
to determine which cuts can be removed,
we scan across every cut,
at each one comparing the value held by every bin bordering the cut on the ``left'' to the value held by its counterpart on the ``right.''
If for every one of these comparisons across the cut, the difference $\Delta p$ between the bin on the left and right is very small,
then the cut is removed.
This process is performed for each variable.
Two examples are considered.

\begin{figure}[htbp!]
\centering
\includegraphics[width=0.45\textwidth]{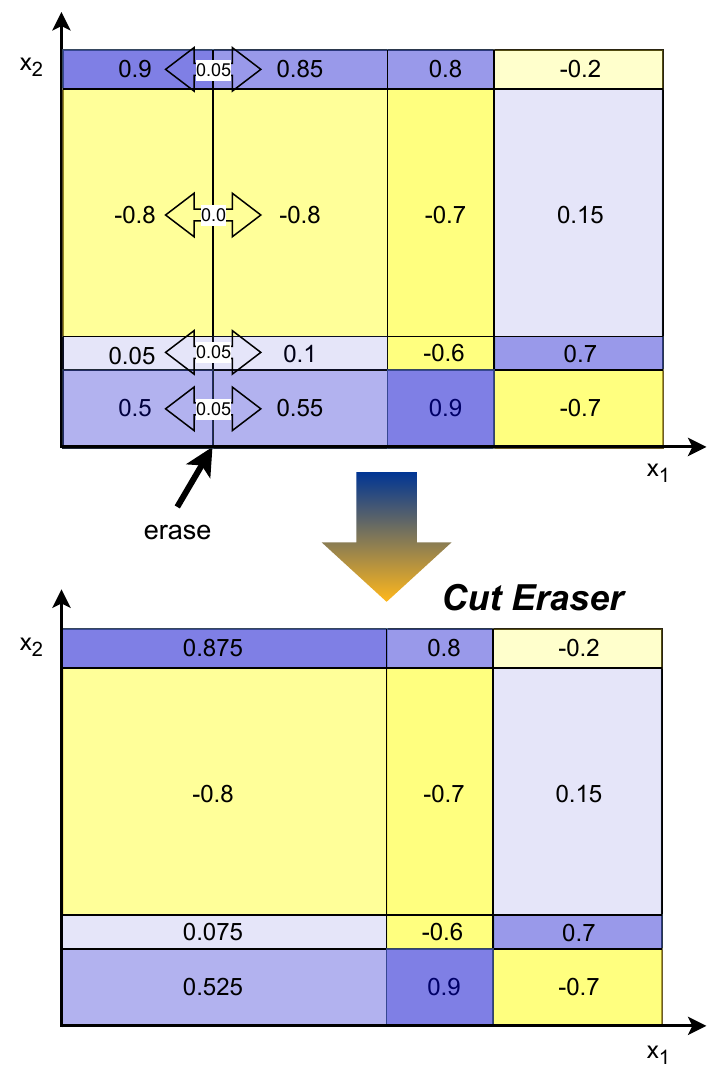}
\caption{
    \label{fig:cut_eraser_ex}
    Example of the \textsc{Cut Eraser} algorithm result for two input variables with mock data.
    The number in each region represents the BDT output score that corresponds to that bin.
    The top plot shows the scores in each bin prior to the cut eraser step.
    Small arrows indicate the differences in the scores of adjacent bins,
    which is $\le 0.05$.
    If the threshold to remove cuts is $\le 0.05$ then the vertical boundary is removed and the result is the bottom plot.
}
\end{figure}

The first example is given for a two-variable case in figure \ref{fig:cut_eraser_ex}.
In this graphic the difference between the bins in $x_1$ for one bin boundary is $\le 0.05$,
which is the example threshold used,
so that boundary is removed after the cut eraser step.

\begin{figure}[htbp!]
\centering
\includegraphics[width=0.85\textwidth]{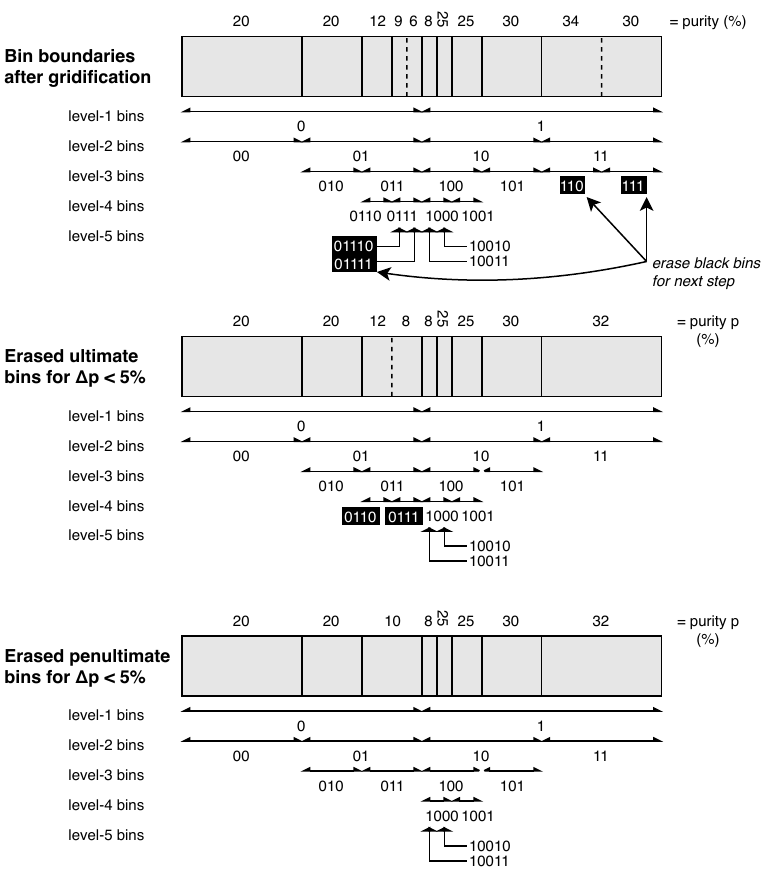}
\caption{
    \label{fig:cut_eraser_algo} 
    Example of the \textsc{Cut Eraser} iterative process for one input variables with mock data.
    The removal occurs in two iterations represented by the top two diagrams,
    followed by the resulting diagram at the bottom.
    The top-most diagram corresponds to the final bin boundaries in the layer $\ell=5$ of figure \ref{fig:bsbe_algo}.
    If the threshold to remove cuts is $< 0.05$ then the vertical boundary is removed and the result is the bottom plot.
    This iterative procedure occurs during the optimization stage in software and not in the resulting firmware.
}
\end{figure}

The second example is given for a one-variable case for the method of binary bit shift binning in figure \ref{fig:cut_eraser_algo}.
The figure continues the example of figure \ref{fig:bsbe_algo},
which uses the BSBE approach.
We have to be more careful with BSBE because it uses all of the cut thresholds after conversion to bit integers.
Therefore,
the cut eraser cannot simply scan left to right as in the first example since it may inadvertently remove a cut in the middle.
So rather than scanning for bin boundaries remove,
we go from the deepest bit levels to the first.
In our figure,
the algorithm starts at level-5 to remove the boundary between bin 01110 and 01111 and level-4 to remove the boundary between bin 110 and 111.
In the second iteration,
the algorithm removes the boundary between bin 0110 and 0111 in layer-4.
All of the bins that are merged are marked with black background.
Finally, the third diagram is the result of the cut eraser.

\section{Details of the firmware design}

\subsection{Cut-based implementation}
\label{appendix:fw_cutbased}

The cut-based classification is a pass-fail algorithm that compares each variable in a set to a set of thresholds,
e.g.,
$x_1 < c_1$ and $c_2 < x_2 < c_2'$ for variables $x_i$ and threshold values $c_i$.
Conceptually this corresponds to the simplest BDT with one decision tree.
However,
because the number of comparisons per variables is limited to the form given in the example above,
the deep-level decomposition of the two \textsc{Bin Engines} of the previous subsection is superfluous.

Instead,
two comparisons are made for each variable using comparators.
The \texttt{AND} of all of the results of the comparators gives the output score of the cut-based approach.
The value of $1$ designates that the event passed the criteria and $0$ designates that the event failed the criteria.
We note that the output score needs to be transformed from the purity formulation to the adjusted purity variant (see table \ref{table:metric}) in order to perform the \textsc{Tree Remover} step.

The cut-based method is implemented in firmware independently of the BDT implementation.
Two if-else statements are used in each dimension to see 
\begin{itemize}
    \itemsep 0pt
    \item If the variable is greater than the lower bound and
    \item If the variable is less than the upper bound.
\end{itemize}
We note that our requirement is $c<x<c'$ rather than $c<x\le c'$,
$c\le x<c'$,
or $c\le x\le c'$,
as the choice is arbitrary and the user can specify the bounds as needed.
These comparisons can be performed for each variable in parallel on an FPGA,
which allow for very low latency values.
As with BDT,
the cut-based method can be converted to $N$-bit integers in advance for faster processing in firmware.

\subsection{Firmware verification and validation}
\label{appendix:valid}

\begin{figure}[htbp!]
\centering
\includegraphics[width=0.95\textwidth]{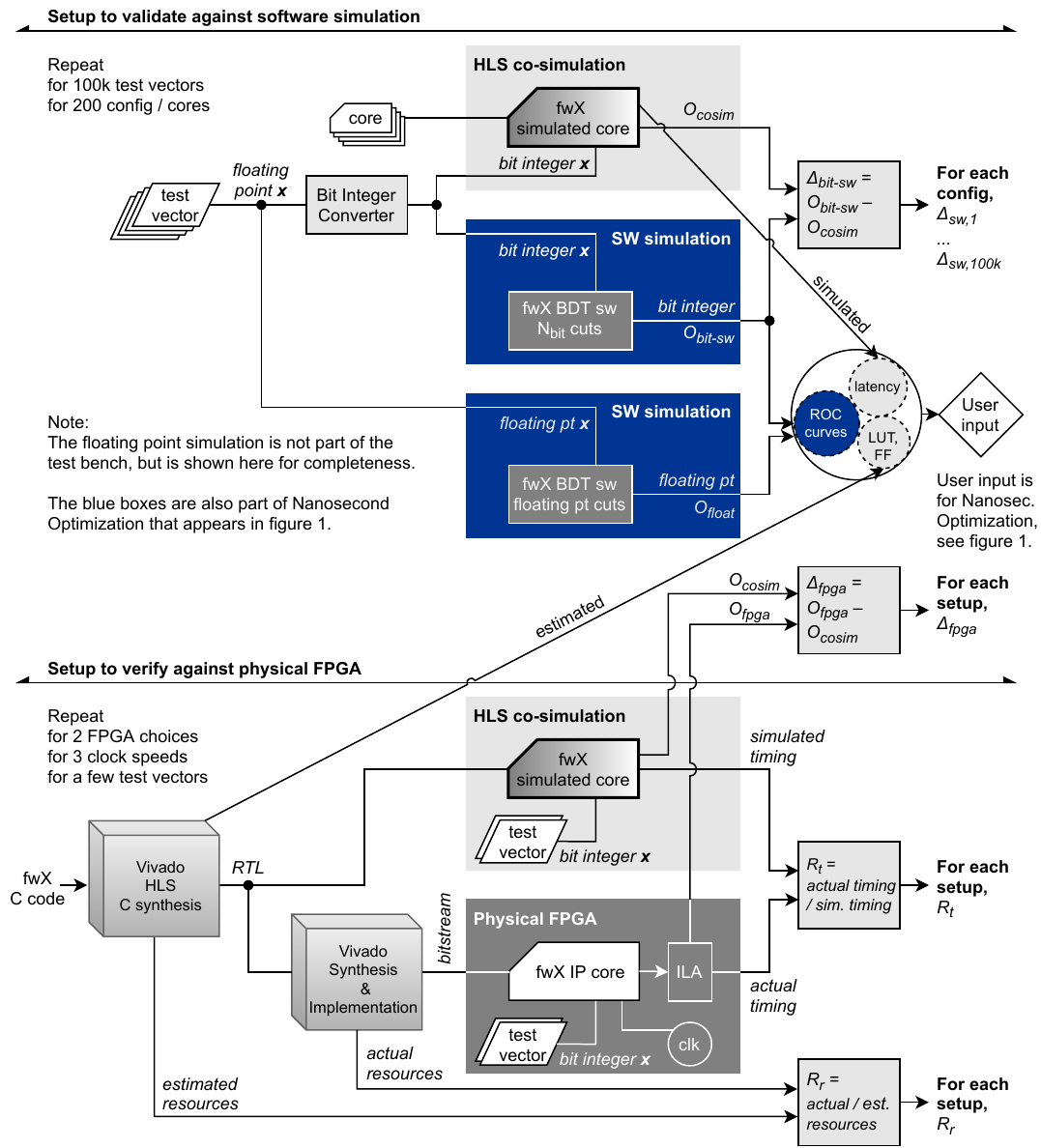}
\caption{
    \label{fig:testbench}
    Diagrams of the test bench.
    The top setup validates the BDT output score
    $O$ with respect to software simulation for a large number of test vectors and configurations.
    The bottom setup verifies the co-simulation results against the physical FPGA.
    The ``core'' in the diagram is the output of Vivado HLS.
    In both setups,
    the HLS co-simulation that is being tested is the same;
    two instances are drawn for figure clarity.
    The dark blue boxes are part of Nanosecond Optimization (figure \ref{fig:fwx}),
    but is included here for completeness.
}
\end{figure}

The test bench for the verification and validation is given in the bottom and top diagrams,
respectively,
of figure \ref{fig:testbench}.\footnote{
    We briefly mention the role of C-simulation.
    C-simulation is a test bench by using the c++ compiler.
    This provides functional testing of the algorithm.
    However,
    being purely based on the c++ compiler,
    it does not offer any insights in the hardware-level information.
    Therefore, C-simulation can be run without synthesis or co-simulation.
    In contrast,
    co-simulation generates an RTL test bench used to verify the RTL output of the HLS C-synthesis.
    This internal test bench is executed and its results are compared to the C-simulation results.
    Therefore, co-simulation necessarily requires C-simulation. Further testing may also be performed using Xilinx Vivado and a custom RTL test bench. This testing may be useful after the generated IP is integrated into a larger design. Because this type of testing is dependent on the use case of the IP, it is not automatically generated by the \fwX\ design flow.
}

\begin{figure}[htbp!]
\centering
\includegraphics[width=0.65\textwidth]{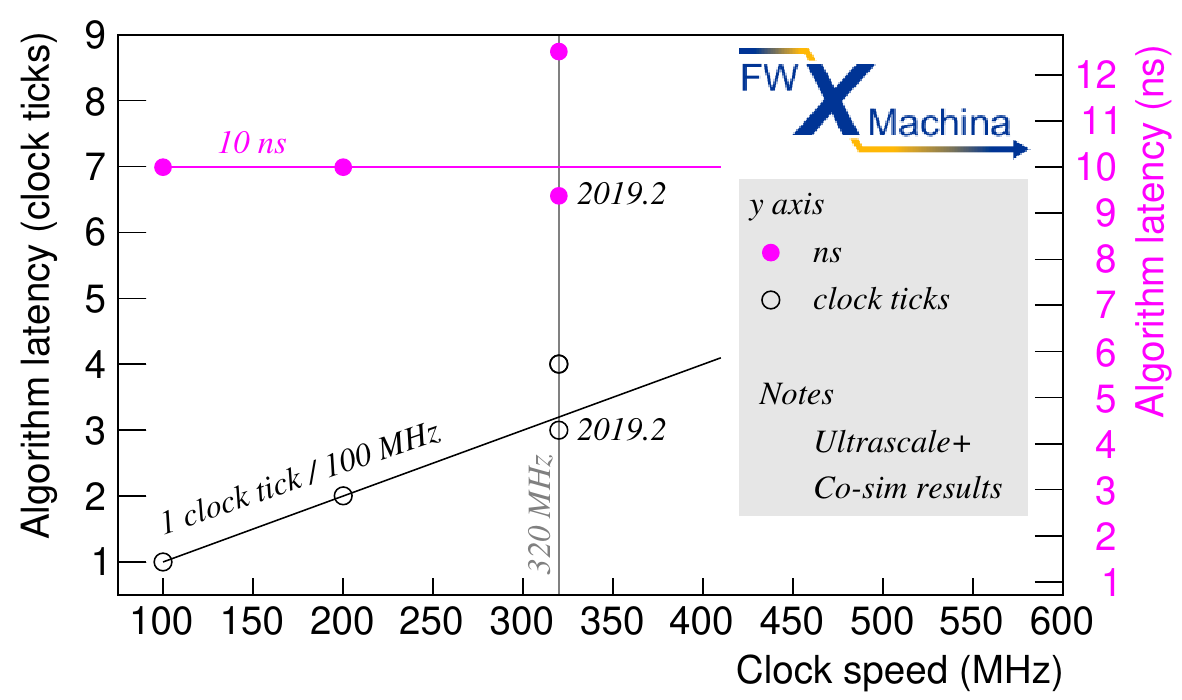}
\caption{
    \label{fig:latency_clock}
    Latency result vs.\ clock speed for the benchmark configuration.
    The number of clock ticks is given on the $y$-axis on the left hand side and the time elapsed in nanoseconds on the right hand side.
    The Vivado HLS version 2018.2 is used for the data points except for the ones for which  version 2019.2 is noted next to the symbol.
    The two data points at $320\,\textrm{MHz}$ show the difference between Vivado HLS versions for the same clock speed.
}
\end{figure}

To further verify the generated design,
we compare the results from firmware simulation with the performance of the IP running in hardware on a physical FPGA.
For the comparison two Xilinx FPGAs are chosen:
\begin{itemize}
    \itemsep 0pt
    \item Virtex UltraScale+ FPGA VCU118 Evaluation Kit (EK-U1-VCU118-G), our benchmark,
    \item Artix-7 FPGA on Zynq-7020 System on Chip (SoC), a smaller FPGA.
\end{itemize}
The Artix-7 has $53\,200$ LUT,
$106\,400$ FF,
$140$ BRAM of 36k blocks,
no URAM,
$220$ DSP.
The clock speed on the Artix-7 is slowed to $100\,\textrm{MHz}$ to meet timing.
Since this change in timing was done on the FPGA and in the \fwX\ design flow,
the result comparisons reflect the accuracy of the simulated results.
Test are done with varying clock speed and Vivado versions.
No change in latency is seen (figure \ref{fig:latency_clock}).

The results for FPGA resource utilization shows better-than-expected results.
Interestingly,
the LUT usage is much lower at $13\%$ of the simulated result.
The FF and BRAM usage is also lower at $68\%$ and $50\%$,
respectively.
This indicates that Vivado is able to significantly optimize the design that is generated by HLS.
It is important to note that the resource usage generated by HLS is estimated.
HLS may generate designs that are not optimized efficiently for implementation.
In that case,
when the design is implemented,
Vivado will remove redundant or unnecessary hardware,
often resulting in lower resource utilization.

\begin{table}[!hbtp]
\caption{
    \label{table:fpga_cost}
    FPGA cost verification against physical FPGA.
    Comparison of the FPGA cost using the bitstream on the FPGA (actual), simulated timing using co-simulation and estimated resources using Vivado HLS (estimated).
    The actual-to-estimated ratios are given as $R$.
    Two FPGA choices and three clock speeds are considered;
    the $320\,\textrm{MHz}$ group of columns represent the benchmark clock.
    For all other configurable parameters,
    see table \ref{table:benchmark}.
    The timing values are reported in units of clock ticks.
    The Xilinx Vivado version used for the actual and estimated columns are noted.
    For the ratios, ``$1$'' signifies no difference.
}
\centering
{\small
\begin{tabular}{
    p{0.10\textwidth}
    p{0.07\textwidth}
    @{/ }
    p{0.06\textwidth}
    @{= }
    p{0.035\textwidth}
    p{0.06\textwidth}
    @{/ }
    p{0.06\textwidth}
    @{= }
    p{0.04\textwidth}
    p{0.06\textwidth}
    @{/ }
    p{0.06\textwidth}
    @{= }
    p{0.04\textwidth}
    p{0.06\textwidth}
    @{/ }
    p{0.06\textwidth}
    @{= }
    p{0.035\textwidth}
}
\hline
Parameter   & \multicolumn{9}{l}{Benchmark FPGA}
            & \multicolumn{3}{l}{Smaller FPGA} \\
\hline
\multicolumn{2}{l}{FPGA setup} \\
\quad Family
            & \multicolumn{9}{l}{Xilinx Virtex Ultrascale+\dotfill}
            & \multicolumn{3}{l}{Xilinx Artix-7\dotfill}
            \\
\quad Model
            & \multicolumn{9}{l}{xcvu9p-flga2104-2L-e\dotfill}
            & \multicolumn{3}{l}{xc7z020-clg400-1\dotfill}
            \\
\quad Speed & \multicolumn{3}{l}{$320\,\textrm{MHz}$\dotfill}
            & \multicolumn{3}{l}{$200\,\textrm{MHz}$\dotfill}
            & \multicolumn{3}{l}{$100\,\textrm{MHz}$\dotfill}
            & \multicolumn{3}{l}{$100\,\textrm{MHz}$\dotfill}
            \\
\quad Period& \multicolumn{3}{l}{$3.125\,\textrm{ns}$\dotfill}
            & \multicolumn{3}{l}{$5\,\textrm{ns}$\dotfill}
            & \multicolumn{3}{l}{$10\,\textrm{ns}$\dotfill}
            & \multicolumn{3}{l}{$10\,\textrm{ns}$\dotfill}
            \\
\quad Vivado& \multicolumn{3}{l}{\footnotesize{2019.2~~\,2019.2}}
            & \multicolumn{3}{l}{\footnotesize{2018.2~~\,2018.2}}
            & \multicolumn{3}{l}{\footnotesize{2018.2~~\,2018.2}}
            & \multicolumn{3}{l}{\footnotesize{2019.1~~\,2019.2}}
            \\
\hline
FPGA cost
            & actual & estim. & $R$
            & actual & estim. & $R$
            & actual & estim. & $R$
            & actual & estim. & $R$
            \\
\quad Latency
            & $3$       & $3$       & $1$
            & $2$       & $2$       & $1$
            & $1$       & $1$       & $1$
            & $4$       & $4$       & $1$
            \\
\quad Interval
            & $1$       & $1$       & $1$
            & $1$       & $1$       & $1$
            & $1$       & $1$       & $1$
            & $1$       & $1$       & $1$
            \\
\quad LUT   & $717$     & $1903$    & $0.4$
            & $717$     & $4015$    & $0.2$
            & $717$     & $4007$    & $0.2$
            & $482$     & $3572$    & $0.1$
            \\
\quad FF    & $147$     & $138$     & $1.1$
            & $147$     & $113$     & $1.3$
            & $147$     & $2$       & $73.$
            & $245$     & $362$     & $0.7$
            \\
\quad BRAM  & $5.5$     & $8$       & $0.7$
            & $5.5$     & $15$      & $0.4$
            & $5.5$     & $15$      & $0.4$
            & $7.5$     & $15$      & $0.5$
            \\
\quad URAM  & $0$       & $0$       & $1$
            & $0$       & $0$       & $1$
            & $0$       & $0$       & $1$
            & NA        & NA        & NA
            \\
\quad DSP   & $2$       & $0$       & NA
            & $2$       & $2$       & $1$
            & $2$       & $2$       & $1$
            & $2$       & $2$       & $1$
            \\
\hline
\end{tabular}
}
\end{table}

The physics performance is validated by considering the pairs of input variable values $\vec{x}$ and the corresponding BDT output score $O_i$.
The output of the HLS co-simulation and the software simulation is called $O_\textrm{cosim}$ and $O_\textrm{bit-sw}$,
respectively.
The differences $\Delta_\textrm{bit-sw}=O_\textrm{est}-O_\textrm{bit-sw}$ are computed to validate the result.
No difference is seen.

As the firmware design operates on bit integers,
the floating point input vector is modified to simulate bit integers for the software evaluation.
This modification occurs in the \textsc{Bit Integer Converter} that serves as an interface between the input test vectors and the \fwX\ module to compute the BDT output score.
The diagram of the dataflow is shown at the top diagram of figure \ref{fig:testbench}.

The test bench for this project is generated in c++ along with the design itself and can be used to evaluate the design on an algorithmic level.
Using {C/RTL} co-simulation,
the synthesized can be evaluated with the same test bench.
\fwX\ generates a unique test bench for the user with every design that it produces,
so the user may validate their own design.

We considered over $200$ different configurations each corresponding to a firmware simulated ``core,''
which is the RTL-level output of HLS C-synthesis.
For each core $10^5$ input data vectors are fed into the test bench.

The output of the HLS co-simulation and software simulation are compared.
We note that the wrapper code converts the test vectors from floating point values to the corresponding bit integer values.
In all of the tests we saw no difference between the firmware output and the bit-integer-simulated software output.

\section{Physics performance AUC results from parameters scans}
\label{appendix:perf_auc}

\begin{figure}[htbp!]
\centering
\includegraphics[width=0.5\textwidth]{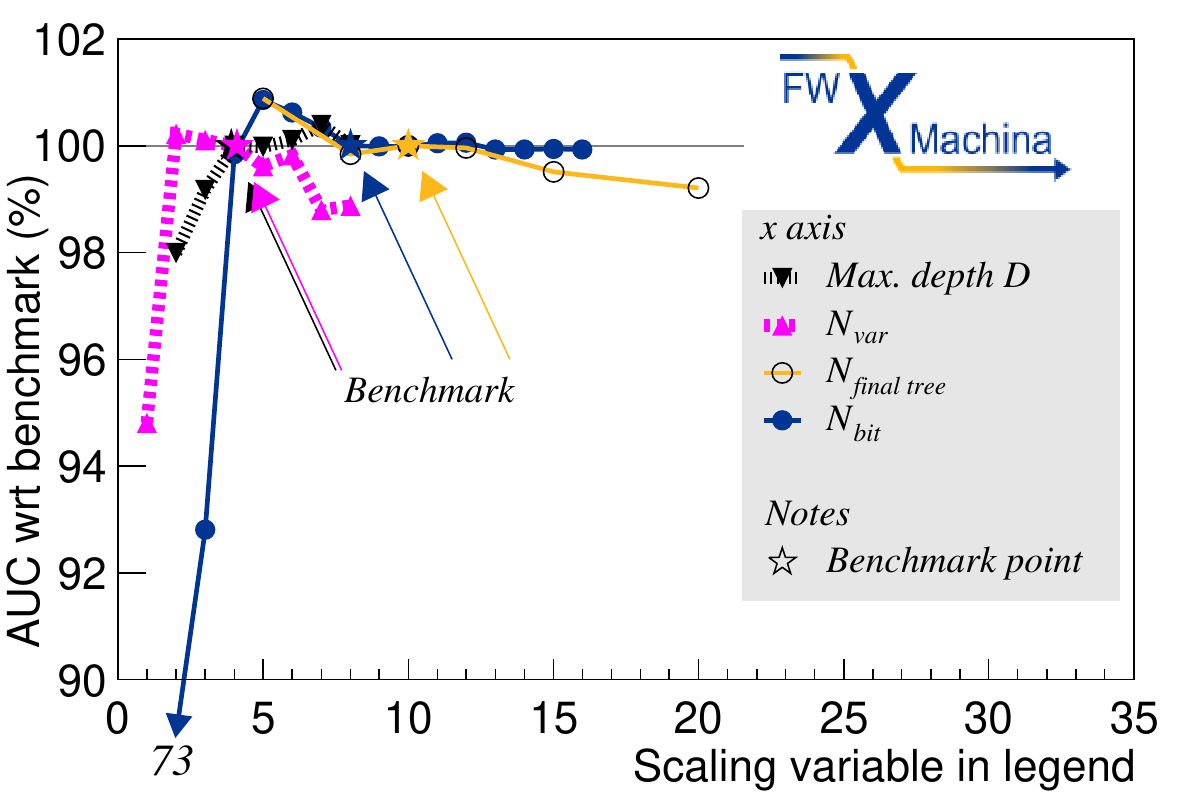}
\caption{
    \label{fig:auc_ey}
    Physics performance measured with area under the ROC curve (AUC) relative to the benchmark point.
    Two trends are seen:
    the rapid fall off below the visible threshold value and the flat with fluctuations above that threshold value.
    The point off the chart has the value $73$ for $N_\textrm{bit}=2$.
    The lines connecting the symbols serve as a visual guide and do not represent interpolations.
    The benchmark point (table \ref{table:benchmark}) is marked by a star.
}
\end{figure}

We report the \emph{relative} AUC in percentage with respect to our benchmark point where we scan the four parameters.
The results are plotted in figure \ref{fig:auc_ey}.
A turn-on effect is observed,
wherein it rapidly plummets below some threshold value while it remains relatively flat above that value.
We note that our benchmark point is well above the threshold values for each parameter.

\section{Study of the number of jet pairs for VBF Higgs vs.\ multijet}
\label{appendix:njet}

The classifier is trained on the highest $m_{jj}$ reconstructed jet pair with VBF $H\rightarrow\textit{invisible}$ as signal and the multijet process as background.
For testing the training step,
the final BDT score for the event is the highest BDT score from all possible jet pairs in the event.

The number of jets ($N_\textrm{jet}$) and the number of jet pairs ($N_\textrm{jet pairs}$ or $J$) per event depend on two external factors.
The first factor is the set of user-defined criteria,
such as the minimum $p_\textrm{T}$ threshold for each jet in the event or a minimum $m_{jj}$ to be considered.
Furthermore, we assume that the list of jet pairs can be sorted based on the $m_{jj}$ such that any pair below a user-defined threshold can be discarded.
The second factor is the user-defined maximum value of $J_\textrm{max}$ to consider,
where if the number of jet pairs $J$ exceeds $J_\textrm{max}$ those with the lowest $m_{jj}$ can be ignored.

For ROC curve comparisons,
we compare the scenario with BDT output scores from all jet pairs, i.e., no limit on $J_\textrm{max}$ against BDT output scores for $J_\textrm{max}=3$ highest $m_{jj}$ pairs.
We also considered a study of $6$ jet pairs,
but negligible difference is seen with respect to the result for $3$.

\begin{figure}[htbp!]
\centering
\includegraphics[width=0.5\textwidth]{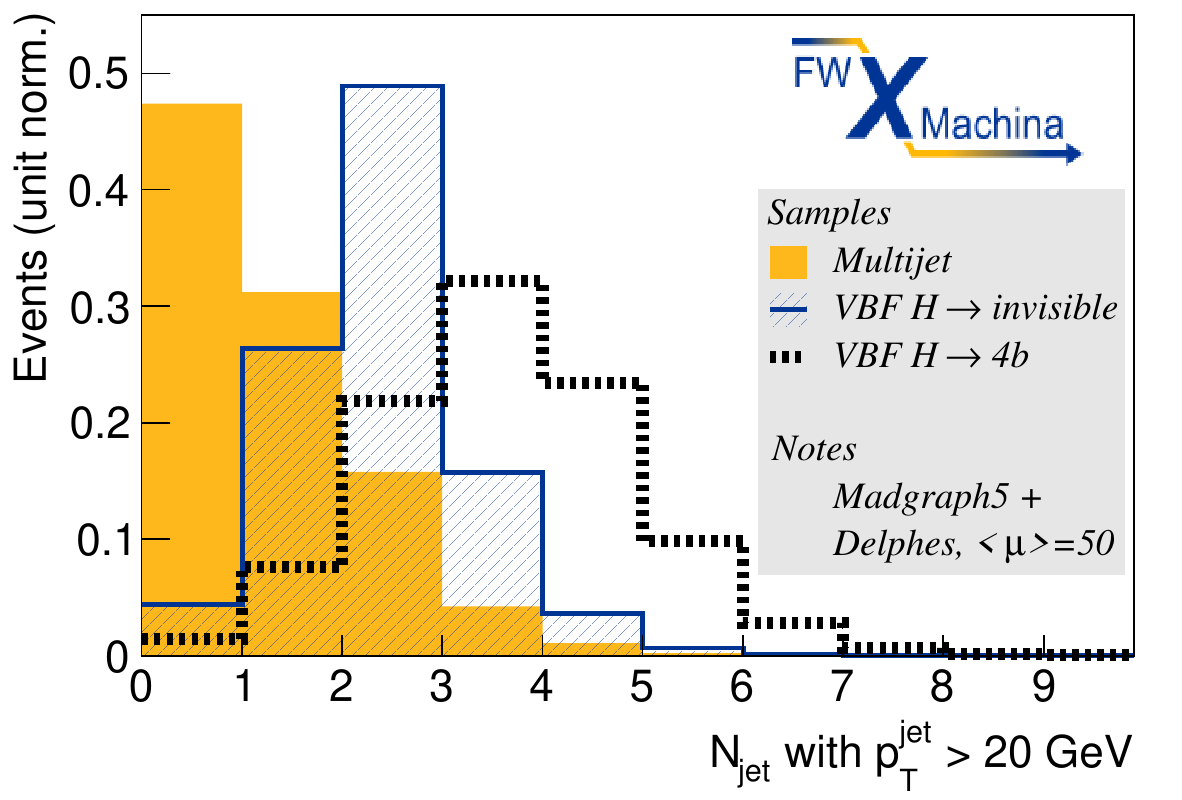}%
\includegraphics[width=0.5\textwidth]{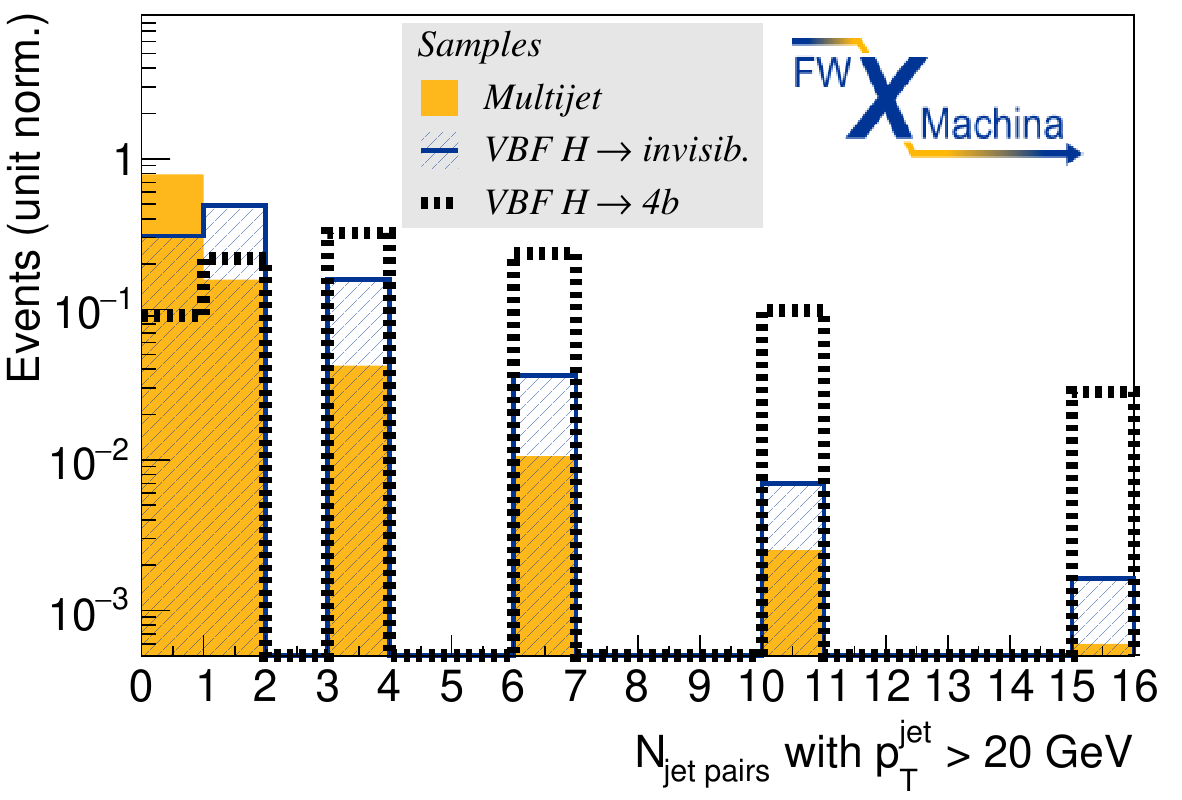}
\caption{
    \label{fig:njet_njetpair}
    Number of jets with $p_\textrm{T}>20\,\textrm{GeV}$ (left) and
    number of jet pairs (right).
    The variable $N_\textrm{jet-pairs}$ is referred to as $J$ in the text.
    The negligible number of events in the overflow bin with $N_\textrm{jet-pairs}\ge 16$ for $4b$ containing $<1\%$ of the events is omitted from the plot.
}
\end{figure}

The distribution of $N_\textrm{jet}$ with $p_\textrm{T}>20\,\textrm{GeV}$ is shown for the background and the signal samples in the left plot of figure \ref{fig:njet_njetpair}.
The number of jet pairs $J$ is the quantity of interest for performance,
which is $J=N_\textrm{jet}(N_\textrm{jet}-1)/2$.
The distribution is shown on the right plot of figure \ref{fig:njet_njetpair}.
As events with no jet pairs are not considered,
only events with $J\ge 1$ are considered.

\section{Benchmark configuration run on hls4ml/Conifer code}
\label{appendix:compare}

We compare the result using \fwX\ with the result using hls4ml/Conifer,
both using the same \TMVA\ output file corresponding to the setup of table \ref{table:benchmark}.
Both results use the same BDT trained by \TMVA\ for our benchmark case with the configuration parameters noted in table \ref{table:compare},
i.e.,
the same physics problem of binary classification,
the same number of variables,
the same number of trees,
and the same clock speed.
We obtained the most recent git commit of hls4ml/Conifer at the time of writing this paper \cite{conifer-github}.\footnote{
    \label{footnote:conifer}
    The results for hls4ml/Conifer has been updated since the initial draft of this paper
    [\href{http://arxiv.org/abs/2104.03408v1}{2104.03408v1}].
    The revised draft [\href{http://arxiv.org/abs/2104.03408v2}{2104.03408v2}]
    reports the results from C synthesis using 
    the latest version of the package \cite{conifer-github} and the associated setup file to execute the code through correspondence with one of the authors of the code.
    This paper updates the comparison using the results from the RTL synthesis and implementation.
}

We note that the application of Nanosecond Optimization (section \ref{sec:nano_opt}) and the use of \textsc{Bin Engines} (sections \ref{sec:fw_bsbe} and \ref{sec:fw_lube}) are unique to the design of \fwX\ and cannot be applied for the inputs to and the configuration of hls4ml/Conifer.
Another notable difference is in the choice of representation of variable precision.
We use $N$-bit integers, 
where $N$ is the number of bits to represent the values ranging from $0$ to $2^N-1$,
whereas hls4ml/Conifer uses fixed point precision.

For the benchmark point,
we use $8$-bit integers.
For hls4ml/Conifer,
rather than using the code's default choice of $\langle 18,8\rangle$---%
which represents $18$ total bits to represent the values with
$8$ of them representing the part before the decimal point and $10$ of them representing the part after the decimal point---%
we make a choice to approximate our $8$-bit integer precision.
For this consideration,
we inspect the distributions of figure \ref{fig:vars_ey} and the ranges of the plotted variables.
For example,
$E_0$ ranges from $0$ to $25\,\textrm{GeV}$ while $f_1$ ranges from $0$ to $1$.
The bins used for these plots can be reasonably approximated by a choice of $32$ bins,
as $16$ bins seem too coarse to separate the two samples.
Therefore,
$5$ bits are chosen for the part above the decimal point to span $0$ to $25\,\textrm{GeV}$ with bins that are approximately $0.8\,\textrm{GeV}$,
while $5$ bits are also chosen for the part below the decimal point to span $0$ to $1$ with bins that are approximately $0.03$ wide.
While a more thorough study is needed,
perhaps by comparing the physics performance scanning over the precision parameter space,
we use the above reasoning to set hls4ml/Conifer's precision to $\langle 10,5\rangle$.

While we make the above choice for the result using hls4ml/Conifer,
we note the following observation regarding variable precision.
If hls4ml/Conifer is run using only $8$ total bits as $\langle 8,n\rangle$ with $n<8$,
for instance,
it would sacrifice accuracy when storing the cuts of certain variables.
And depending on the split between integer and decimal bits,
it could significantly impact the accuracy of the output scores.

Results from two tests are reported.
First,
the result from the hls4ml/Conifer code using ap\_ufixed$\langle 10,5\rangle$,
which represents unsigned values,
is $15$ clock ticks for latency and $1$ clock tick for interval.
Second,
the result using ap\_fixed$\langle 18,8\rangle$,
the default out-of-the-box setting,
gives similar results.
The results for \fwX\ and hls4ml/Conifer using ap\_ufixed$\langle 10,5\rangle$ are summarized in table \ref{table:compare}.
The table shows the FPGA cost from the results using RTL synthesis and implementation (called ``actual'' in table \ref{table:benchmark}),
where the design is programmed on the FPGA via the bitstream file.
The hls4ml/Conifer-to-\fwX\ ratio for the latency is about five while
the ratios for LUT and FF are about five and ten, respectively.
The interval is the same while RAM and DSP usage is zero or negligible for both projects.

Regarding the hls4ml/Conifer results that we report,
it is likely that the FPGA cost from the out-of-the-box code can be reduced,
as no problem-specific optimization was done on our side.

\begin{table}[hbtp!]
\caption{
    \label{table:compare}
    Comparison of the \fwX\ result (from table \ref{table:benchmark})
    and one from hls4ml/Conifer using the out-of-the-box code \cite{conifer-github}.
    The test uses the benchmark configuration listed in table \ref{table:benchmark},
    i.e.,
    the same BDT configuration from \TMVA\ trained on the same training samples.
    The first two groups of rows show the parameters for ML training, FPGA, and firmware.
    The bottom group of rows shows the FPGA cost.
    See the text for the details regarding the choice of ap\_ufixed$\langle 10,5\rangle$,
    as well as the results using other choices for the precision.
    }
\centering
{\small
\begin{tabular}{
    p{0.24\textwidth}
    p{0.245\textwidth}
    p{0.245\textwidth}
    p{0.15\textwidth}
}
    \hline
Parameter
                    & \fwXmachina
                    & hls4ml/Conifer
                    & Comments
                    \\
\hline
\multicolumn{2}{l}{ML training setup} \\
    \quad Training software
                    & \TMVA
                    & \TMVA
                    & same
                    \\
    \quad Physics problem
                    & electron vs.\ photon
                    & electron vs.\ photon
                    & same
                    \\
    \quad Training samples
                    & from ref.\ \cite{Mendeley:2017}
                    & from ref.\ \cite{Mendeley:2017}
                    & same
                    \\
        \quad No.\ of event classes
                    & $2$
                    & $2$
                    & same
                    \\
    \quad No.\ of training trees
                    & $100$
                    & $100$
                    & same
                    \\
    \quad Max.\ depth
                    & $4$
                    & $4$
                    & same
                    \\
    \quad No.\ of input variables
                    & $4$
                    & $4$
                    & See figure \ref{fig:vars_ey} \\
    \quad Other \TMVA\ parameters
                    & \TMVA\ defaults
                    & \TMVA\ defaults
                    & same \\
    \quad Nanosec.\ Optimization
                    &  Flattened 
                    \& merged to $10$ final trees,
                    without \textsc{Tree Remover} or \textsc{Cut Eraser}
                    & N/A
                    & Unique to \texttt{\textsc{fwX}} \\
\hline
\multicolumn{2}{l}{FPGA and firmware setup} \\
    \quad Chip family
                    & Xilinx Virtex Ultrascale+
                    & Xilinx Virtex Ultrascale+
                    & same
                    \\
    \quad Chip model& xcvu9p-flga2104-2L-e
                    & xcvu9p-flga2104-2L-e
                    & same
                    \\
    \quad Vivado HLS version
                    & 2019.2
                    & 2019.2
                    & same
                    \\
    \quad Clock speed, period
                    & $320\,\textrm{MHz}$, $3.125\,\textrm{ns}$
                    & $320\,\textrm{MHz}$, $3.125\,\textrm{ns}$
                    & same
                    \\
    \quad Precision
                    & ap\_int$\langle 8\rangle$
                    & ap\_ufixed$\langle 10,5 \rangle$
                    & See text
                    \\
    \quad \textsc{Bin Engine}
                    & BSBE
                    & N/A
                    & Unique to \texttt{\textsc{fwX}} \\
\hline
\multicolumn{2}{l}{FPGA cost} \\
    \multicolumn{4}{l}{\quad Actual timing values and resource usage by RTL synthesis and implementation} \\
    \qquad Latency  & $3$ clock ticks, $9.375\,\textrm{ns}$
                    & $15$ clock ticks, $46.875\,\textrm{ns}$
                    & -
                    \\    
    \qquad Interval & $1$ clock ticks, $3.125\,\textrm{ns}$
                    & $1$ clock tick, $3.125\,\textrm{ns}$
                    & same
                    \\

    \qquad LUT      & $717$, $0.06\%$ of total
                    & $3834$, $0.3\%$ of total
                    & -
                    \\
    \qquad FF       & $147$, $< 0.01\%$ of total
                    & $1966$, $< 0.1\%$ of total
                    & -
                    \\
    \qquad BRAM 18k & $5.5$, $0.1\%$ of total
                    & $0$
                    & -
                    \\
    \qquad URAM     & $0$
                    & $0$
                    & same
                    \\
    \qquad DSP      & $2$, $0.03\%$ of total
                    & $0$
                    & -
                    \\
\hline
\end{tabular}
}
\end{table}


\FloatBarrier 
 
\addcontentsline{toc}{section}{References}
\bibliographystyle{JHEP}

\end{document}